\documentclass[12pt]{article}
\usepackage[latin1]{inputenc}
\usepackage{natbib}
\usepackage[a4paper,left=2.5cm,right=2.5cm,top=2cm,bottom=1.5cm,footnotesep=.75cm,headheight=13.6pt]{geometry}
\usepackage{setspace}
\usepackage{amsmath, mathtools, amsthm}
\usepackage{graphicx}
\usepackage{caption}
\usepackage{subcaption}
\usepackage{epigraph}
\usepackage{xcolor}
\usepackage{authblk}

\title{Temperature in the Iberian Peninsula: Trend, seasonality, and heterogeneity\footnote{Partial financial support from ``la Caixa'' Foundation, grant LCF/PR/SR20/52550012-Climate change and economic challenges for the Spanish society, and from the Spanish Government grant PID2022-139614NB-C22 (MINECO/FEDER) is gratefully acknowledged. We are also grateful to Jouni Helske for his help with some of the codes used in this paper and to participants at the 69th International Symposium of Forecasters (July, 2022), Rome-Waseda Time Series Symposium (October, 2022), 16th International Conference on Computational and Financial Econometrics (December, 2022), 6th International Conference on Econometrics and Statistics (August, 2023), and 35th Mexican National Statistical Forum (September, 2023). The usual disclaimer applies.}}

\author[1]{C. Vladimir Rodr\'iguez-Caballero\thanks{Corresponding author. E-mail: vladimir.rodriguez@itam.mx}}
\author[2]{Esther Ruiz}
\affil[1]{Department of Statistics, ITAM (Mexico)}
\affil[2]{Department of Statistics, Universidad Carlos III de Madrid (Spain)}

\date{}

\begin{document}

\maketitle

\begin{abstract}
In this paper, we propose fitting unobserved component models to represent the dynamic evolution of bivariate systems of centre and log-range temperatures obtained monthly from minimum/maximum temperatures observed at a given location. In doing so, the centre and log-range temperature are decomposed into potentially stochastic trends, seasonal, and transitory components. Since our model encompasses deterministic trends and seasonal components as limiting cases, we contribute to the debate on whether stochastic or deterministic components better represent the trend and seasonal components. The methodology is implemented to centre and log-range temperature observed in four locations in the Iberian Peninsula, namely, Barcelona, Coru\~{n}a, Madrid, and Seville. We show that, at each location, the centre temperature can be represented by a smooth integrated random walk with time-varying slope, while a stochastic level better represents the log-range. We also show that centre and log-range temperature are unrelated. The methodology is then extended to simultaneously model centre and log-range temperature observed at several locations in the Iberian Peninsula. We fit a multi-level dynamic factor model to extract potential commonalities among centre (log-range) temperature while also allowing for heterogeneity in different areas in the Iberian Peninsula. We show that, although the commonality in trends of average temperature is considerable, the regional components are also relevant.
\end{abstract}



\setcounter{page}{1}

\newpage

\doublespacing

\epigraph{What we find more difficult to talk about is our deep dissatisfaction with the ability of our models to inform society about the pace of warming, how this warming plays out regionally, and what it implies for the likelihood of surprises.}{\textit{Palmer and Stevens (2019)}}

\section{Introduction}

Climate change can be defined as the variation in the joint probability distribution that describes the state of the atmosphere, oceans, and freshwater, including ice; see Hsiang and Kopp (2018). According to the latest fifth and sixth assessment reports of the International Panel for Climate Change (IPCC, 2014, 2023), one of the most critical aspects of climate change is global warming, described by the evolving distribution of temperature.

The stand of the literature using econometric models to describe the evolution of temperature focuses on average temperature, which is obviously an important characteristic of the distribution.\footnote{Although we focus on econometric models, there is extensive literature based on deterministic climate models; see, for example, Diebold and Rudenbush (2022b) for a comparison between both types of models in the context of measuring ice volume in the Arctic.} However, average temperature alone is not enough to reflect complicated climate variations. In an early paper, Katz and Brown (1992) show that the frequency of extreme weather events is relatively more dependent on any changes in the variability than on the mean of temperature, with this sensitivity being relatively greater the more extreme the event. Consequently, policy makers should not rely on scenarios of future temperature involving only changes in means but should also consider evolving variability of temperature; see, for example, Phella, Gabriel and Martins (2024), who fit quantile regressions with time-varying parameters to generate scenarios for extreme temperatures and their relation with anthropogenic emissions.

Our paper contributes to the important literature on econometric time series modelling of temperatures in two main ways. The first contribution is methodological. Instead of analysing only average temperature, we use the rich joint information contained in the minimum/maximum temperature interval. By doing so, we add to the information of the average temperature, represented by the centre of the interval, the information about the log-range between the maximum and minimum temperature and, consequently, information associated to extreme temperatures. 

 
Furthermore, the proposed model decomposes the centre (average) and the range (variability) of temperature into potentially stochastic trends, seasonal and transitory components. This specification encompasses deterministic trends as a limiting case and, consequently, allows to distinguish between deterministic and stochastic behaviour in the long-run trend of average temperature. It also allows us to model the strong pattern of seasonality associated with temperature and determine whether it changes over time.   

In particular, at a given location, we propose using a state space representation of the potentially non-stationary and seasonal system of centre/log-range temperature; see Harvey (1997) about the advantages of using unobserved components models instead of the widely used VAR models based on differencing and cointegration and Pretis and Hendry (2013), who suggest using a state-space approach due to the conflicting results often encountered when unit root and cointegration tests are used in the context of monthly data. The unobserved component models framework allows us to decide about the nature of the trends and seasonal components of the centre and log-range temperature, using the Kalman filter and smoothing (KFS) algorithms to extract the components together with measures of their uncertainty. Structural time series models have been implemented before in the context of average temperatures by, for example, Bloomfield (1992), Woodward and Gray (1993), Visser and Molenaar (1995), Zheng and Bahsher (1999) and Stern and Kaufmann (2000) for annual data, and by Good \textit{et al}. (2007) and Proietti and Hillebrand (2015), and Hillebrand and Proietti (2017), for monthly data. However, instead of analysing point average temperatures, we propose unobserved component models for interval temperatures. Our methodology is closely related with the smoothing procedure proposed by Maia and Carvalho (2011), who fit two independent models to the mid-points and ranges of the intervals and estimate the smoothing parameters by minimising the interval sum of one-step-ahead forecast errors; see Harvey and Jaeger (1993), who show that the steady-state Kalman filter for the local linear trend model takes the form of Holt's recursions with suitably defined smoothing constants. However, Maia and Carvalho (2011) do not consider the seasonal component. Furthermore, instead of fitting two independent models for the centre and log-range, we propose modelling them jointly to allow for potential relations between them. 

On top of modelling the system of centre and log-range temperature at a given location, we also consider modelling them simultaneously at a large number of locations. To deal with the heterogeneity in the joint evolution of centre and log-range temperature observed at various locations, we propose using a multi-level Dynamic Factor Model (ML-DFM), which is estimated using the KFS algorithms. The ML-DFM allows to represent the joint evolution of a large system of time series of centre/log-range temperature, assuming that some of their trends are common to all locations while others may be common to some subsets of locations. Consequently, ML-DFM can represent some commonality in the evolution of temperatures while allowing, at the same time, for some idiosyncratic movements that explain the heterogeneity often observed when looking at temperatures at different locations. Marotta and Mumtaz (2023) also propose a ML-DFM for annual average temperature. However, our proposal considers not only modelling the average temperature but also its variability at several locations. Furthermore, we propose using data-driven procedures to identify regions with specific regional common factors. 

Our second contribution is the empirical analysis of monthly 
centre and log-range temperatures observed at 68 locations spread throughout the Iberian Peninsula over nearly a century from January 1930 to December 2020. This analysis is particularly interesting because of the severe heat waves experienced in Southern Europe in recent years; see, for example, Kew \textit{et al}. (2019). After analysing minimum/maximum temperatures separately in four selected locations, namely Barcelona, Coru\~{n}a, Madrid and Seville, we show that the main characteristics of the centre temperature in each of these four cities can be represented by a smooth stochastic trend. The fact that the slope of the trend changes over time can explain why some authors find upward trends while others conclude about the hiatus in warming (global mean temperature has not risen significantly over the last two decades). Furthermore, we observe some heterogeneity in the trends of center temperatures in these four locations, with different slopes of the trend at the end of the observation period, December 2020. In particular, the slope of the trend is larger in Barcelona and smaller in Seville. We also show that the seasonal component of centre temperature is stochastic. When looking at the log-range, we conclude that it can be represented by a stochastically evolving level with stochastic seasonality. As expected, at the end of the sample period, the log-range is clearly larger in the two locations in the interior of the Iberian Peninsula, Madrid and Seville, and smaller in the two coastal cities, Barcelona and Coru\~{n}a. Another important conclusion from the empirical analysis of these four cities is that the centre and log-range temperature are not correlated in each location, allowing us to model both characteristics separately.

After fitting separate models to the system of centre/log-range temperature in each city, we consider a multivariate ML-DFM for the simultaneous analysis of such series of temperatures observed at 68 locations in the Iberian Peninsula. First, we show that the commonality in the long-run trend of average temperature is strong, with the regional factors being somehow weaker, although also relevant for understanding the long-run behaviour of average temperature. However, when looking at log-range temperature, we find that global and regional factors are equally important.

The rest of this paper is organized as follows. In Section \ref{section:survey}, we briefly survey the literature using econometric models to describe the dynamic evolution of temperatures and the still open debates. Section \ref{section:methodology} describes the methodology proposed in this paper to model centre and log-range temperature. In particular, we describe how using state space models and the Kalman filter allows us to estimate trends and seasonal components at a given location. We also describe the ML-DFM and the implementation of KFS to extract the common factors. Section \ref{section:data} describes the data and the main empirical characteristics of the 
centre/log-range systems at Barcelona, Coru\~{n}a, Madrid and Seville. Section \ref{section:univariate} fits state-space models to extract trends and seasonal components in each of these four locations of the Iberian Peninsula. Section \ref{section:multivariate} is devoted to the simultaneous analysis of centre and log-range temperature in a large number of locations spread over the Iberian Peninsula using ML-DFMs. Finally, Section \ref{section:conclusions} concludes the paper.

\section{Econometric modelling of temperature: Open debates}
\label{section:survey}

When describing global warming,  the interest is on the evolution of the distribution of temperature either at a given location or jointly at several locations. However, regardless of whether the analysis is carried out by looking at temperature observed at a particular location or simultaneously at several locations, the stand literature focuses on analysing the evolution of average temperature, i.e. the central tendency of the distribution, based on a large variety of statistical and econometric approaches, periods and frequency of observation, and locations. Many of these studies find an upward trend in average temperature. Increasing trends are found by, for example, Deng and Fu (2019), who compare several methods for extracting cycles from daily temperatures, and Barbosa, Scotto and Alonso (2011), who analyse daily temperatures in several locations in Central Europe. These latter authors use quantile regressions to fit linear trends to the quantiles of average daily temperature. Note that, by doing this, they are analysing the cross-sectional distribution of average temperature but not the distribution of temperature itself. Similarly, Scotto, Barbosa and Alonso (2011) use Extreme Value Theory (EVT) to analyse the extremes of the distribution of daily average temperature in Europe and the spatial distribution of extreme events; see also Wang \textit{et al}. (2021), who use Generalized EVT to look for fingerprints on temperatures. A similar approach is considered by Gadea Rivas and Gonzalo (2020, 2022), who use regression analysis to estimate quantiles of the full cross-sectional distribution of average temperatures or the distribution of average daily temperatures within a given year, and by 
 Chang \textit{et al}. (2020), who construct densities of global average temperature in the Southern Hemisphere (measured as anomalies concerning average temperatures over the base period 1961-90) and conclude that there is persistence in the mean, while non-stationarity is less evident in the variance. Kruse-Becher (2023) also considers cross-sections of temperatures to model their averages and ranges with adaptive methods that are robust under structural changes. Finally, Marotta and Mumtaz (2023) analyse the presence of global and regional common factors in annual temperatures observed in 160 countries from 1900 to 2020. They found an upward trend in the global factor. Similarly, Bogalo, Poncela and Senra (2024) also find an upward trend when analysing average monthly temperatures in 12 cities of Southern Europe observed monthly from January 1950 to December 2022. Note that although most evidence is about average temperature having an upward increasing trend, it is essential to point out that several authors describe what is known as the hiatus in warming; see, for example, Schmidt, Shindell and Tsigaridis (2014), Pretis, Mann and Kaufmann (2015), Estrada and Perron (2017), Medhaug \textit{et al}. (2017) and Miller and Nam (2020). 

Although the literature focused on modelling average temperature is relevant, there is some agreement that more is needed to reflect the complicated climate variations. Consequently, policy makers should rely on future climate scenarios involving more than only changes in means; see, for example, the recent scenario analysis by Phella, Gabriel and Martins (2024). In this direction, on top of modelling average temperature, several authors consider modelling the evolution of range temperature, computed as the difference between the maximum and minimum temperatures within a given period. These studies often find a downward trend in temperature variability, although the evidence is not as clear as for the positive trend of average temperature. For example, Vose, Easterling and Gleason (2005), Dupuis (2014), Qu, Wan and Hao (2014) and Meng and Taylor (2022) show that increases in minimum temperatures have been more relevant than increases in maximum temperatures in the globe (the first), different regions of the US (the second and third), and four cities in Spain (the last). Xu \textit{et al}. (2013) analyse minimum and maximum daily temperatures in 825 stations in China observed from 1951 to 2020 and conclude that the diurnal temperature range has significantly decreased at 49\% of the stations, with significant increases being identified at 3\% of them. Diebold and Rudebusch (2022a) propose modelling the average and range temperature using separate regressions with deterministic trends, seasonal dummies and their interactions. The two univariate models are fitted to average and range temperature observed daily in selected cities in the US from 1960 to 2017. Instead of considering separate regressions, Meng and Taylor (2022) propose two alternative methods to model Interval-Valued Time Series (IVTS) of minimum and maximum temperature. First, they propose modelling the two-time series separately by fitting models with deterministic trends and seasonal components and allowing for interactions between minimum and maximum temperatures. Alternatively, IVTS of minimum and maximum temperature are modelled by a bivariate VARMA-MGARCH model. Using daily data in four Spanish cities observed from 1951 to 2015, their results also suggest a decrease in the diurnal temperature range and an increase in trend. Note that they do not restrict the model to avoid crossing minimum  and maximum temperature at a given moment. The authors recognise that, although unlikely, the temperatures may cross.

On top of whether or not temperature variability should be modelled together with the evolution of average temperature, several important debates about modelling the dynamic characteristics of average and range temperature are very active these days. First, there is a growing literature about the nature of the trend in temperature; see, for example, the discussions by Coggin (2012) and Proietti and Hillebrand (2015). Within this literature, one of the most important controversies is whether trends in temperatures (or other climatological variables) should be modelled by assuming that they are deterministic or stochastic. For example, Fatichi \textit{et al}. (2009) examine trends of daily average temperatures recorded in 26 stations in Tuscany (Italy) and conclude that a deterministic trend can be regarded as the most appropriate model only for a subset of 9 stations. Kaufmann, Kauppi and Stock (2010), Kaufmann \textit{et al}. (2013) and Chang \textit{et al}. (2020) are also among those who support the presence of stochastic trends in temperatures. However, Gao and Hawthorne (2006), Gay, Estrada and S\'anchez (2009) and Gadea Rivas and Gonzalo (2020, 2022) argue that trend-stationary processes better characterise temperature. Furthermore, Chen, Gao and Vahid (2022) state that deterministic non-linear trends can represent global temperature. Seidel and Lanzante (2004), Gil-Alana (2008), Estrada, Gay and S\'anchez (2010), Pretis and Allen (2013), Estrada, Perron and Mart\'inez-L\'opez (2013), Estrada and Perron (2017), Friedrich \textit{et al}. (2020), Kim \textit{et al}. (2020) and Gadea-Rivas, Gonzalo and Ramos (2023) further propose models with deterministic trends with breaks, while Friedrich, Smeekes and Urbain (2020) and Friedrich \textit{et al}. (2020) propose smooth non-parametric trends.\footnote{It is essential to note that it is challenging to establish the nature of trend in a time series with a finite sample and, consequently, there is still a methodological debate about the power of tests for trend-stationarity or difference-stationarity; see, for example, the general discussions by Diebold and Senhadji (1996), Phillips (2005, 2010) and Rao (2010) and the comments by Stern and Kaufmann (2000) and Pretis and Hendry (2013) in the particular context of climate variables.} There are also several proposals for modelling temperatures based on long-memory models; see, for example, Baillie and Chung (2002), Gil-Alana (2005), Ventosa-Santaularia, Heres and Mart\'inez-Hern\'andez (2014), Mangat and Reschenhofer (2020) and Vera-Vald\'es (2021).

Another important debate when addressing the evolution of temperature has to do with its seasonal variation, which is the most prominent source of climate variability, so climate change could also be reflected in it; see Pezzulli, Stephenson and Hannachi (2005) and Proietti and Hillebrand (2015) for its importance. Bogalo, Poncela and Senra (in press) show that seasonality is the main feature of monthly average temperatures observed in 12 southern European cities, capturing 88.3\% of the total variability, while the trend accounts for 1\% of the total variability. Consequently, there is also an interest in knowing the most appropriate model for the seasonal components of temperature. Several authors have suggested that the seasonal pattern in temperatures varies over time with more intense warming in winters than in summers; see Harvey and Mills (2003), Vogelsang and Franses (2005), Cohen \textit{et al}. (2012), Proietti and Hillebrand (2015), Hillebrand and Proietti (2017), and Bogalo, Poncela and Senra (2024). Furthermore, Hillebrand and Proietti (2015) also conclude that seasonal warming trends may vary across locations, while Bogalo, Poncela and Senra (in press) conclude that seasonal variations are primarily common.

Finally, among the hazards encountered when modelling climate-related time series, Pretis and Hendry (2013) point out the spatial variation of temperature trends, suggesting that there may be unmodelled heterogeneity. Dupuis (2014) and Scotto, Barbosa and Alonso (2011) also find heterogeneity, analysing minimum and maximum temperatures observed in 12 locations in the South-western US and extremes in average daily temperature in Europe. Recently, Estrada, Kim and Perron (2021) contribute to this debate by showing that the response of high latitudes to increases in radiative forcing is much larger than elsewhere in the world, with warming being more than twice the global average. Other authors find different trends in average temperature depending on the location are Chang \textit{et al}. (2020), Holt and Ter\"asvirta (2020) and Gadea Rivas and Gonzalo (2022).\footnote{It is also important to remark that, although in this paper we focus on urban stations, it could also be interesting to extend the analysis to non-urban ones; see, for example, Hausfather \textit{et al}. (2013) for an analysis of the impact of urbanization on temperature trends and the discussion by Estrada and Perron (2021).} The heterogeneity in climate change may have important consequences for policy makers; see Kaufmann \textit{et al}. (2017), who argue that the scepticism about climate change could partially be caused by its spatial heterogeneity, and Zaval \textit{et al}. (2014) and Binelli, Loveless and Schaffuer (2023), who causally link perceived changes in local temperature to changes in global warming beliefs. Furthermore, Holt and Ter\"asvirta (2020) 
find evidence of co-shifting of hemispheric temperatures, which aligns with earlier evidence from Kaufmann and Stern (1997) on the co-movement of hemispheric temperature evolution and the anthropogenic character of climate change. Recently, Gao, Linton and Peng (2024) propose a nonparametric panel model for climate data with seasonal and spatial variation allowing the trends to vary by location and season. Marotta and Mumtaz (2023) specify a DFM with stochastic volatility to model annual temperatures in 160 countries around the world, assuming that there are global and country-specific factors. They conclude that the common factor explains a considerable portion of the average temperature globally.

\section{State space models for centre and log-range temperature}
\label{section:methodology}

In this section, we describe the statistical methodologies proposed to represent centre/log-range temperature, first, at a given location and, second, at several locations simultaneously. The approach used for the first case is based on unobserved components time series models and the use of the KFS algorithms to extract the unobserved trends and seasonal components. On the other hand, the methodology designed to study several locations is based on ML-DFM with the factors extracted using KFS.

\subsection{Models for temperature intervals at a given location}



Denote by $X_t=\left(C_t, R_t \right)^{\prime}$, the $2 \times 1$ vector of the centre and log-range temperature observed at a particular location at time $t$, which is modelled as the sum of the vector of trends, $\mu_t=\left(\mu_{1t}, \mu_{2t} \right) ^{\prime}$, the vector of seasonal components, $\gamma_t=\left(\gamma_{1t}, \gamma_{2t} \right) ^{\prime}$, and the vector of irregular components, $\varepsilon_t=\left(\varepsilon_{1t}, \varepsilon_{2t} \right) ^{\prime}$. Consider the following bivariate Frequency-Specific Basic Structural Model (FS-BSM), with stochastic trends and seasonal components
\begin{subequations}
\label{SSM} 
\begin{align} 
X_t & =\mu_t + \gamma_t + \varepsilon_t, \label{SSM1}\\
\mu_t & =\mu_{t-1}+\beta_{t-1}+\eta_t, \label{SSM2}\\  
\beta_t & =\beta_{t-1}+\zeta_{t}, \label{SSM3}\\
\gamma_t & = \sum_{j=1}^{6} \gamma^{(j)}_{t},\label{SSM4}\\
\gamma^{(j)}_{t} & =\gamma^{(j)}_{t-1}cos \lambda_j + \gamma^{*(j)}_{t-1} sin \lambda_j + \omega^{(j)}_{t},\label{SSM5}\\
\gamma^{*(j)}_{t} & =-\gamma^{(j)}_{t-1} sin \lambda_j + \gamma^{*(j)}_{t-1} cos \lambda_j + \omega^{*(j)}_{t},\label{SSM6}                               
\end{align}
\end{subequations}
where $\lambda_j=\frac{\pi j}{6}$, $j=1,...,6$, are the seasonal frequencies in radians and $\beta_t=\left(\beta_{1t}, \beta_{2t} \right) ^{\prime}$ is the vector of time-varying slopes of the trends. $\varepsilon_t$ is assumed to be white noise with covariance matrix $\Sigma_{\varepsilon}$ with the elements in the main diagonal, denoted by $\sigma^2_{1\varepsilon}$ and $\sigma^2_{2\varepsilon}$, representing the variances of the transitory component of the centre and log-range, respectively. The off-diagonal element of $\Sigma_{\varepsilon}$, denoted by $\sigma_{12\varepsilon}$, represents the covariance between the transitory components of the centre and log-range temperature. The vectors of noises of the levels, $\eta_t=(\eta_{1t}, \eta_{2t})$, of the slopes, $\zeta_t=\left(\zeta_{1t}, \zeta_{2t} \right) ^{\prime}$, and of the seasonal components, $\omega^{(j)}_{t}=\left(\omega^{(j)}_{1t}, \omega^{(j)}_{2t} \right)$ and  $\omega^{*(j)}_{t}=\left(\omega^{*(j)}_{1t}, \omega^{*(j)}_{2t} \right)$, are also assumed to be white noises with covariance matrices $\Sigma_{\eta}$, $\Sigma_{\zeta}$, $\Sigma^{(j)}_{\omega}$ and $\Sigma^{(j)}_{\omega^*}$, respectively. The notation for the elements of these matrices and their interpretations are analogous to those used for the elements of $\Sigma_{\epsilon}$. Note that $\gamma_t^{*}$ appears as a matter of construction, and its interpretation is not particularly important; see Harvey (1989) for a more detailed description. It is assumed that $\Sigma_{\omega}^{(i)}=\Sigma^{(i)}_{\omega^{*}}$ and that all disturbances in the model, $\varepsilon_{t}$, $\eta_t$, $\zeta_t$, $\omega^{(j)}_t$, and $\omega^{*(j)}_t$ are mutually and serially uncorrelated at all lags and leads. Note that the covariances in the matrices $\Sigma_{\varepsilon}, \Sigma_{\eta}$, $\Sigma_{\zeta}$ and $\Sigma_{\omega}^{(i)}$ being all equal to zero, implies that centre and log-range temperature can be modelled separately by fitting univariate FS-BSM models to each of them.

In order to interpret the elements of the FS-SBM in (\ref{SSM}), consider its univariate analogue, with $X_t$ being a scalar observation. If $\sigma^2_{\eta}=\sigma^2_{\zeta}=0$, then model (\ref{SSM}) implies a deterministic trend. If $\sigma^2_{\zeta}=0$ with $\sigma^2_{\eta}>0$, then the trend is stochastic with a fixed slope, $\beta_t=\beta$. Finally, if $\sigma^2_{\eta}=0$ with $\sigma^2_{\zeta}>0$, the trend is slowly changing, with its slope having a smooth evolution. A smooth trend model can be shown to underpin the detrending filter of Hodrick and Prescott (1997); see Harvey and Jaeger (1993). With respect to the seasonal component, model (\ref{SSM}), which implies a seasonal component with six different variances, is described in detail by Hindrayanto \textit{et al}. (2013). The BSM popularized by Harvey (1989) is obtained when the variances of the seasonal shocks are equal to each other, i.e. $\sigma^{2(j)}_{\omega}=\sigma^2_{\omega}$, for $j=1,...,6$. Finally, note that given the need for parsimony, and in the context of a univariate FS-BSM, Hindrayanto \textit{et al}. (2013) propose to reduce the number of seasonal variances to two, which are denoted by $\sigma^2_{I}$ and $\sigma^2_{II}$. In our context, we reduce the number of seasonal covariance matrices to $\Sigma^{(I)}_{\omega}=\Sigma^{(1)}_{\omega}$ and $\Sigma^{(II)}_{\omega}=\Sigma^{(2)}_{\omega}=...=\Sigma^{(6)}_{\omega}$. The trigonometric specification of the seasonal component is rather popular when modelling climate-related monthly data; see Campbell and Diebold (2005) and Dupuis (2012, 2014) for applications modelling the seasonal pattern of temperatures and Friedrich, Smeekes and Urbain (2020) and Friedrich \textit{et al}. (2020) for seasonal patterns of methane emissions.

It is important to note that model (\ref{SSM}) allows both the trend and seasonal components of temperature centre and log-range to evolve stochastically with deterministic components obtained as limiting cases. In fact, if $\sigma^2_{\zeta_1}=0$ ($\sigma^2_{\zeta_2}=0$), then the slope of the trend for the centre (log-range) is constant and, therefore, $\beta_{11}=...=\beta_{1T}=\beta_1$ ($\beta_{21}=...=\beta_{2T}=\beta_2$). If further, $\sigma^2_{\eta_1}=0$ ($\sigma^2_{\eta_2}=0$), then the trend is deterministic. Similarly, if the variances of the seasonal components, $\sigma^{2(I)}_{\omega1}$ and $\sigma^{2(I)}_{\omega2}$, and $\sigma^{2(II)}_{\omega1}$ and $\sigma^{2(II)}_{\omega2}$ are zero, then the seasonal components of centre and log-range temperature are, respectively, deterministic.

If the parameters of the FS-SBM model in (\ref{SSM}) were known, the KFS algorithms can be implemented to extract one-step-ahead, filtered and smoothed estimates of the levels and seasonal components of the temperature centre and log-range, together with their associated Mean Squared Errors (MSEs), which can be used to obtain prediction bounds for the estimated components. 
However, the variances and covariances involved in model (\ref{SSM}) are unknown and can be estimated by Maximum Likelihood (ML) based on the Kalman filter; see Harvey (1989) for a detailed description.

As discussed above, one crucial debate when modelling temperatures is whether trends are deterministic or stochastic. In the context of model (\ref{SSM}), with $X_t$ being a scalar,\footnote{As we will see latter, at each location, temperature center and log-range are mutually uncorrelated and, consequently, they can be modelled separately by fitting model (\ref{SSM}) with either $X_t=C_t$ or $X_t=R_t$.} testing for a deterministic trend can be carried out as proposed by Nyblom and Harvey (2000). Three different tests can be implemented depending on the particular model assumed for the trend under the alternative. First, assuming that there is not a seasonal component and the irregular component is white noise, if $\sigma^2_{\zeta} =0$, i.e. the slope is constant, $\beta_t=\beta, \forall t$, one can test for a deterministic trend, i.e. $H_0: \sigma^2_{\eta}=0$ against $H_1:\sigma^2_{\eta}>0$, using the following statistic
\begin{equation}
\label{eq:test1}
RW=\frac{1}{T \hat{\sigma}_e^2}\sum_{t=1}^T \left[\sum_{r=1}^t e_r\right]^2, 
\end{equation}
where $e_t$ are the Ordinary Least Squares (OLS) residuals and $\hat{\sigma}_e^2 = \frac{1}{T} \sum_{t=1}^T e_t^2$. Under the null, if it is assumed that there is no drift, i.e. $\beta=0$ and $e_t=X_t-\bar{X}$, RW statistic has an asymptotic Cramer von Mises (CvM) distribution with one degree of freedom for which the 5\% critical value is 0.461. Alternatively, if there is a drift, i.e. $\beta\neq 0$, the statistic in (\ref{eq:test1}) is denoted as RWD and $e_t$ are residuals from a regression of $X_t$ on a constant and time. In this case, the RWD has an asymptotic second-level CvM distribution, with the 5\% critical value being 0.148.

In the context of the Integrated Random Walk (IRW), i.e. when $\sigma^2_{\eta}=0$, testing for a deterministic trend implies testing for $H_0: \sigma^2_{\zeta}=0$. In this case, one should use the following statistic
\begin{equation}
\label{eq:test2}
IRW=\frac{1}{T^4 \hat{\sigma}^2_e} \sum_{t=1}^T \left[ \sum_{s=1}^t \sum_{r=1}^2 e_r\right]^2. 
\end{equation} 
The critical values of the RW and IRW tests are reported by Harvey (2001). Note that, in the case of the IRW test, convergence to the asymptotic critical values is relatively slow and, consequently, using the asymptotic critical values can lead to the IRW test being oversized. Nyblom and Harvey (2000) show that the IRW test appears to have little or no power advantage over the RWD test.


Finally, testing for a deterministic seasonal component can be carried out using the CvM seasonality test proposed by Harvey (2001) and Busetti and Harvey (2003). The CvM test statistic can be constructed using the one-step-ahead prediction errors from the model written as a state space model under the null hypothesis and with the nuisance parameters estimated under the alternative. For $H_0: \sigma^{2(j)}_{\omega}=0$, the statistic follows a CvM distribution with two degrees of freedom for $j=1,...,5$ and with one degree of freedom for $j=6$; see Harvey (2001) for the critical values. Furthermore, Hindrayanto \textit{et al}. (2013) show that, in the model with two seasonal variances, the test for $H_0: \sigma^{2(II)}_{\omega}=0$ leads to a CvM test with nine degrees of freedom. 




\subsection{Modelling at several locations: Multi-level Dynamic Factor Models}

Consider now that the centre and log-range temperature has been observed at each moment of time $t$ at $N$ locations. If  at each location, such series were mutually uncorrelated, then one could model separately the system of centres and the system of log-ranges. Consider, for instance, the system of centres.\footnote{A similar analysis can be carried out for the system of log-ranges.} Given that our main interest is separating the common and heterogeneous behaviour of the trends in the centre temperature at different locations, we consider the deseasonalised centres.\footnote{Each centre is deseasonalised using the seasonal component estimated at each location separately.} Denote by $Y_t=\left(y_{1t},...,y_{Nt} \right) ^{\prime}$ the $N\times 1$ vector of deseasonalised centres, which is assumed to be decomposed into a vector of common components (the common evolution of centre temperature) and a vector of idiosyncratic components (the heterogeneity in the trends of centre temperature at different locations) as follows
\begin{equation}
\label{eq:DFM}
Y_t=PF_t+U_t,
\end{equation}
where $P$ is the $N\times r$ matrix of factor loadings, $F_t = \left( F_{1t},...,F_{rt} \right)^{\prime}$ is the $r\times 1$ vector of underlying unobserved factors at time $t$, and $U_t$ is the $N \times 1$ vector of idiosyncratic components, which are allowed to be weakly cross-sectionally correlated but uncorrelated with the factors, $F_t$. 
Given the nature of the data in this paper, we can assume that centre temperature may be organized in blocks corresponding to different geographical areas. These blocks imply zeros in the loading matrix $P$ as not all variables in $Y_t$ load on all $r$ factors in the DFM. 
In this case, it is more appropriate to extract the factors from a multi-level DFM obtained after imposing the adequate zero restrictions on the matrix of loadings, $P$. Consequently, we consider a multi-level DFM in which there is a factor, which is common to the full cross-section of centres and a number of factors, each common to locations in a particular geographical area. The number of such regions is specified using prior information about the data structure. Consider, for example, that there are five geographical areas with the variables in $Y_t$ divided in five blocks, $Y_t=\left(Y_{1t}, Y_{2t}, Y_{3t}, Y_{4t}, Y_{5t} \right)^{\prime}$, with $Y_{1t}$ being an $N_1 \times 1$ vector of temperatures in the first region and $Y_{2t}, Y_{3t}, Y_{4t}$ and $Y_{5t}$ being vectors of temperatures in the second, third, fourth and firth regions with dimensions $N_2, N_3, N_4$ and $N_5$, respectively. The multi-level DFM with $r_g=1$ global factor and $r_s=5$ regional factors (one per region) is given by
\begin{equation}
\label{eq:multilevel_def}
Y_t= \begin{bmatrix}
Y_{1t}\\
Y_{2t}\\
Y_{3t}\\
Y_{4t} \\
Y_{5t}
\end{bmatrix}=\begin{bmatrix}
p_{11} & p_{12} & 0 & 0 & 0 & 0 \\
p_{21} & 0 & p_{23} & 0 & 0 & 0 \\
p_{31} & 0 & 0 & p_{34} & 0 & 0\\
p_{41} & 0 & 0 & 0 & p_{45} & 0 \\
p_{51} & 0 & 0 & 0 & 0 & p_{56}
\end{bmatrix} \begin{bmatrix}
F_{gt}\\
F_{s1t}\\
F_{s2t}\\
F_{s3t}\\
F_{s4t}\\
F_{s5t}
\end{bmatrix} + U_t,
\end{equation}
where $F_{gt}$ is a pervasive factor that loads in all centre temperature in the system and $F_{s1t}$, $F_{s2t}$, $F_{s3t}$, $F_{s4t}$ and $F_{s5t}$ are regional factors loading in the variables $Y_{1t}$, $Y_{2t}$, $Y_{3t}$, $Y_{4t}$, and $Y_{5t}$, respectively. For $j=1,...,6$, $p_{ij}$ are $N_i \times 1$ vectors of loadings, and $U_t$ is the $N \times 1$ vector of idiosyncratic components defined as in the DFM in (\ref{eq:DFM}). Finally, if, for example, the global common factor is assumed to be an integrated random walk, then it is given by\footnote{The global factor could be modelled assuming other alternative specifications for the trend.}
\begin{subequations}
\begin{align}
\label{eq:factor}
F_{gt}=F_{gt-1} + \beta_{t-1}\\
\beta_t=\beta_{t-1} + \xi_t, 
\end{align}
\end{subequations}
where $\xi_t$ is a white noise with variance $\sigma^2_{\xi}$. If the regional factors are assumed to be stationary, then centre temperatures are assumed to be cointegrated, while if they are non-stationary, then temperatures at different regions are not cointegrated unless the regional factors are themselves cointegrated. In this paper, we assume that the regional factors are stationary and given by the following AR($1$) model
\begin{equation}
\label{eq:factorregional}
F_{sit}=\phi_{1i} F_{sit-1} + \eta_{it},
\end{equation}
where $\eta_{it}$ are mutually independent white noises with variances $\sigma^2_{\eta_i}$. 

Assuming stationary factors, Banbura, Giannone and Reichlin (2011) propose estimating the parameters of the multi-level DFM in (\ref{eq:multilevel_def}) by ML using the EM algorithm proposed by Doz, Giannone and Reichlin (2012). In this paper, we slightly modify their algorithm to allow for the global factor to be an integrated random walk and implement instead the two-step procedure of Doz, Giannone and Reichlin (2011); see the Appendix for details.


\section{Empirical characteristics of temperature in the Iberian Peninsula}
\label{section:data}

In this section, we describe the main empirical characteristics of centre and log-range temperature in the Iberian Peninsula.

\subsection{The data}

\begin{figure}[]
\caption{Map of the Iberian Peninsula and Mediterranean islands belonging to Spain. The locations marked with blue bullets correspond to geographical coordinates where minimum and maximum temperatures are observed. The red bullets represent the locations of the main cities in Spain and Portugal, which are the capitals of their corresponding provinces.}
\label{fig:Spain1}
\begin{center}
\includegraphics[width=0.65\textwidth]{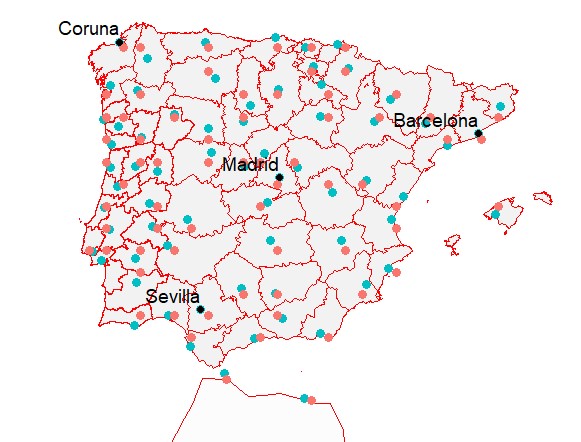}
\end{center}
\end{figure}

One very popular database for temperature-related variables is the Climate Research Unit gridded Time Series (CRUTS) maintained by the University of West Anglia, which is corrected to avoid inhomogeneities; see Mitchell and Jones (2005) and Harris \textit{et al}. (2020) for descriptions, Chang \textit{et al}. (2020) and Marotta and Mumtaz (2023) for recent applications using this database and Wijngaard, Klein Tauk and K\"onnen (2003) for a discussion about homogeneity in European temperature series.\footnote{https://sites.uea.ac.uk/cru/data} Intervals of minimum ($y^{min}_t$) and maximum ($y^{max}_t$) temperatures (measured in centigrades) are observed monthly from January 1930 to December 2020 ($T=1092$) in 68 locations in Spain.\footnote{The minimum and maximum temperatures are monthly means of the individual daily minimum and maximum temperatures. They are not the overall minimum or maximum temperatures recorded each month. Furthermore, we have not considered temperatures in some locations in the Atlantic Ocean because they were somewhat irregular. Similarly, we found some irregularities in data recorded before 1930. Alternatively, one can use the database E-OBS provided by the European Climate Assessment under the Copernicus project of the European Commission with daily observations since 1950; see, for example, Meng and Taylor (2022) for an application using the E-OBS data set, which can be found at https://cds.climate.copernicus.eu/cdsap\#!/dataset/insitu-gridded-observations-europe?tab=overview. However, the E-OBS database may have issues related to inhomogeneity.} Figure \ref{fig:Spain1} represents the Iberian Peninsula and some Spanish islands in the Mediterranean Sea and the locations selected for the analysis.

To analyse in detail the main empirical characteristics of temperature in the Iberian Peninsula over the last and present centuries, we select four locations representing different climates. The first location selected is Barcelona, a highly populated city situated on the Spanish Mediterranean coast with a Mediterranean climate. The second location is Coru\~{n}a, a small city with an Atlantic climate on the Atlantic Northwest Spanish coast. The third location considered is Madrid, Spain's largest city, which is in the centre of the Iberian Peninsula, with a continental climate. Finally, we consider temperature in Seville, in the south of Spain, which has the warmest summer in continental Europe among all cities with a population over 100,000 people; see Figure \ref{fig:Spain1} for a map of the Iberian Peninsula and the locations of the four cities considered.\footnote{Meng and Taylor (2022) also consider minimum and maximum temperatures in Madrid and Seville. Instead of analysing Barcelona and Coru\~{n}a, they consider C\'aceres and Albacete. Furthermore, note that they consider daily data, for a shorter period from 1951 to 2015.}

\subsection{Descriptive statistics}

\begin{figure}[ht]
\caption{Seasonal polar plots of centre and log-range temperature in Barcelona, Coru\~{n}a, Madrid and Seville.}
     \begin{subfigure}[b]{0.5\textwidth}
         \centering
         \includegraphics[width=1.0\textwidth]{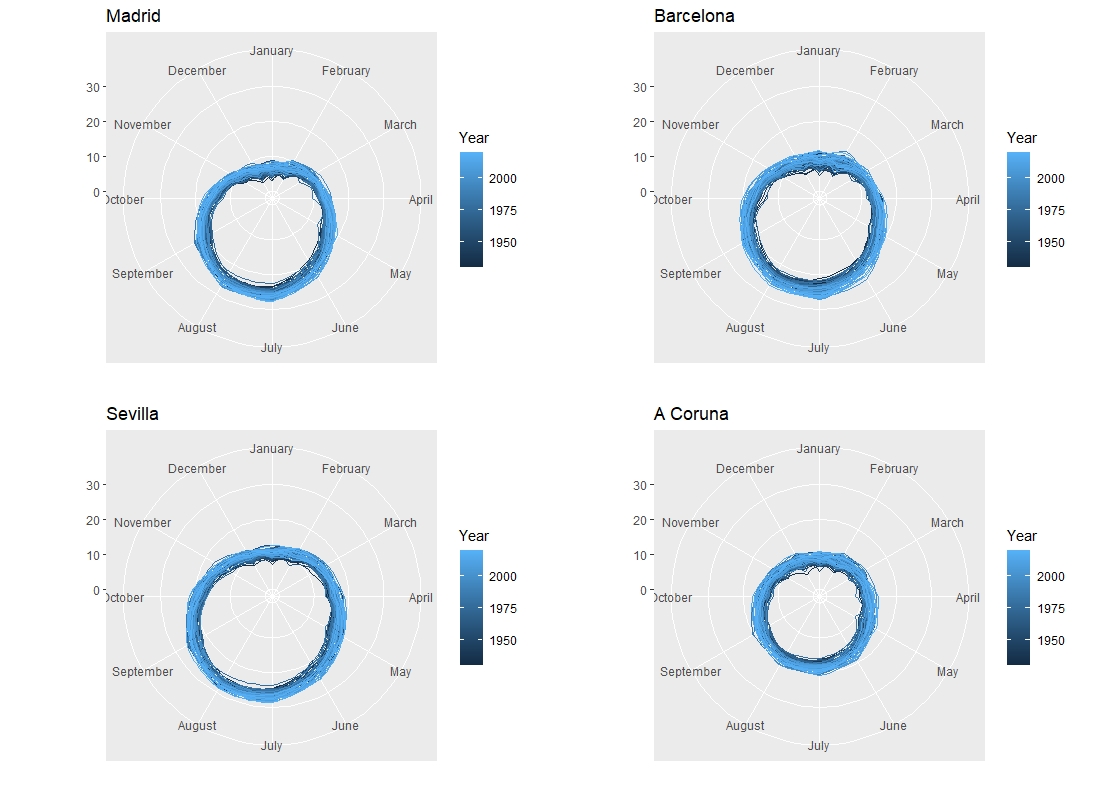}
         \caption{centre}
         \label{fig:centre}
     \end{subfigure}
     \begin{subfigure}[b]{0.5\textwidth}
         \centering
         \includegraphics[width=1.0\textwidth]{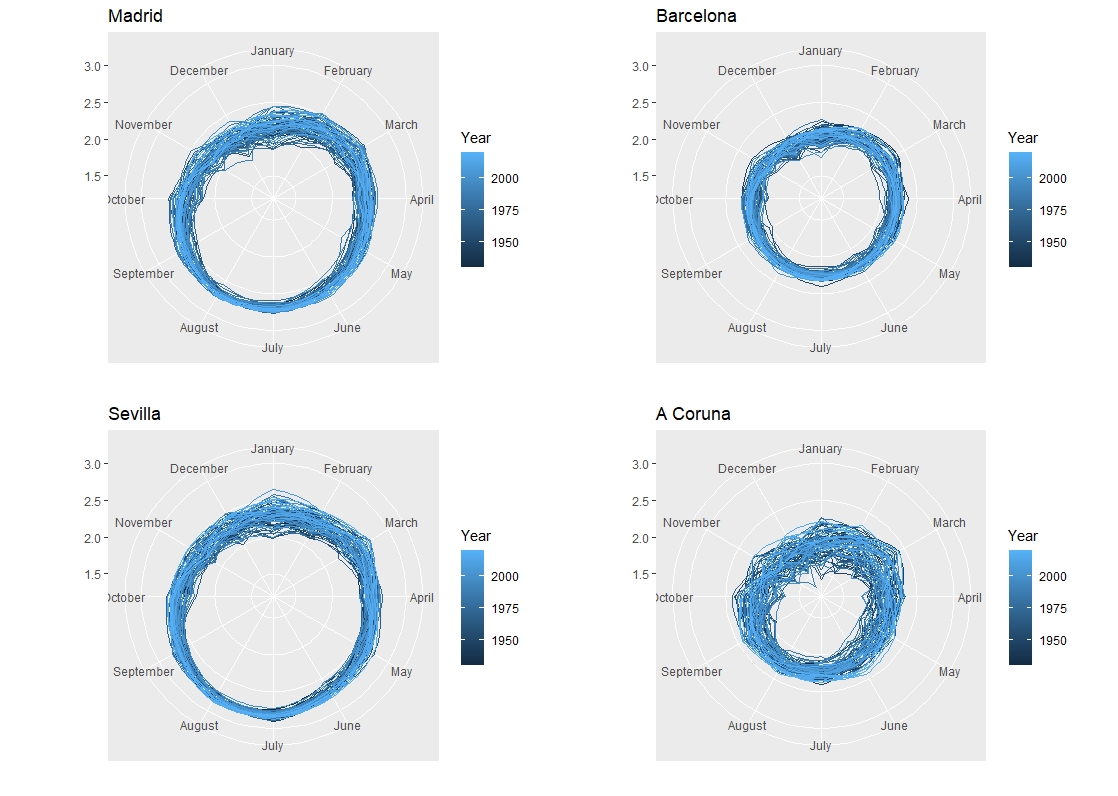}
         \caption{Log-range}
         \label{fig:range}
     \end{subfigure}
\label{fig:polar}
\end{figure}

We describe the main empirical properties of the centre and log-range temperature, $C_t=\frac{y^{max}_t+y^{min}_t}{2}$ and $R_t=ln(y^{max}_t-y^{min}_t)$, respectively. Instead of standard time plots, Figure \ref{fig:polar} displays seasonal polar plots, which help to visualize their strong seasonal patterns.\footnote{The first author developed all codes in R needed to obtain the empirical results in this paper.} First, consider the plots of the centres. We can conclude that the seasonal patterns observed in Barcelona and Seville are very similar, with the largest seasonal variations. The slightest seasonal variations are observed in Coru\~{n}a. Second, Figure \ref{fig:centre} shows that, as expected, the annual variations of centre temperatures are more considerable in the two coastal cities (Barcelona and Coru\~{n}a) than in the cities located in the interior of the Iberian Peninsula (Madrid and Seville); see also the sample standard deviations of the annual variations reported in Table \ref{tab:descriptive}. Moreover, Figure \ref{fig:centre} also suggests the possibility of climate change in the four locations considered, with centre temperatures being larger in more recent years than at the beginning of the 20th Century. 

With respect to the log-range temperatures, the patterns in the four locations are rather different. First, Figure \ref{fig:range} suggests periodic heteroscedasticity with the dispersion being larger in winter than in summer months; see Dupuis (2014) for the same conclusion when looking at minimum temperatures in South-western US and Meng and Taylor (2022) for minimum and maximum temperatures in the four cities in Spain mentioned above. Also, note that this pattern is more pronounced in the cities located in the interior of the Iberian Peninsula than in those at the coast. With respect to Coru\~{n}a, the dispersion is smaller than in the rest of locations but with a considerable variability during the years analysed. Finally, the dispersion has decreased over time in Barcelona while it seems to increase in the other three cities.

Summarising, Figure \ref{fig:polar} suggests the presence of possible trends and strong seasonal patterns, some of which may have changed over the last century. Furthermore, we can also observe some heterogeneity in the patterns of centre and log-range temperatures in the four locations considered. 

Table \ref{tab:descriptive} reports several descriptive statistics of the annual differences of the centre and log-range at each of the four locations. In particular, it reports the sample mean, standard deviation, skewness and kurtosis, the last two together with their $p$-values obtained according to the results of Bai and Ng (2005). Table \ref{tab:descriptive} also reports the $p$-values of the Bai and Ng (2005) test for normality in the presence of temporally correlated data and the Box-Pierce test for joint serial uncorrelation of the first 12 lags. First, we can observe that the sample means of the annual variations of the centre temperature are positive and approximately the same, 0.02, in the four locations considered. However, their standard deviations differ, being largest in Barcelona and smallest in Seville. It is also important to note that normality seems to be adequate to represent the probabilistic distribution of the annual variations of the centre temperature. When looking at the annual variations of the log-range temperature, we observe that they have the largest standard deviation in Coru\~{n}a. In this case, normality is rejected in all locations except Barcelona due to excess of kurtosis.

\begin{table}[ht!]
\caption{Descriptive statistics of annual differences of centre and log-range temperature: sample mean (Mean), standard deviation (St. dev.), skewness and kurtosis together with the corresponding $p$-values of the Bai and Ng (2005) tests of the last two quantities. $p$-values of the normality test (BN) and the Box-Pierce test for 12 lags (Q(12)) are also reported.}
\label{tab:descriptive}
\begin{center}
\resizebox{450pt}{!}{
\begin{tabular}{l c c c c c c c c c c c c c c c c c}
& Barcelona & Coru\~{n}a & Madrid & Seville &&& Barcelona & Coru\~{n}a & Madrid & Seville\\
\hline
& \multicolumn{4}{c}{Centre} &&& \multicolumn{4}{c}{Log-range}\\
\hline
Mean & 0.02 & 0.02 & 0.02 & 0.02 &&& 0.00 & 0.00 & 0.00 & 0.00 \\
St. dev. & 1.62 & 1.52 & 1.50 & 1.42 &&& 0.10 & 0.20 & 0.12 & 0.11 \\
Skewness & 0.01 & 0.02 & 0.09 & 0.01 &&& -0.05 & -0.10 & -0.26 & -0.16  \\
& \small{(0.12)} & \small{(0.42)} & \small{(0.12)} & \small{(0.46)} &&&\small{(0.74)} &\small{(0.87)} &\small{(0.97)} &\small{(0.90)}\\
Kurtosis & 3.13 & 2.85 & 2.87 & 2.71 &&& 3.32 & 3.55 & 4.54 & 4.33 \\
& \small{(0.27)} & \small{(0.79)} & \small{(0.77)} & \small{(0.97)} &&&\small{(0.12)} &\small{(0.02)} &\small{(0.00)} &\small{(0.00)}\\
BN & 0.48 & 0.56 & 0.17 & 0.04 &&& 0.16 & 0.00 & 0.00 & 0.00 \\
Q(12) & 0.00 & 0.00 & 0.00 & 0.00 &&& 0.00 & 0.00 & 0.00 & 0.00\\
\end{tabular}
}
\end{center}
\end{table}



\begin{figure}[]
\caption{Correlation map between the centres and the log-ranges temperatures in the 68 locations in the Iberian Peninsula.}
\label{fig:correlation}
\includegraphics[width=1\textwidth]{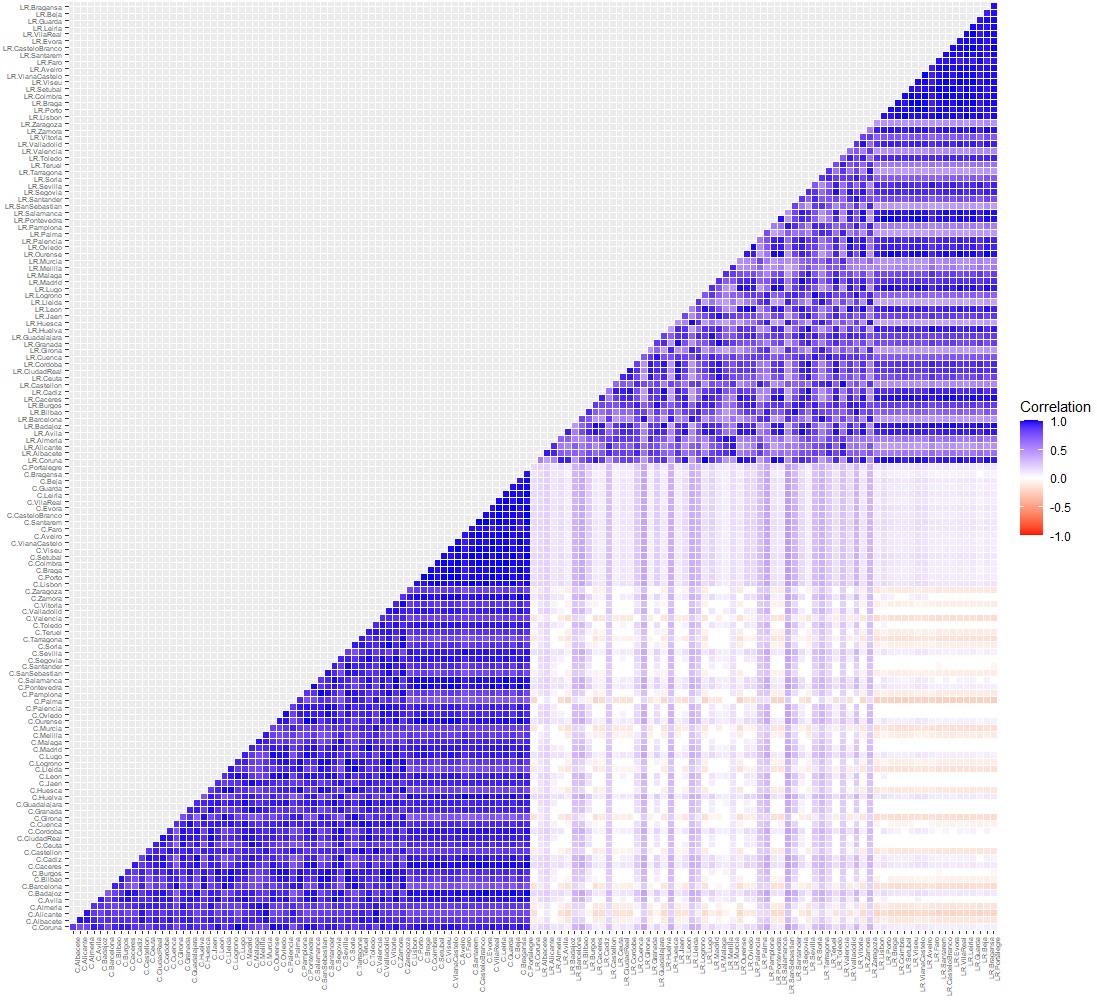}
\end{figure}
\begin{figure}[]
\caption{Correlation map of centre temperatures (left) and map of the Iberian Peninsula with the resulting clusters (right).}
\label{fig:correlation_C}
\begin{center}
\includegraphics[width=0.45\textwidth]{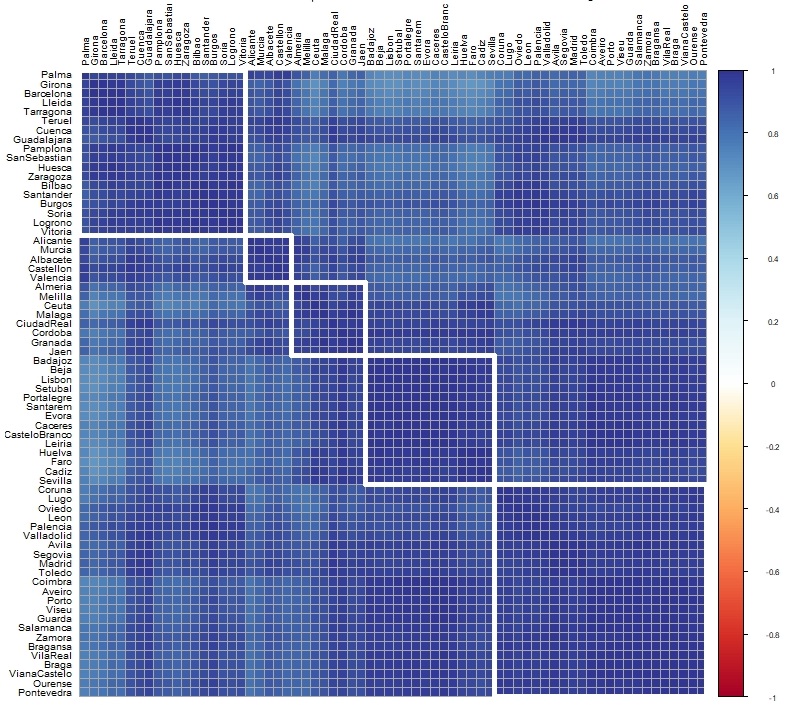}
\includegraphics[width=0.45\textwidth]{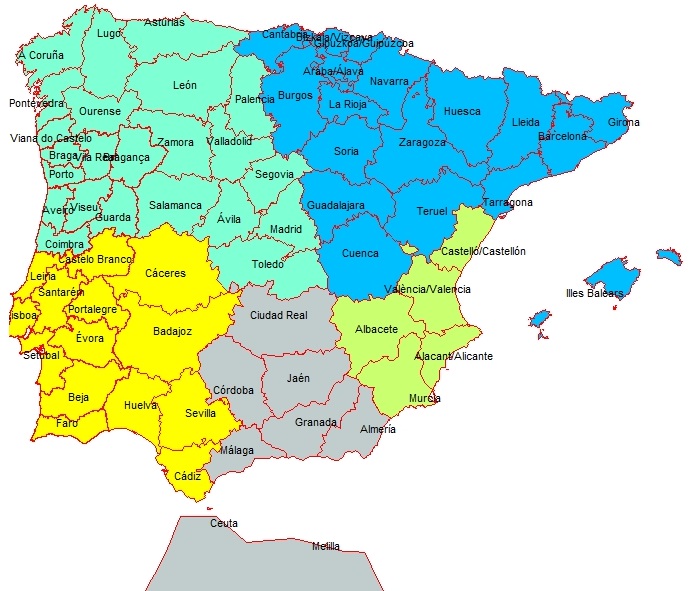}
\end{center}
\end{figure}
\begin{figure}[]
\caption{Correlation map of log-range temperatures (left) and map of the Iberian Peninsula with the resulting clusters (right).}
\label{fig:correlation_LR}
\begin{center}
\includegraphics[width=0.45\textwidth]{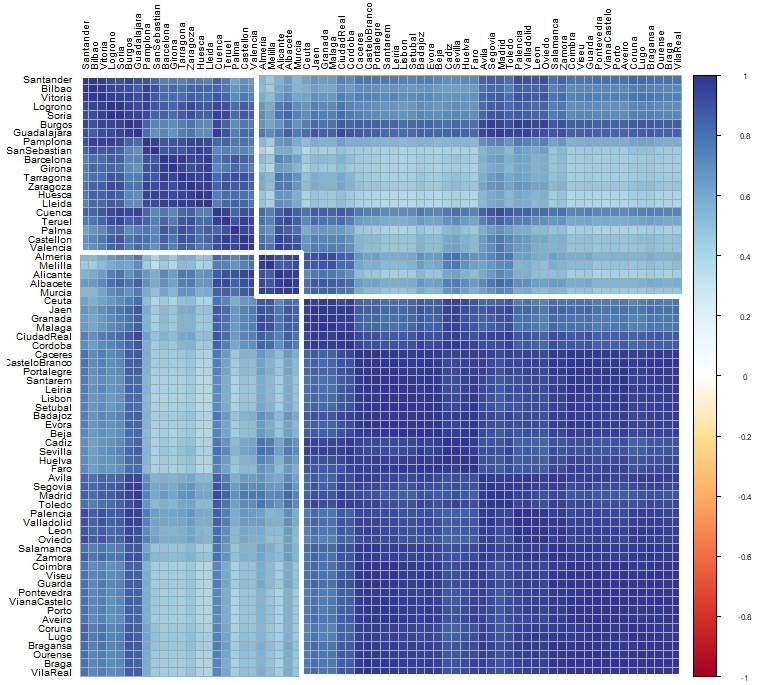}
\includegraphics[width=0.45\textwidth]{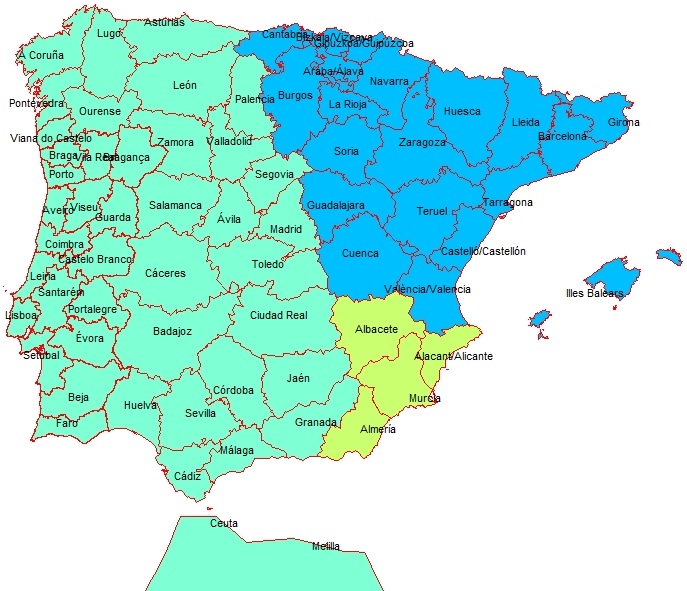}
\end{center}
\end{figure}

Finally, we analyse correlations between centres and log-ranges at all 68 locations to further investigate heterogeneity in the evolution of temperature. Figure \ref{fig:correlation} plots the map of pairwise correlations between centres and log-ranges at all 68 locations. The main conclusion from this figure is that, at each location, centres and log-ranges do not seem to be mutually correlated and, consequently, they can be modelled separately. However, there are relevant correlations among the centres at different locations and among the log-ranges at different locations. Consequently, we analyse the presence of clusters among the centres and among the log-range temperatures. Figures \ref{fig:correlation_C} and \ref{fig:correlation_LR} plot the correlation maps for each of them at all 68 locations in the Iberian Peninsula. Using the complete linkage method for hierarchical clustering, which defines the distance between two clusters as the maximum distance between their components, we can identify five clusters in the centre temperature and three in the log-ranges; see the maps in Figures \ref{fig:correlation_C} and \ref{fig:correlation_LR}.

\section{Empirical modelling of centre and log-range temperature at selected locations}
\label{section:univariate}

The results of fitting model (\ref{SSM}) separately in the four locations considered, Barcelona, Coru\~{n}a, Madrid and Seville, are reported in the Online Appendix. The reported estimates show that, in concordance with the descriptive analysis carried out in the previous section, and regardless of the particular location considered, the shocks of the trend, seasonal and irregular components of the centre and log-range are uncorrelated.\footnote{All calculations are carried out using the KFAS library developed by Helske (2017) in the R programming environment.} Consequently, in this section, we fit separate DF-BSM to centre and log-range at each location.

\begin{figure}[]
\caption{Diagnostics of standardized residuals of the SSM model fitted for centre (left column) and log-range (right column) temperature in Barcelona together with sample autocorrelations and Q-Q plot.}
\label{fig:BarcelonaDiag_0}
\begin{center}
\includegraphics[width=0.45\textwidth]{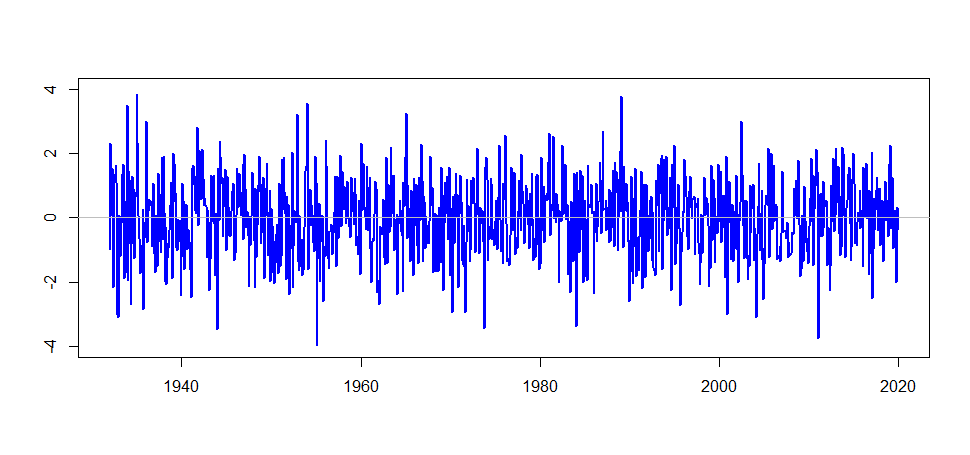}
\includegraphics[width=0.45\textwidth]{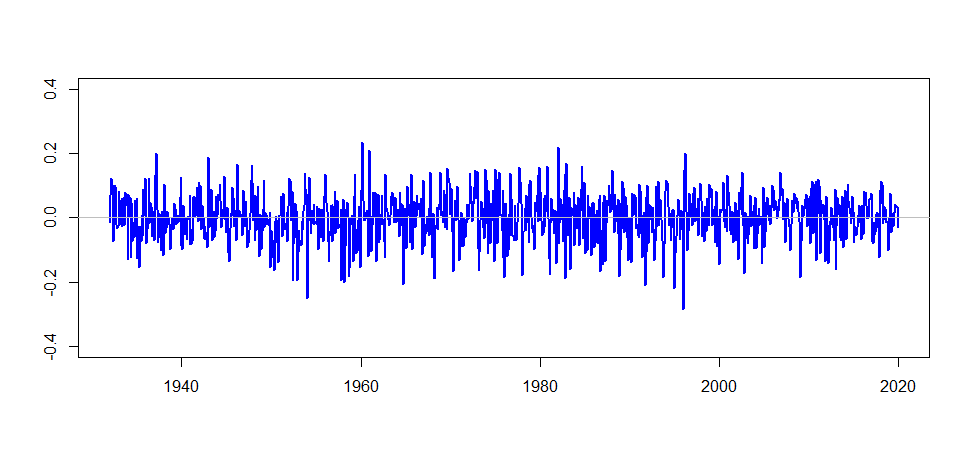}
\includegraphics[width=0.40\textwidth]{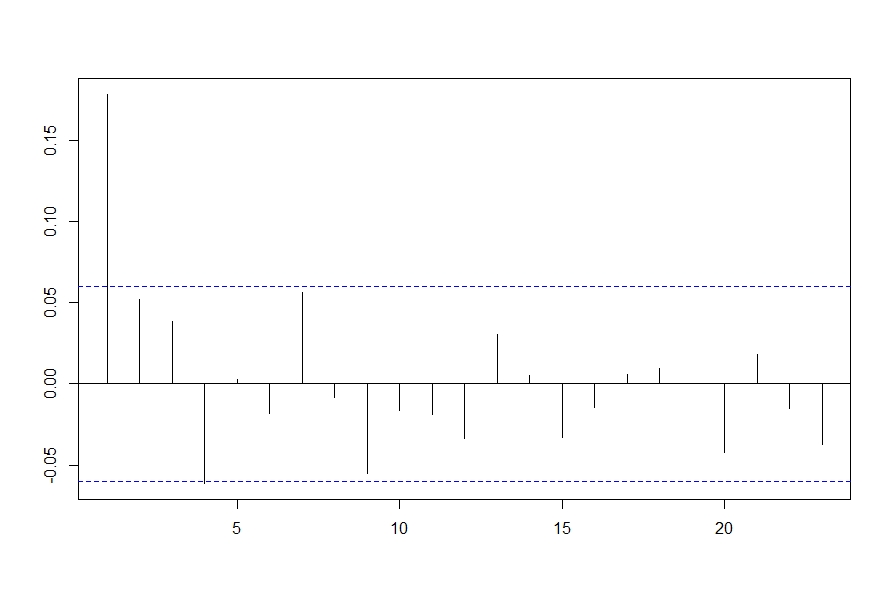}
\includegraphics[width=0.40\textwidth]{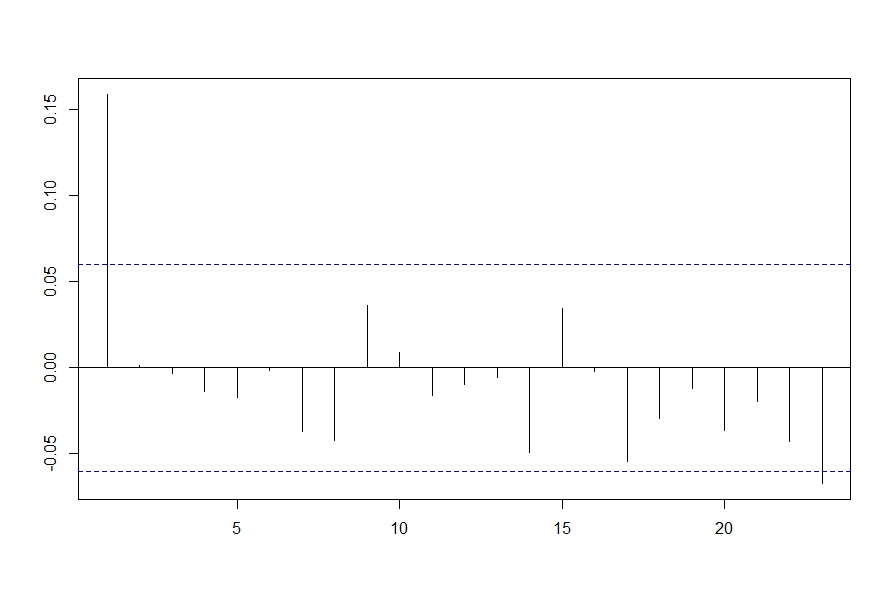}
\includegraphics[width=0.40\textwidth]{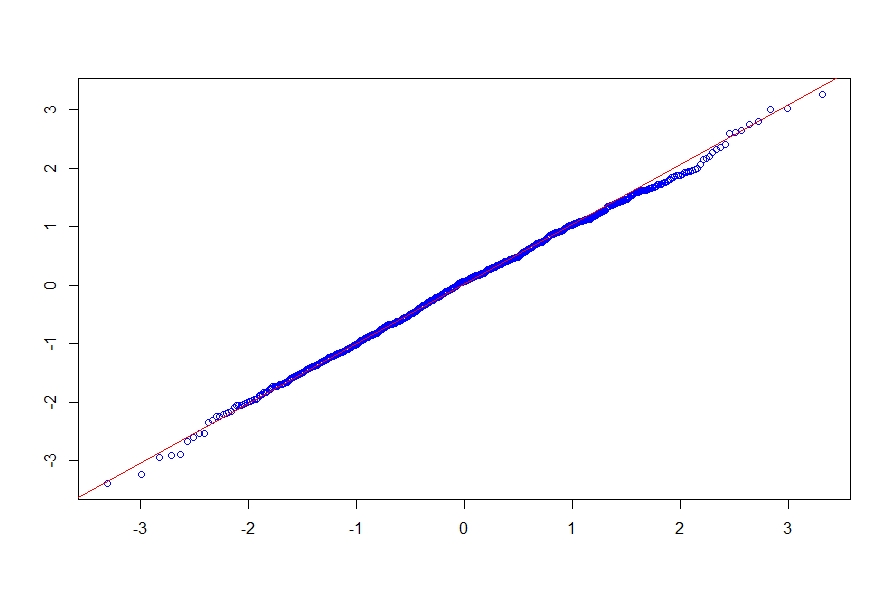}
\includegraphics[width=0.40\textwidth]{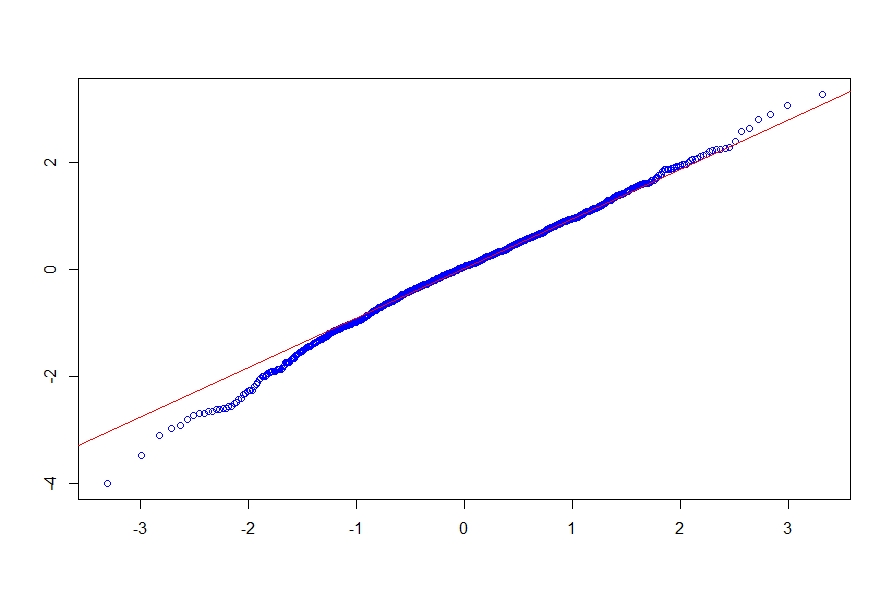}
\end{center}
\end{figure}

\begin{figure}[]
\caption{Diagnostics of standardized residuals of the SSM model fitted for centre (left column) and log-range (right column) temperature in Coru\~{n}a together with sample autocorrelations and Q-Q plot.}
\label{fig:CorunaDiag_0}
\begin{center}
\includegraphics[width=0.45\textwidth]{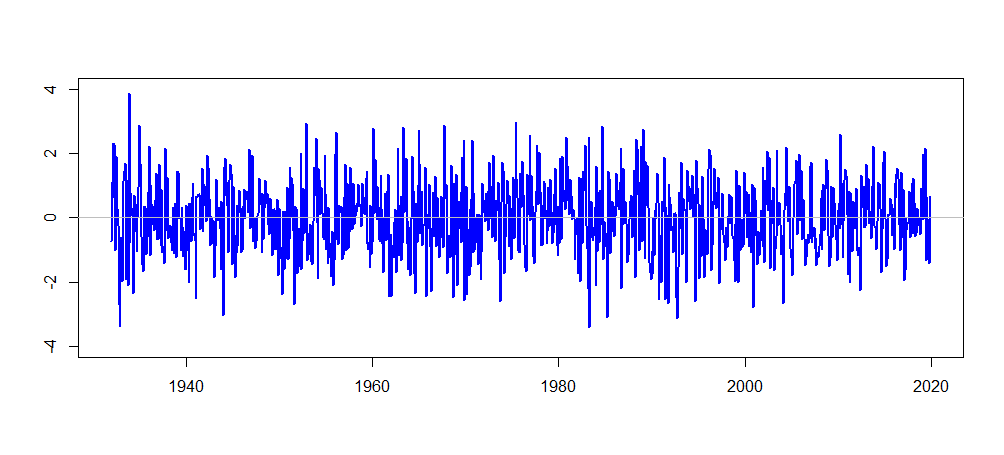}
\includegraphics[width=0.45\textwidth]{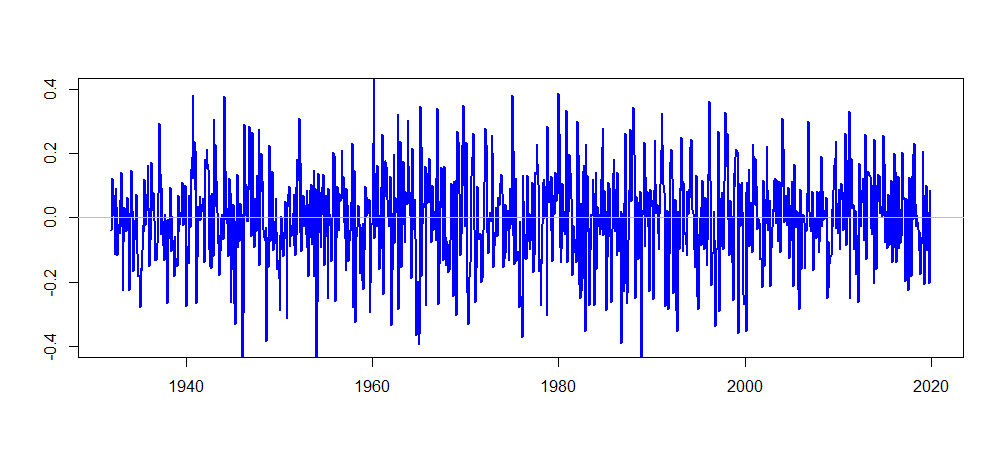}
\includegraphics[width=0.40\textwidth]{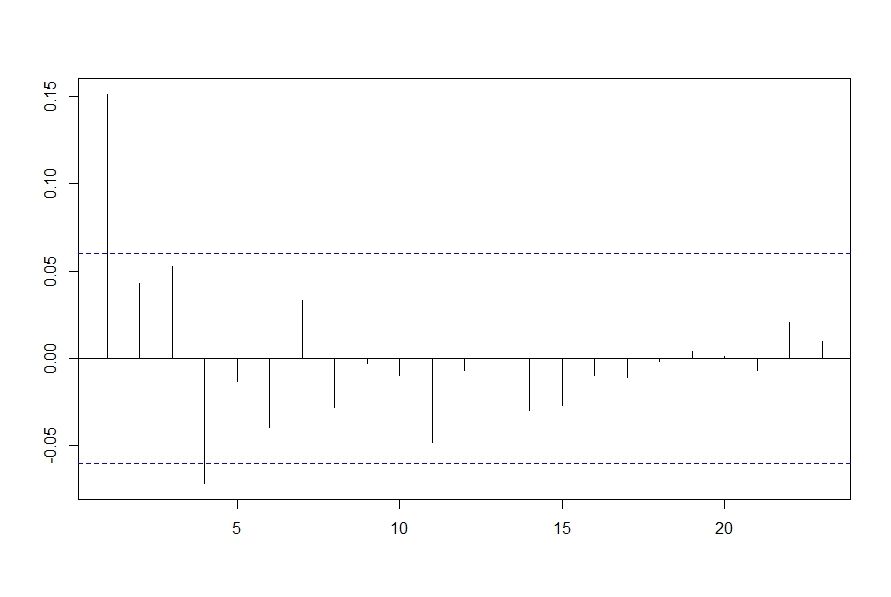}
\includegraphics[width=0.40\textwidth]{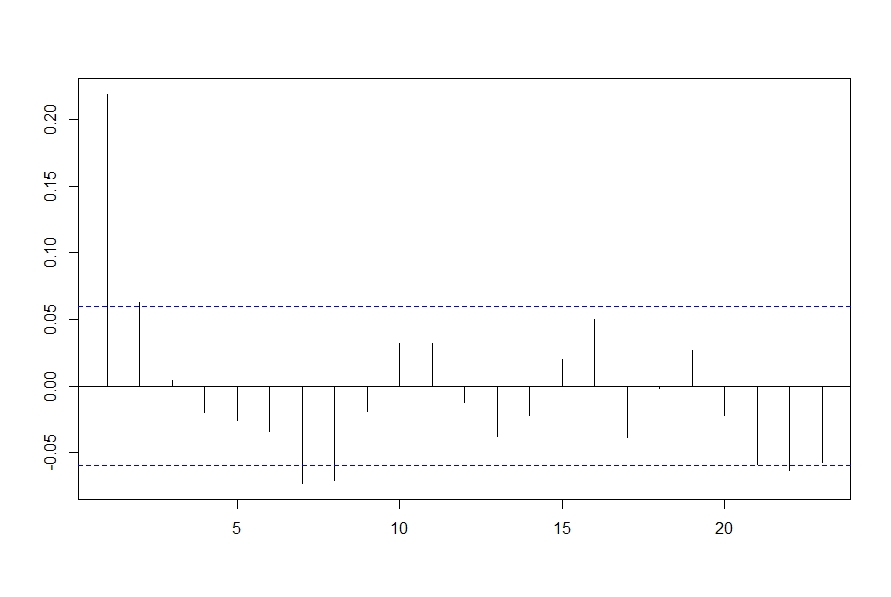}
\includegraphics[width=0.40\textwidth]{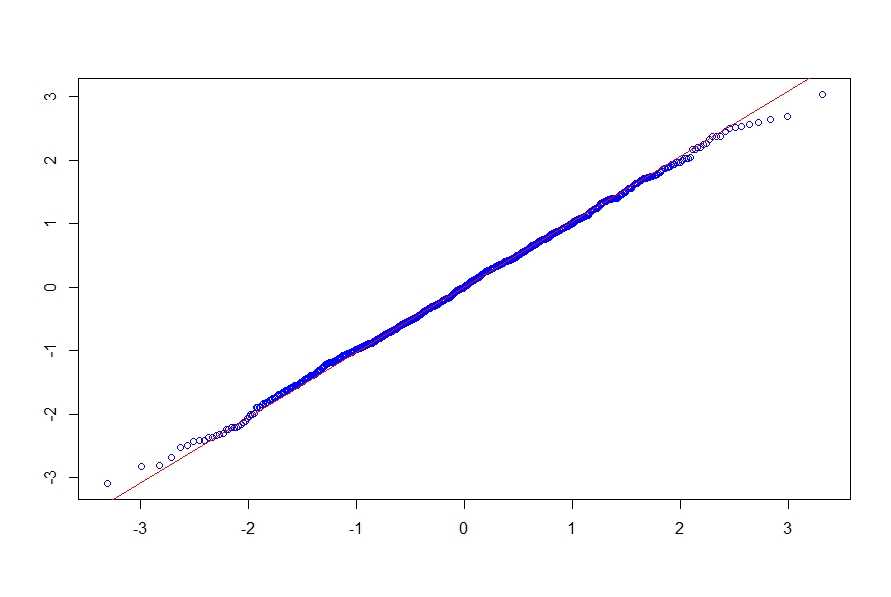}
\includegraphics[width=0.40\textwidth]{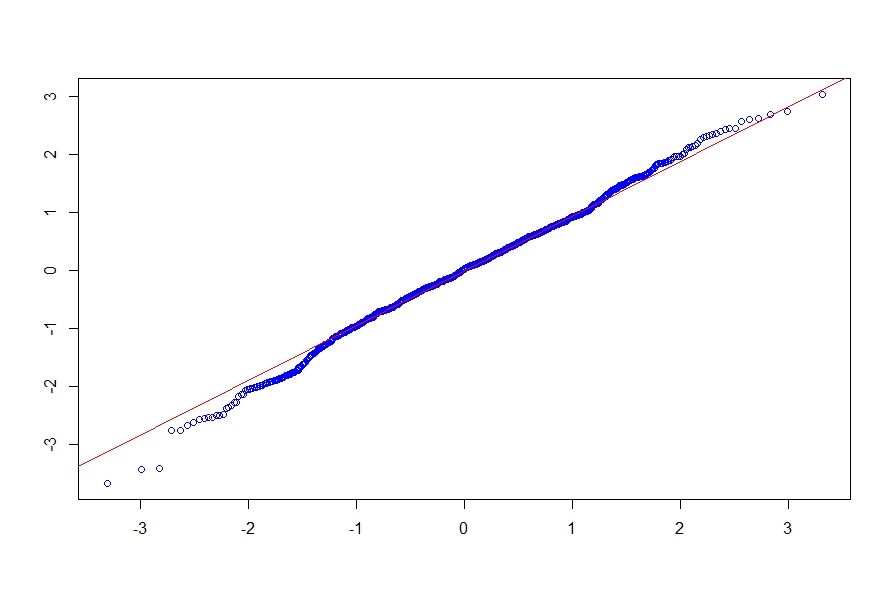}
\end{center}
\end{figure}

\begin{figure}[]
\caption{Diagnostics of standardized residuals of the SSM model fitted for centre (left column) and log-range (right column) temperature in Madrid together with sample autocorrelations and Q-Q plot.}
\label{fig:MadridDiag_0}
\begin{center}
\includegraphics[width=0.45\textwidth]{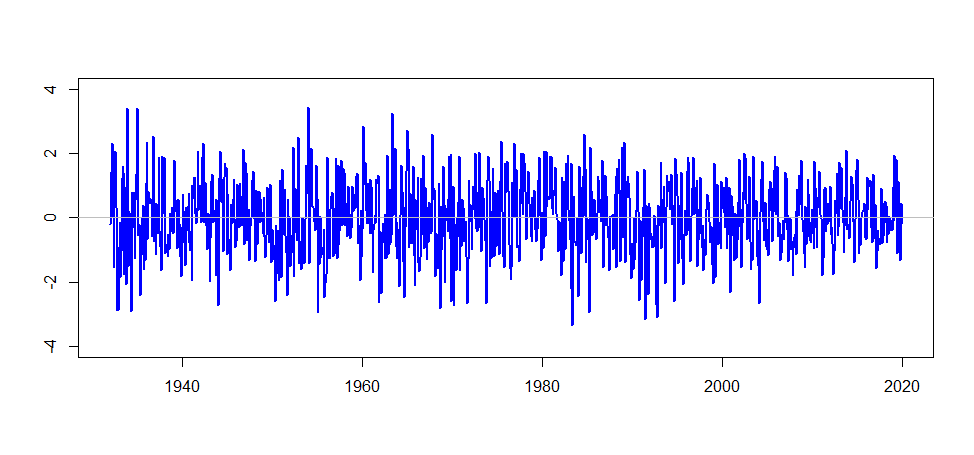}
\includegraphics[width=0.45\textwidth]{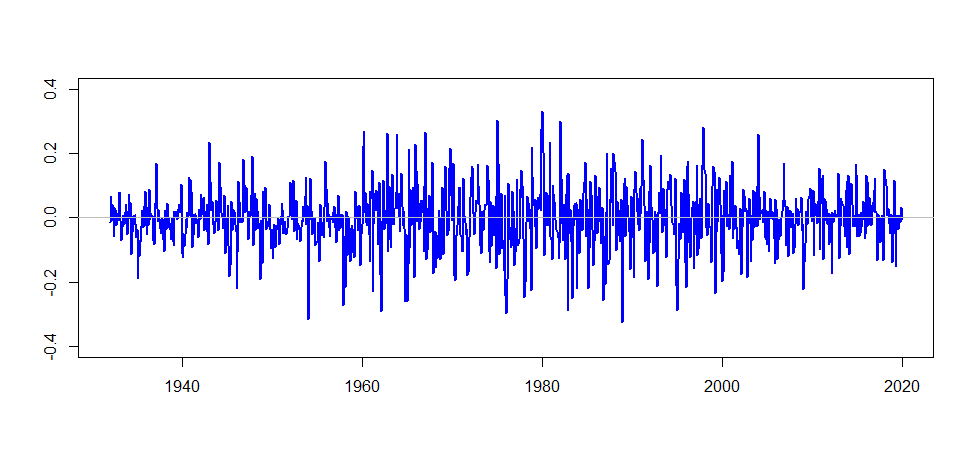}
\includegraphics[width=0.40\textwidth]{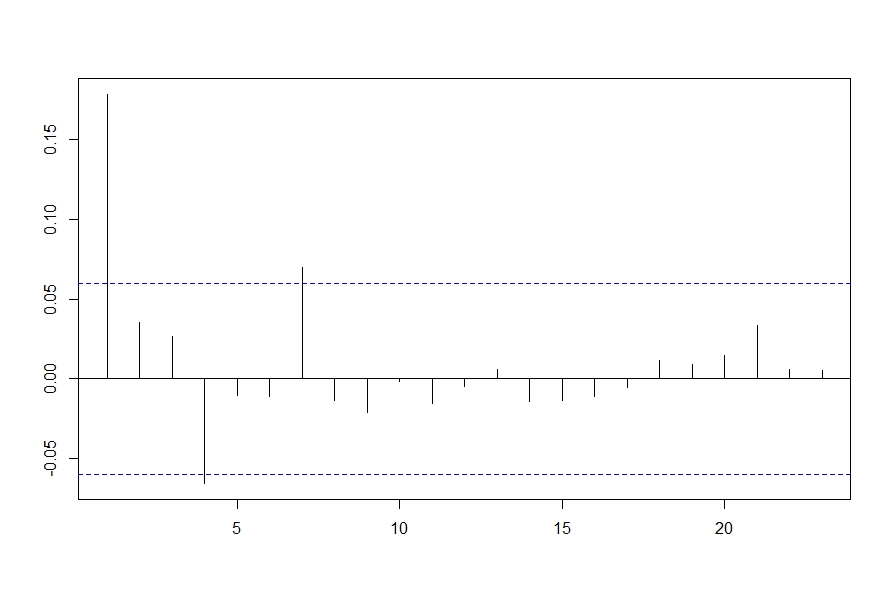}
\includegraphics[width=0.40\textwidth]{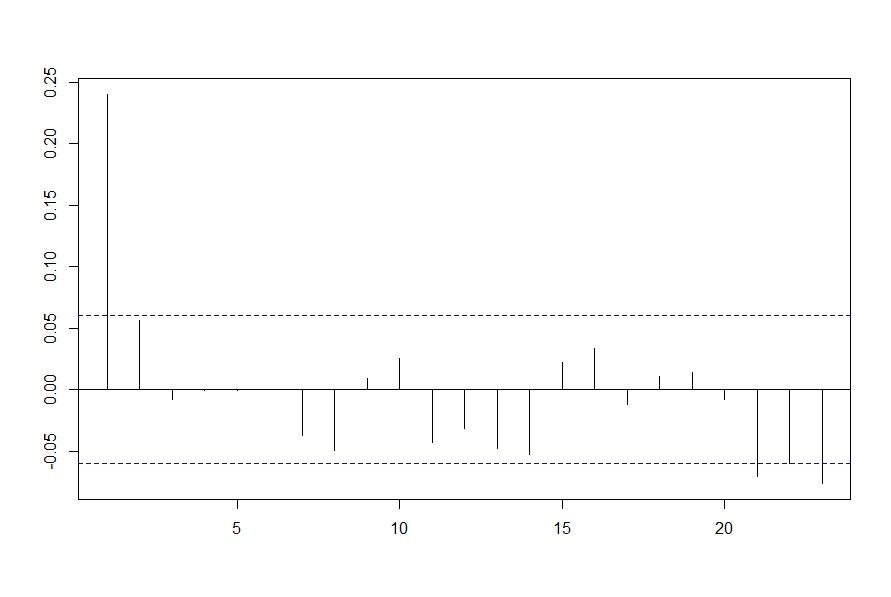}
\includegraphics[width=0.40\textwidth]{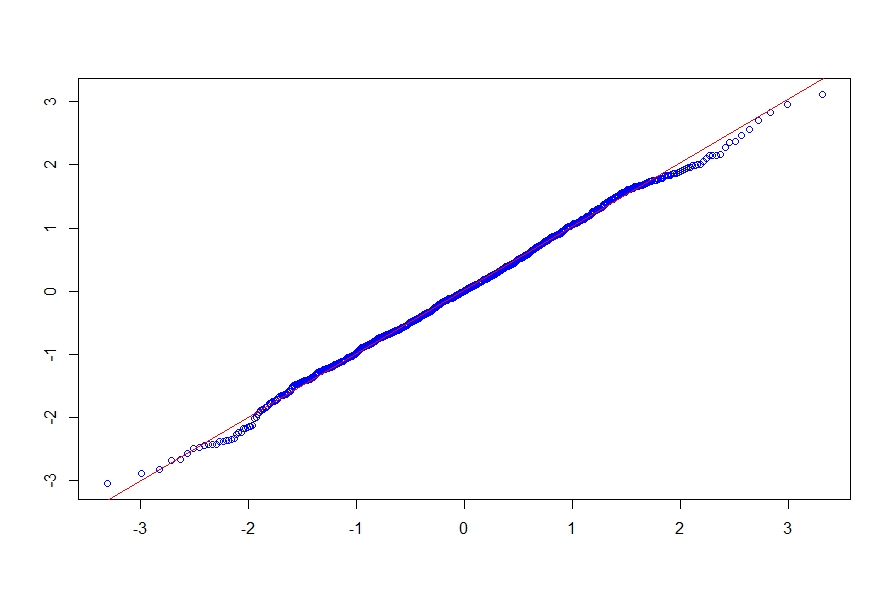}
\includegraphics[width=0.40\textwidth]{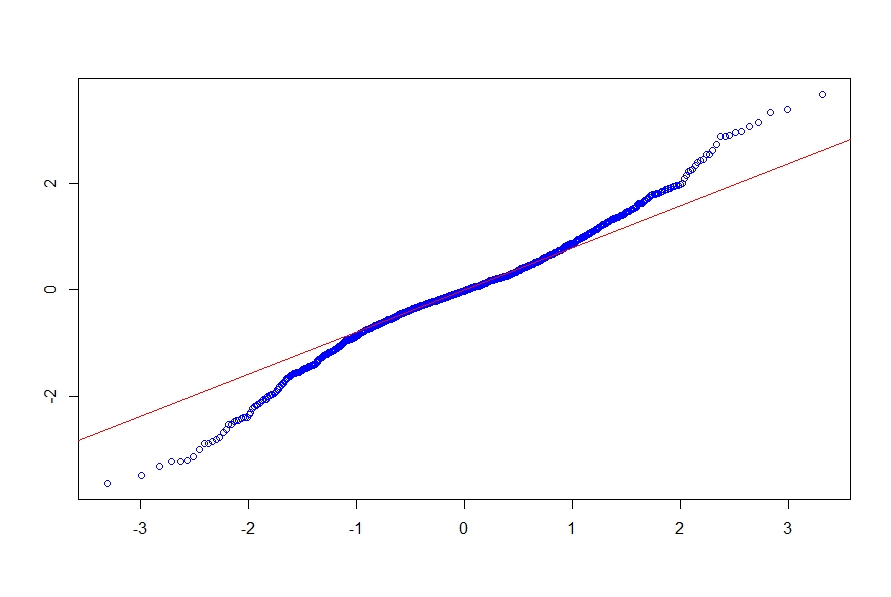}
\end{center}
\end{figure}

\begin{figure}[]
\caption{Diagnostics of standardized residuals of the SSM model fitted for centre (left column) and log-range (right column) temperature in Seville together with sample autocorrelations and Q-Q plot.}
\label{fig:SevilleDiag_0}
\begin{center}
\includegraphics[width=0.45\textwidth]{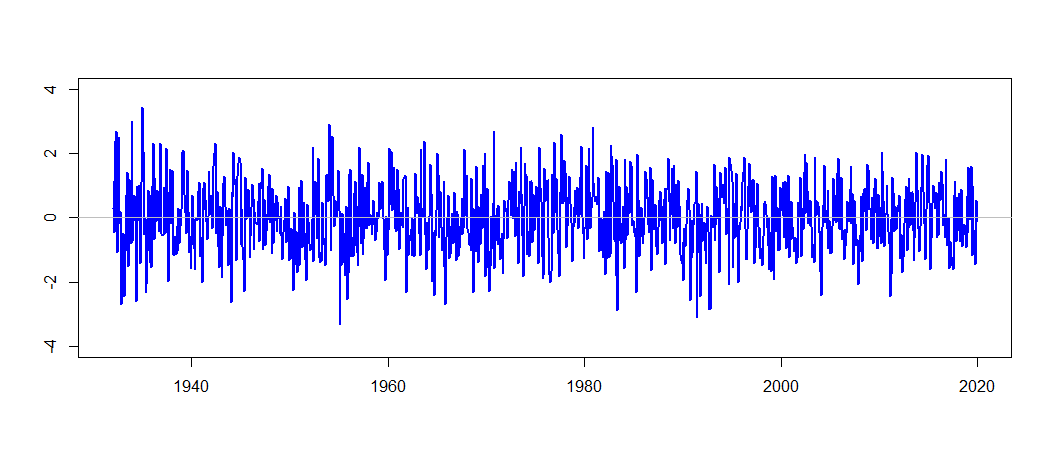}
\includegraphics[width=0.45\textwidth]{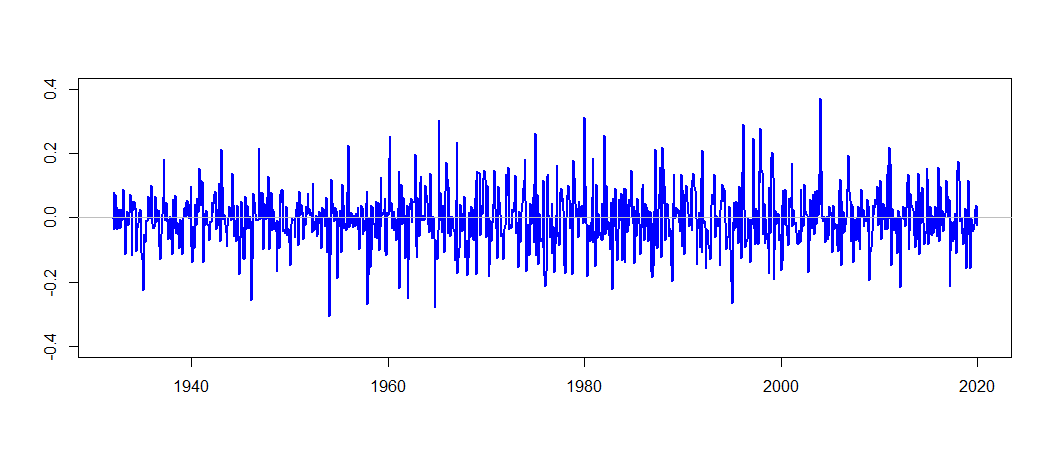}
\includegraphics[width=0.40\textwidth]{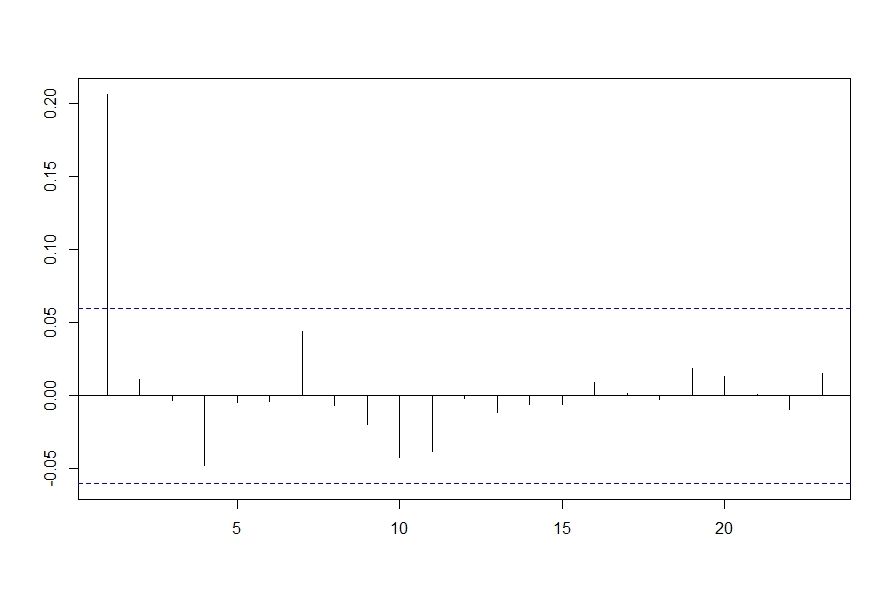}
\includegraphics[width=0.40\textwidth]{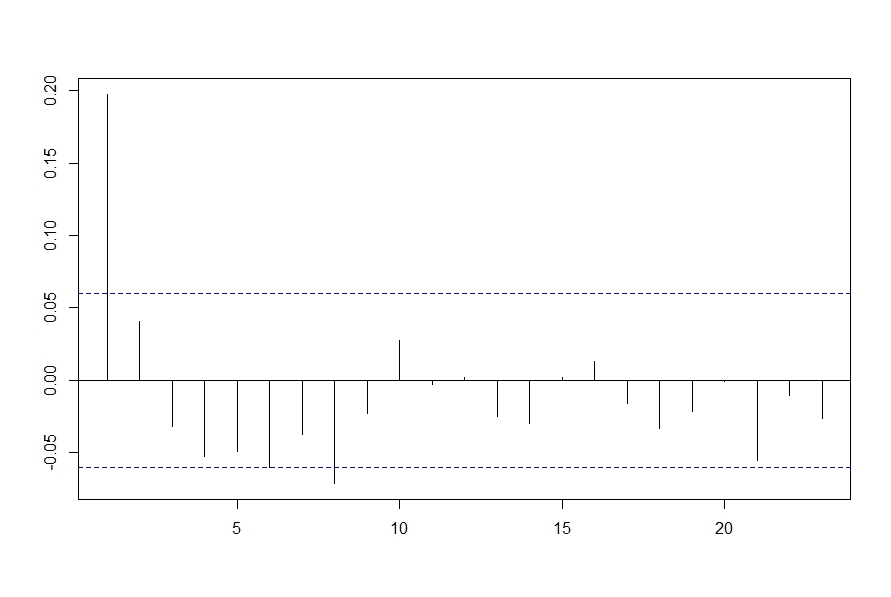}
\includegraphics[width=0.40\textwidth]{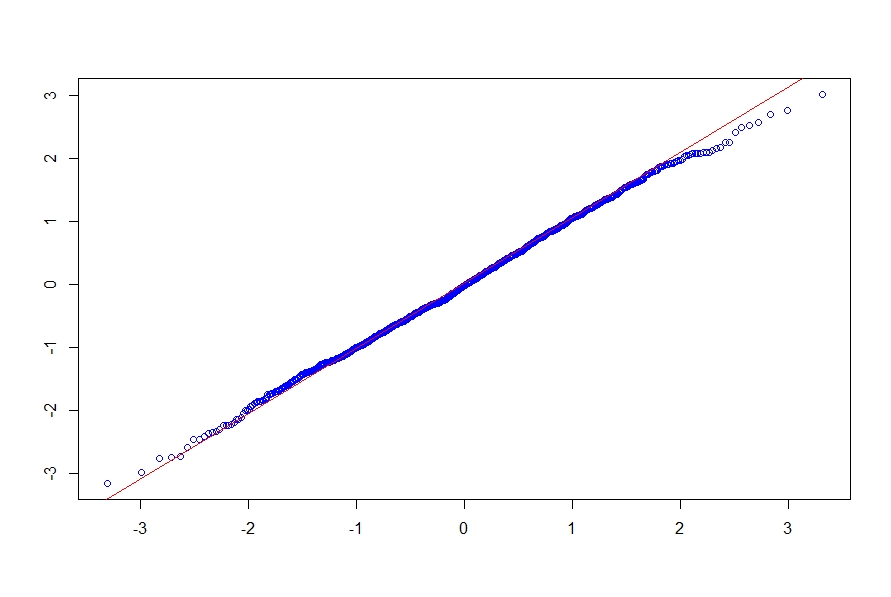}
\includegraphics[width=0.40\textwidth]{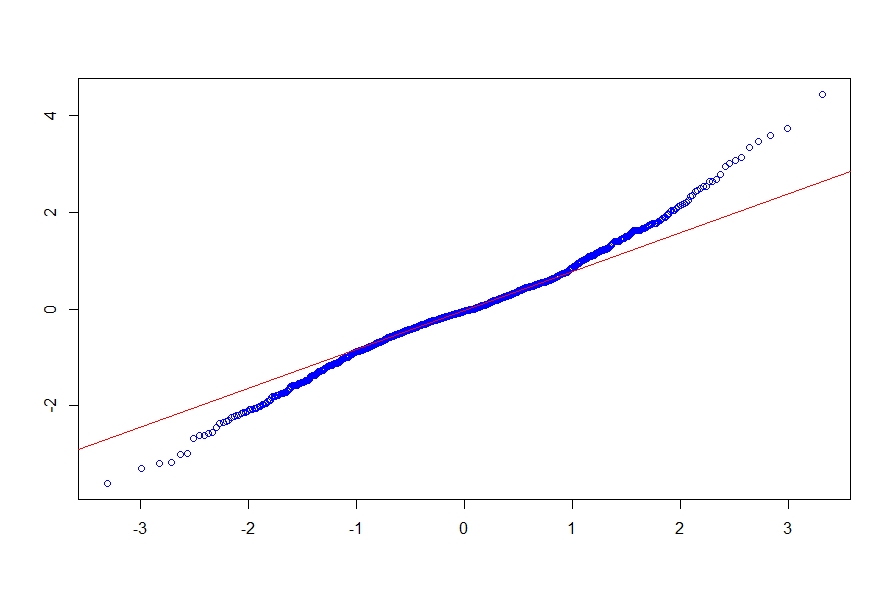}
\end{center}
\end{figure}

The estimated parameters of model (\ref{SSM}) are reported in Table \ref{tab:estimation_0}, which also reports the RWD and IRW statistics as well as those for the null of the seasonal variances being zero.\footnote{Note that, based on previous analysis not reported to save space, the seasonal variances of the last five frequencies are assumed to be equal.} Some diagnostic statistics can be found in Figures \ref{fig:BarcelonaDiag_0} to \ref{fig:SevilleDiag_0}. We observe that, regardless of whether we look at centre or log-range temperature and of the location, the residuals are uncorrelated. Furthermore, as expected, although normality seems to be an appropriate approximation for centre temperature, log-ranges have heavy tails, mainly in the two interior locations.

\begin{table}[ht!]
\caption{Estimation results of separate state space models fitted to centre and log-range temperature in four locations in the Iberian Peninsula: i) Estimated variances together with statistics of tests for deterministic components: RWD, IRW, $H_{0I}:\sigma^{2(I)}_{\omega}=0$ and $H_{0II}:\sigma^{2(II)}_{\omega}=0$; ii) Estimated components at the end of the sample period together with standard deviations in parenthesis.}
\label{tab:estimation_0}
\begin{center}
\begin{tabular}{lcccccc}
\hline
& Centre & Log-range &&& Centre & Log-range \\
\hline
&\multicolumn{2}{c}{Barcelona} &&& \multicolumn{2}{c}{Coru\~{n}a}\\
\hline
$\sigma^2_{\varepsilon}$ & 1.264 & 0.004 &&& 1.141 & 0.018 \\
$\sigma^2_{\eta}$ & $1.14\times 10^{-16}$ & $4.20 \times 10^{-6}$ &&& $1.50 \times 10^{-16}$ & $5.12 \times 10^{-5}$ \\
RWD & 0.097 & $1.028^{***}$ &&& 0.081 & $1.846^{***}$\\
$\sigma^2_{\zeta}$ & $7.85 \times 10^{-8}$ & $1.28 \times 10^{-33}$ &&& $1.24 \times 10^{-7}$ & $1.62 \times 10^{-33}$ \\
IRW & $0.808^{**}$ & 0.004 &&& $1.554^{***}$ & 0.003 \\
$\sigma^{2(I)}_{\omega}$ & $3.2 \times 10^{-4}$ & $7.53 \times 10^{-28}$ &&& $3.5 \times 10^{-4}$ & $9.22 \times 10^{-28}$\\
$H_{0I}$ & $1.129^{***}$ & $1.213^{***}$ &&& $0.928^{***}$ & $0.795^{***}$ \\
$\sigma^{2(II)}_{\omega}$ & $6.6 \times 10^{-25}$ & $6.26 \times 10^{-168}$ &&& $9.8 \times 10^{-25}$ & $2.11 \times 10^{-167}$\\
$H_{0II}$ & $3.521^{***}$ & 1.421 &&& $2.249^{**}$ & $2.675^{**}$ \\
$\mu_T$ & 17.294 & 2.199 &&& 14.298 & 2.177 \\
        & (0.03) & (0.00)&&& (0.03) & (0.00)\\
$\beta_T$ & 0.005 & $3.15 \times 10^{-5}$ &&& 0.004 & 0.000\\
          & (0.00) & (0.00)               &&& (0.00) & (0.00)\\
\hline
& \multicolumn{2}{c}{Madrid} &&& \multicolumn{2}{c}{Seville}\\
\hline
$\sigma^2_{\varepsilon}$ & 1.111 & 0.007 &&& 1.001 & 0.006 \\
$\sigma^2_{\eta}$  & $1.52\times 10^{-16}$ & $5.84\times 10^{-6}$ &&& $2.03 \times 10^{-16}$ & $6.31 \times 10^{-5}$ \\
RWD & 0.131 & $4.299^{***}$ &&& 0.181 & $6.282^{***}$\\
$\sigma^2_{\zeta}$ & $1.13\times 10^{-7}$ & $1.41 \times 10^{-33}$ &&& $1.10 \times 10^{-7}$ & $1.36 \times 10^{-33}$  \\
IRW & $1.912^{***}$ & 0.108 &&& $3.271^{***}$ & 0.077\\
$\sigma^{2(I)}_{\omega}$ & $ 6.3\times 10^{-4}$ & $8.14 \times 10^{-28}$ &&&$7.58 \times 10^{-4}$ & $7.89 \times 10^{-28}$ \\
$H_{0I}$ & 0.173 & 0.190 &&& 0.116 & $1.476^{***}$\\
$\sigma^{2(II)}_{\omega}$ & $9.81\times 10^{-25}$ & $1.00 \times 10^{-167}$ &&& $1.45 \times 10^{-24}$ & $8.27 \times 10^{-164}$ \\
$H_{0II}$ & $2.765^{**}$ & 1.317 &&& $2.261^{**}$ & $2.980^{***}$\\
$\mu_T$ & 15.887 & 2.484 &&& 18.894 & 2.572 \\
        & (0.03) & (0.00)&&& (0.03) & (0.00)     \\
$\beta_T$ & 0.003 & 0.000 &&& 0.003 & 0.000\\
          & (0.00) & (0.00) &&& (0.00) & (0.00) \\
\hline
\end{tabular}\newline\\
\end{center}
	{\small \textit{Notes:} * significant at 10\% level; ** significant at 5\% level; *** significant at 1\% level.\\} 
\end{table}%

Several important conclusions about the nature of the trend and seasonal component of the centre (average) and log-range (variability) temperature in each of the four locations considered can be extracted from the estimated variances reported in Table \ref{tab:estimation_0}. First, we can observe that, regardless of the location, the trend of the centre temperature is represented by an integrated random walk with $\hat{\sigma}^2_{\eta}$ being not significantly different from zero, while $\hat{\sigma}^2_{\xi} \neq 0$. This implies a stochastic trend with a smooth evolution. This result concurs with the works that conclude that the trend of average temperature is stochastic. Furthermore, according to the integrated random walk model, the slope of the trend can evolve over time. 

\begin{figure}[]
\caption{Estimated trend of centre temperature together with 95\% confidence intervals: Barcelona (top left), Coru\~{n}a (top right), Madrid (bottom left), Seville (bottom right).}
\label{fig:Trend_centre}
\begin{center}
\includegraphics[width=0.47\textwidth]{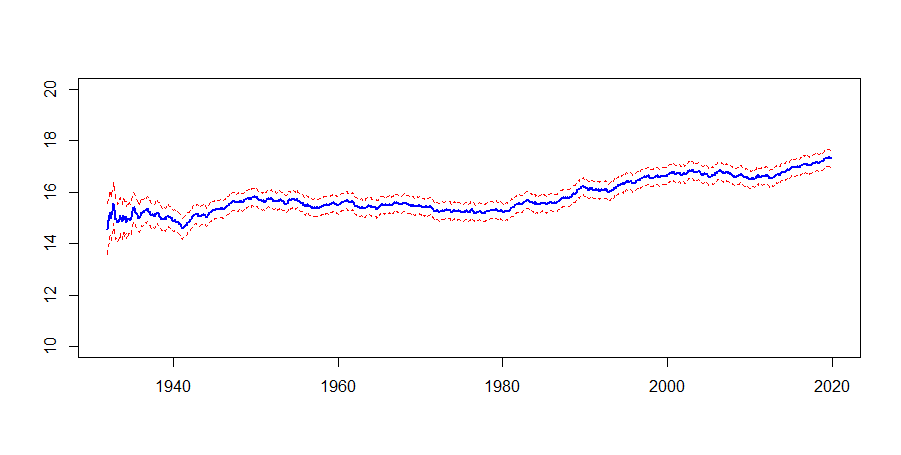}
\includegraphics[width=0.47\textwidth]{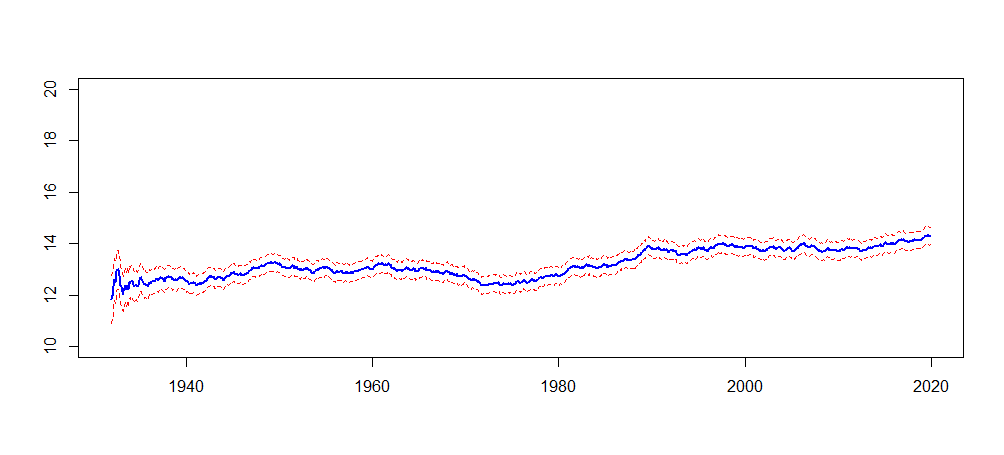}
\includegraphics[width=0.47\textwidth]{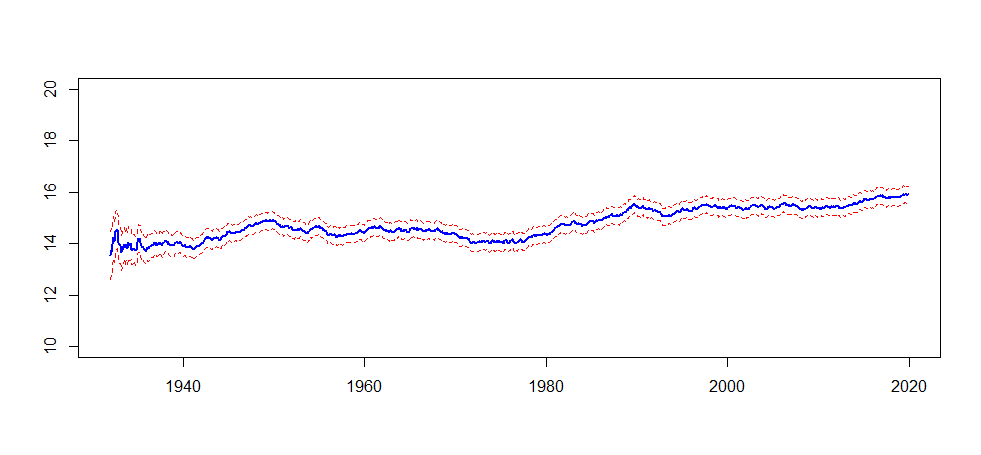}
\includegraphics[width=0.47\textwidth]{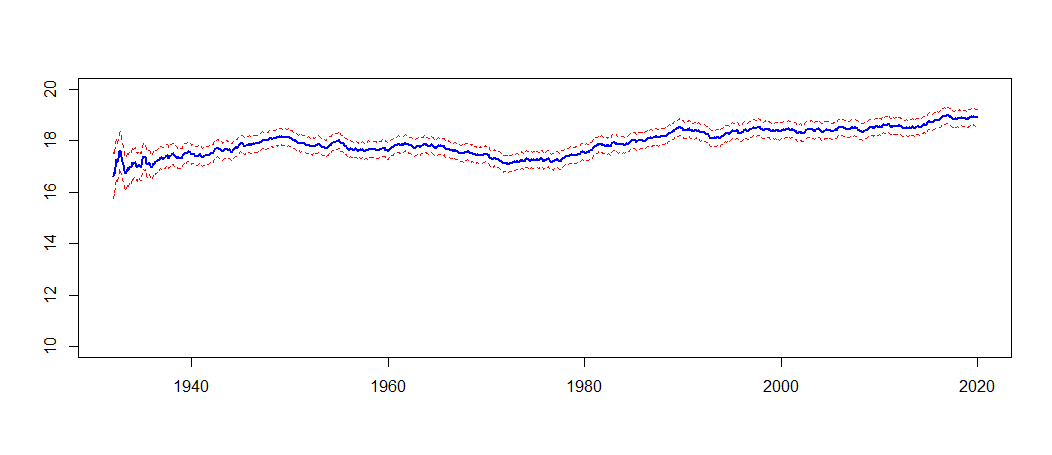}
\end{center}
\end{figure}

Figure \ref{fig:Trend_centre} displays the estimated (smoothed) trends of centre temperature in the four selected locations together with 95\% bounds obtained using the MSEs delivered by the Kalman filter. The figure also shows that there is a smoothly increasing trend over the last 90 years. Consider, for example, the centre temperature at Barcelona, which starts around $15^\circ C$ in 1930 while the estimated underlying level in December 2020 is $17.298^\circ C$. Note that even if we consider the uncertainty in the estimated trend, the increase of 2 degrees in the centre temperature over the last 90 years is clearly significant. We can also observe in Figure \ref{fig:Trend_centre} that the slope of the trend is changing over the sample period, and, except for Barcelona, it is smaller since 2000. This deceleration is in concordance with the hiatus of average temperature described above. However, note that there is a recent acceleration since the second decade of the XXI Century, which is more pronounced in Barcelona and less in Seville. Table \ref{tab:estimation_0} also reports the estimated trends and slopes at the end of the sample period, December 2020, for the centre at each location. Note that, in December 2020, the estimated level of the trend is largest in Seville (18.896) and Barcelona (17.294) followed by Madrid (15.889) and Coru\~{n}a (14.301). Even more important is to observe that the estimated slopes of the trend are larger in the coastal locations (0.004) and smaller at the interior locations, Madrid (0.003) and Seville (0.002), with these two latter slopes being not significant.

\begin{figure}[]
\caption{Estimated seasonal component of centre temperature together with 95\% confidence intervals: Barcelona (top left), Coru\~{n}a (top right), Madrid (bottom left), Seville (bottom right).}
\label{fig:Seasonal_centre}
\begin{center}
\includegraphics[width=0.47\textwidth]{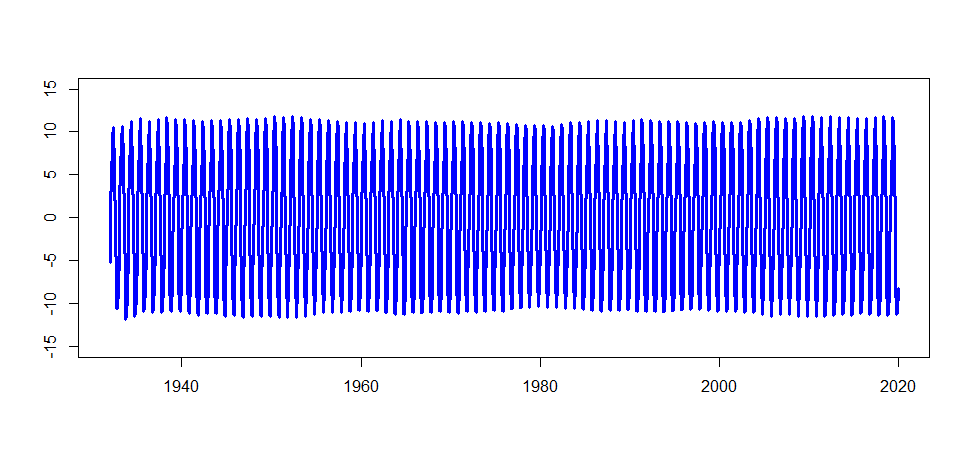}
\includegraphics[width=0.47\textwidth]{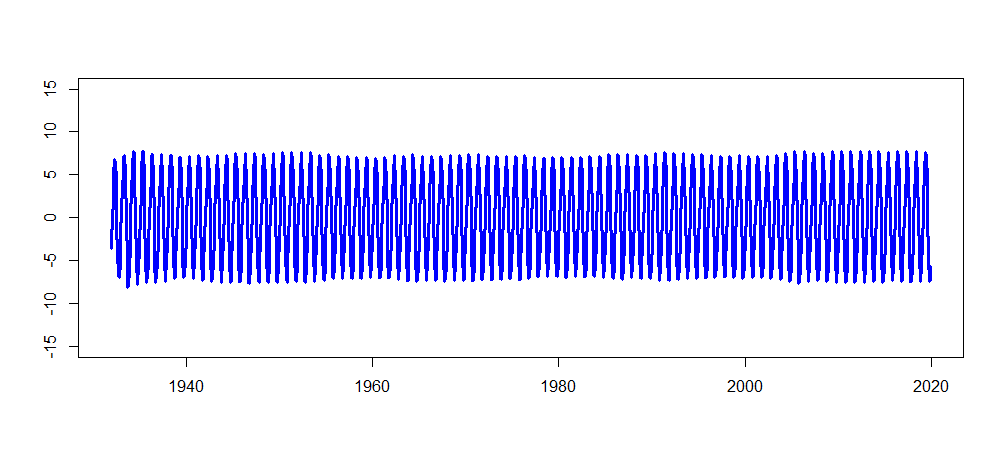}
\includegraphics[width=0.47\textwidth]{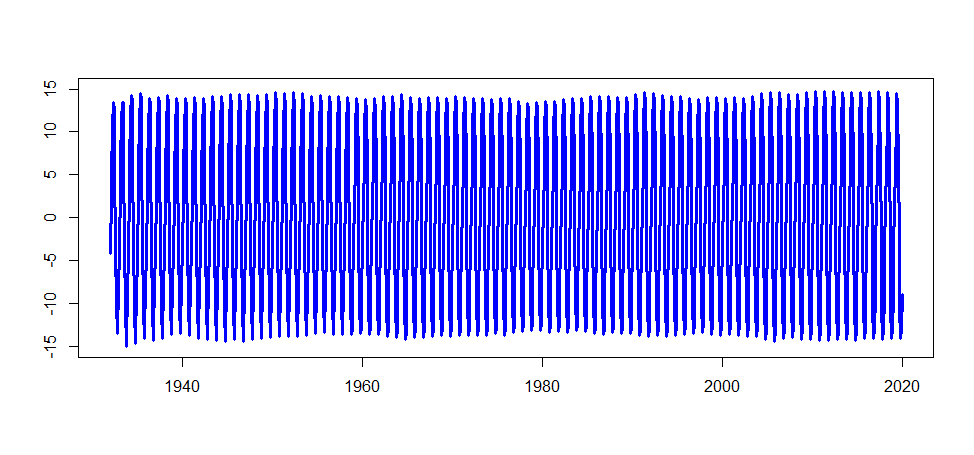}
\includegraphics[width=0.47\textwidth]{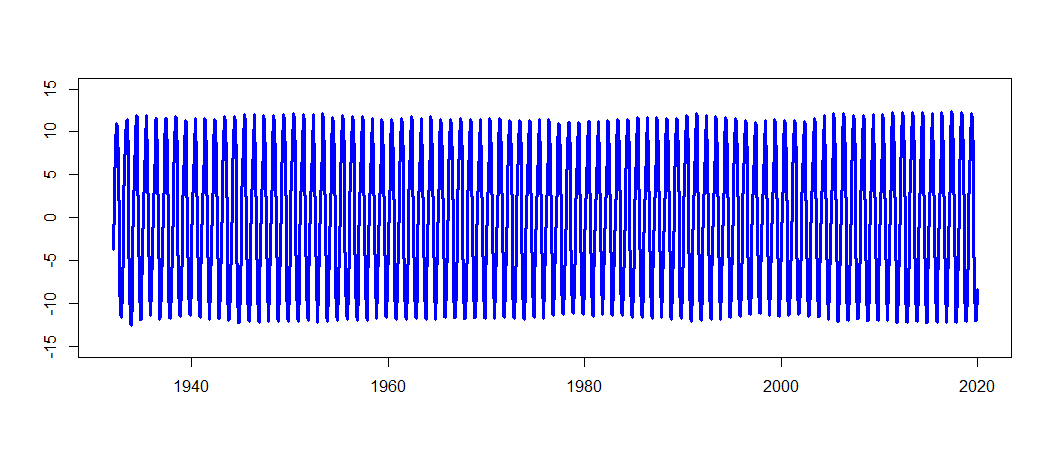}
\end{center}
\end{figure}

Second, we look now at the results for the seasonal component of the centre. Regardless of the location, we observe that the seasonal component is stochastic, with the corresponding variances significantly different from zero. The evolution of the estimated smoothed seasonal component of the centre temperature in each of the four locations is plotted in Figure \ref{fig:Seasonal_centre}. Our results are in line with those works that support an stochastic evolution of the seasonality of temperatures, even if, as observed in Figure \ref{fig:Seasonal_centre}, this evolution is very smooth; see also the results by Bogalo, Poncela and Senra (2024), who also conclude that there are changes in the seasonal pattern.

\begin{figure}[]
\caption{Estimated trend component of log-range temperature together with 95\% confidence intervals: Barcelona (top left), Coru\~{n}a (top right), Madrid (bottom left), Seville (bottom right).}
\label{fig:Trend_LR}
\begin{center}
\includegraphics[width=0.47\textwidth]{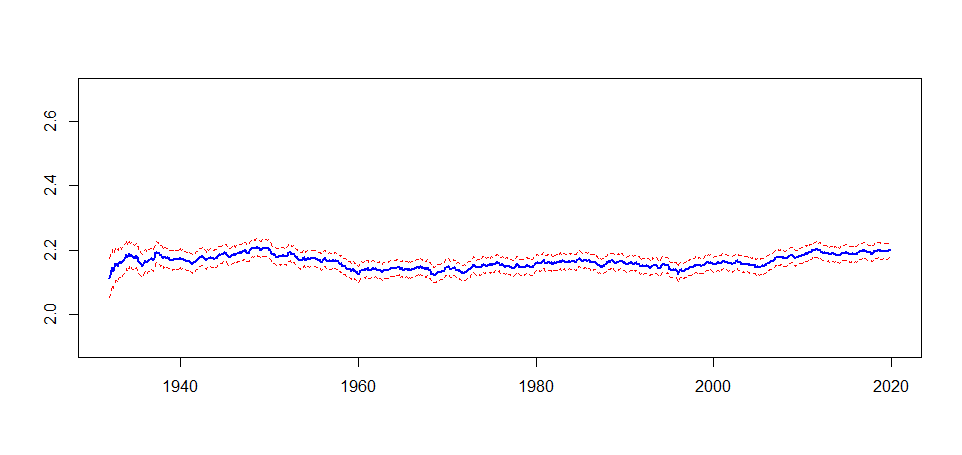}
\includegraphics[width=0.47\textwidth]{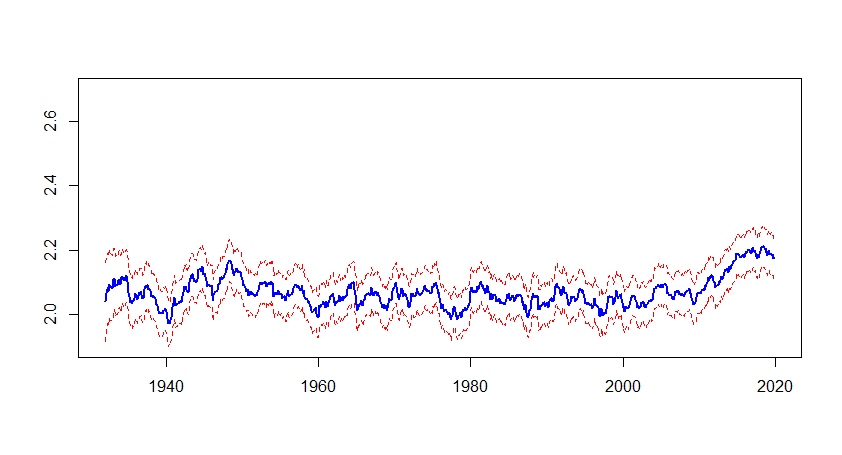}
\includegraphics[width=0.47\textwidth]{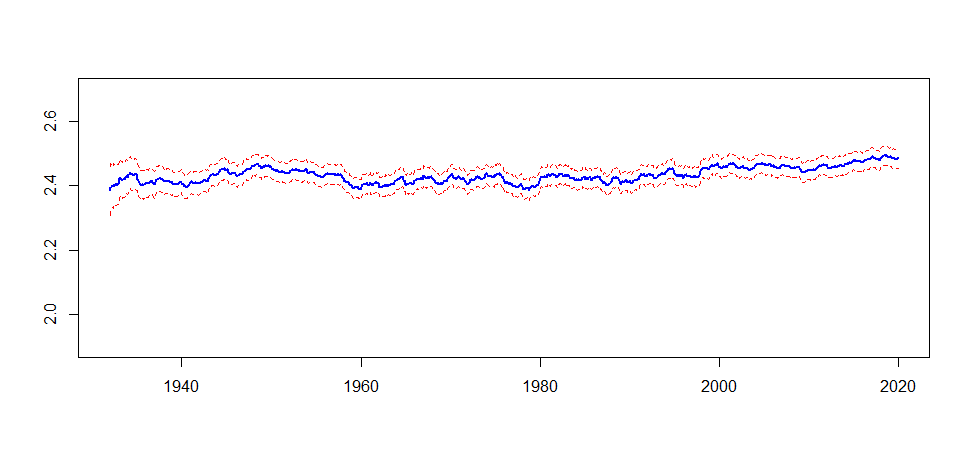}
\includegraphics[width=0.47\textwidth]{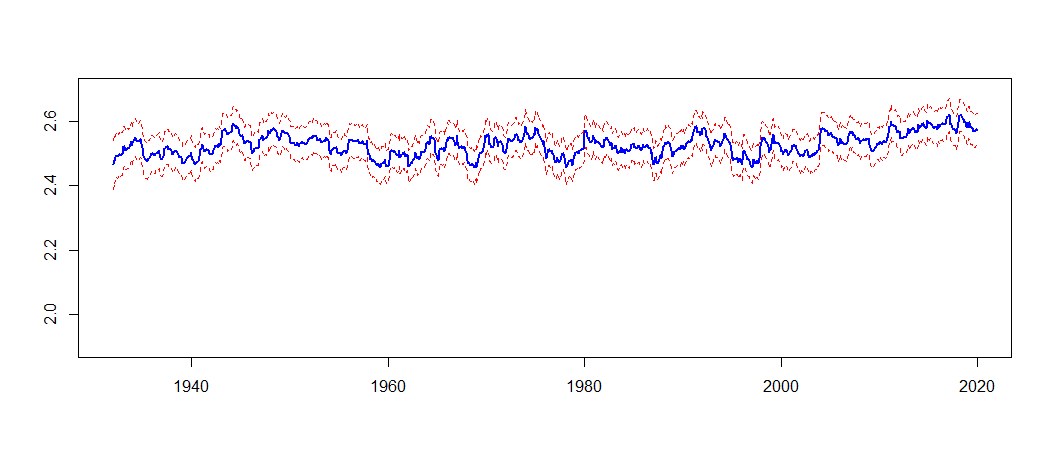}
\end{center}
\end{figure}

However, the trend of the log-range has a constant slope equal to zero, implying that $\mu_t$ is a random walk. Henceforth, the level of log-range is non-stationary implying that minimum and maximum temperatures are not cointegrated. This finding is relevant because it shows that coldest and warmest months over time do not follow the same stochastic pattern in the long run. Figure \ref{fig:Trend_LR} plots the smoothed estimated level of the log-range temperature, and shows a random evolution without a particular trend. Table \ref{tab:estimation_0} reports the estimated value of $\mu_T$ and shows that it is larger in Madrid and Seville (2.48 and 2.57, respectively) than at the coastal cities (2.20 at Barcelona and 2.18 at Coru\~{n}a). The dispersion between minimum and maximum temperatures has increased in recent years.

\begin{figure}[]
\caption{Estimated seasonal component of log-range temperature together with 95\% confidence intervals: Barcelona (top left), Coru\~{n}a (top right), Madrid (bottom left), Seville (bottom right).}
\label{fig:Seasonal_LR}
\begin{center}
\includegraphics[width=0.47\textwidth]{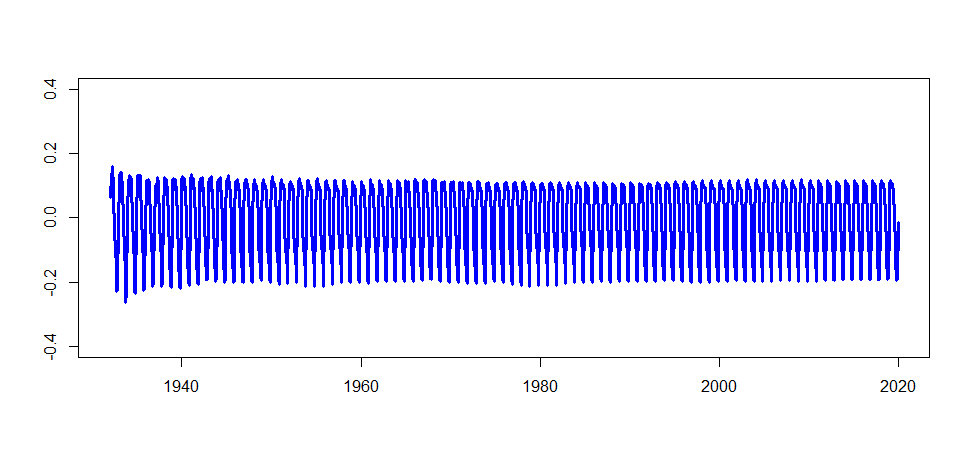}
\includegraphics[width=0.47\textwidth]{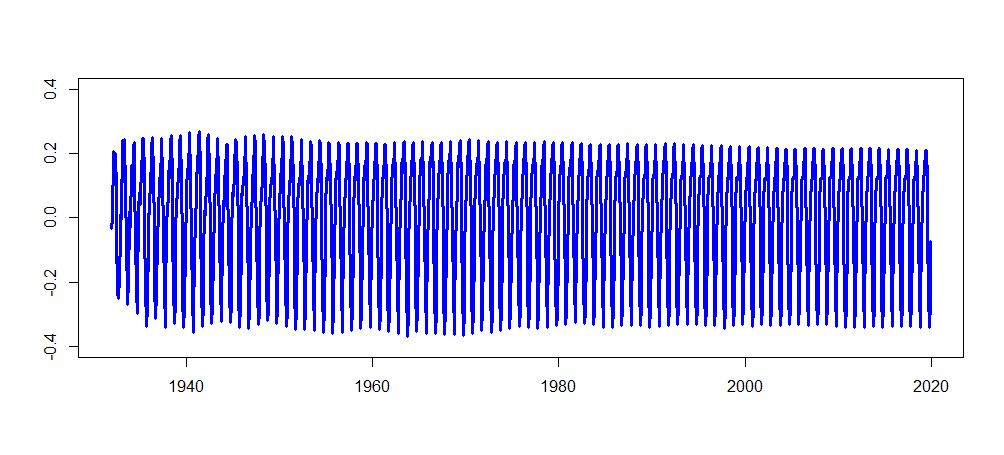}
\includegraphics[width=0.47\textwidth]{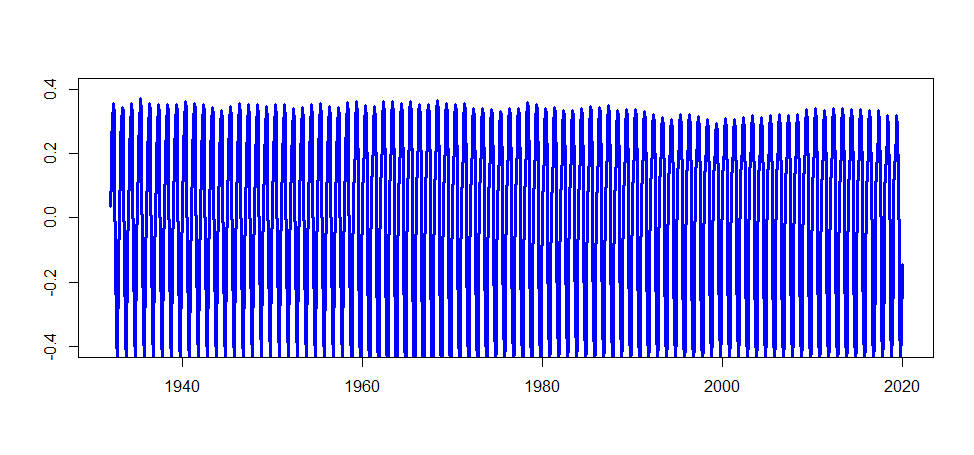}
\includegraphics[width=0.47\textwidth]{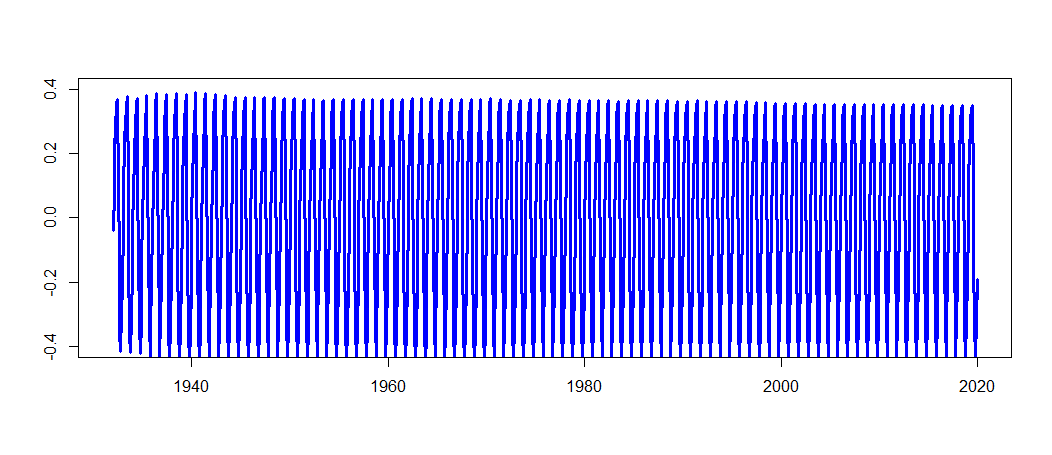}
\end{center}
\end{figure}

Finally, regardless of the location, the seasonal component of the dispersion between minimum and maximum temperature is strong and has a stochastic evolution with the corresponding variances being significantly different from zero; see Figure \ref{fig:Seasonal_LR}. 

\section{Multivariate modelling of centre and log-range temperature in the Iberian Peninsula}
\label{section:multivariate}

It is well known that local climate can exacerbate the impact of global warming; see, for example, Estrada and Perron (2021). Consequently, disentangling the local and global drivers of warming can guide local adaptation policies. In this section, we fit a multi-level DFM to the centre and log-range temperature observed at 68 locations in the Iberian Peninsula. As discussed in previous sections, the seasonal component is ubiquitous. Therefore, we extract the seasonal component from the centre and log-range temperature before analysing the common patterns in temperature trends.\footnote{The seasonal component is extracted from each series after fitting a structural model by subtracting the estimated filtered seasonal component.} 
\subsection{Muti-level DFM for centre temperature}

We first analyse the degree of comovement in the trends of centre temperature in the Iberian Peninsula. We fit the ML-DFM that decomposes each centre temperature into (i) a global component that is common to all centre temperature, (ii) one common component specific to each of the five regions as determined in Figure \ref{fig:correlation_C}, and (iii) purely idiosyncratic components. According to the results in the univariate analysis, this common component is assumed to follow an integrated random walk. The regional specific components are assumed to follow stationary autoregressive models; see a detailed description of the ML-DFM fitted to the centre temperature in the Appendix. 


The ML-DFM is estimated by ML implemented using the expectation-maximization (EM) algorithm proposed by Doz, Giannone and Reichlin (2012); see Delle Chiaie, Ferrara and Giannone (2022) for the implementation of the algorithm to a ML-DFM. The EM algorithm is modified to take into account that the global common component is an integrated random walk; see the Appendix for details. Figure \ref{fig:gobal_factor} plots the estimated global factor (filtered estimates). We can observe that during the sample period, the global factor has a smoothly evolving trend with a positive slope in the more recent years. 
Figure \ref{fig:gobal_factor} also displays the estimated loadings in each of the locations in the sample and shows that they are similar and close to one.

\begin{figure}[]
\caption{First row: Global common factor of centre temperature during the entire sample period in black, and (standardised) deseasonalised monthly centre temperature at 68 locations in the Iberian Peninsula in lightblue (first row). Second row: Estimated loading in each location.}
\label{fig:gobal_factor}
\begin{center}
\includegraphics[width=0.95\textwidth]{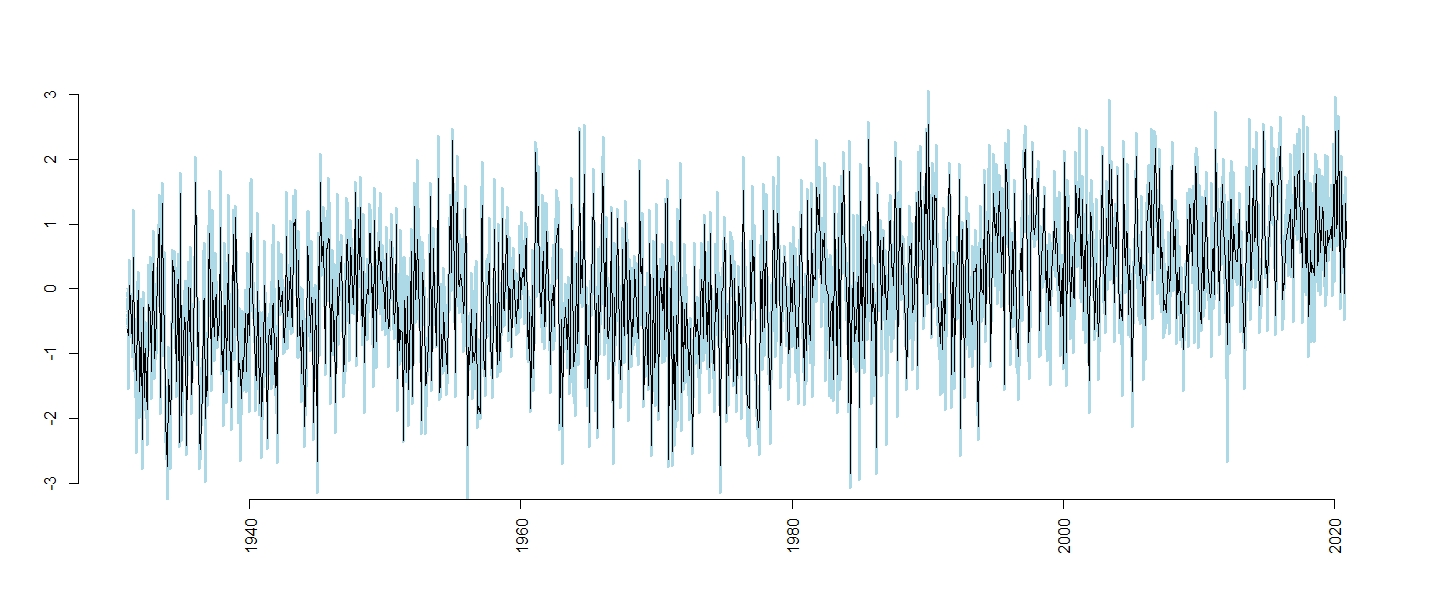}
\includegraphics[width=0.80\textwidth]{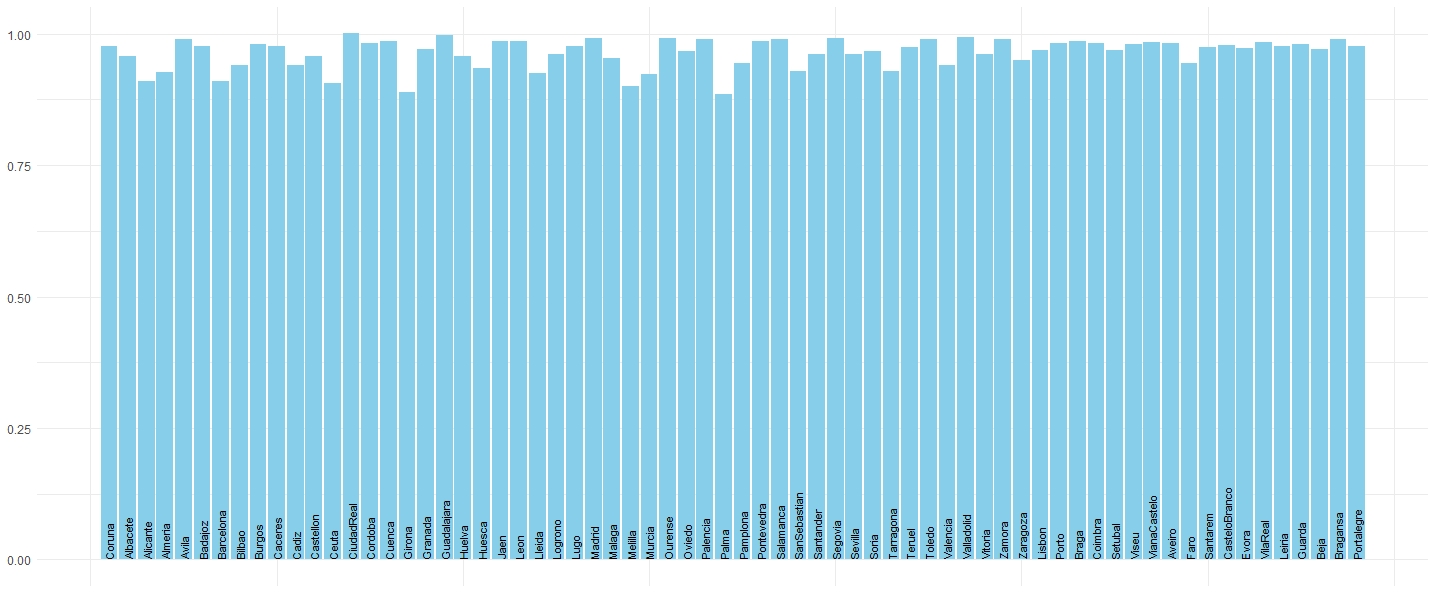}
\end{center}
\end{figure}

\begin{figure}[]
\caption{Regional common factors of centre temperature during the full sample period (top panel). The bottom panel shows loadings of the regional factors (red) and the corresponding loadings of the global factors (blue).}
\label{fig:regional_factors}
\begin{center}
\includegraphics[width=0.80\textwidth]{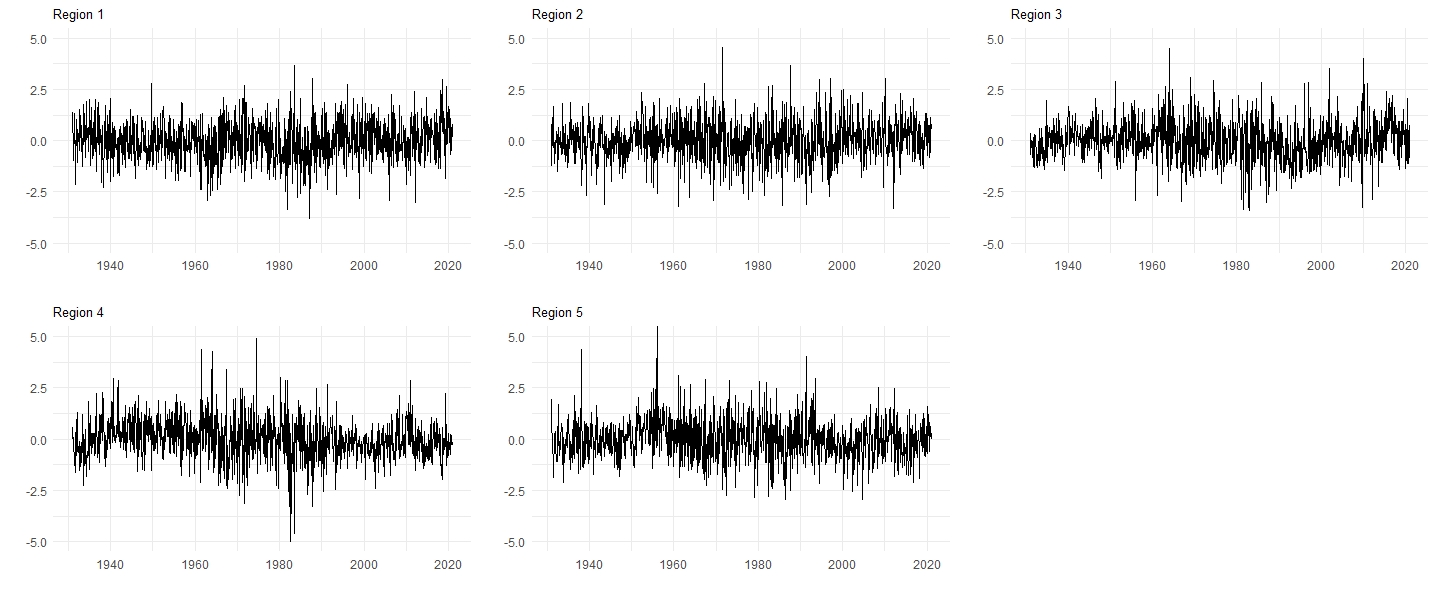}
\includegraphics[width=0.80\textwidth]{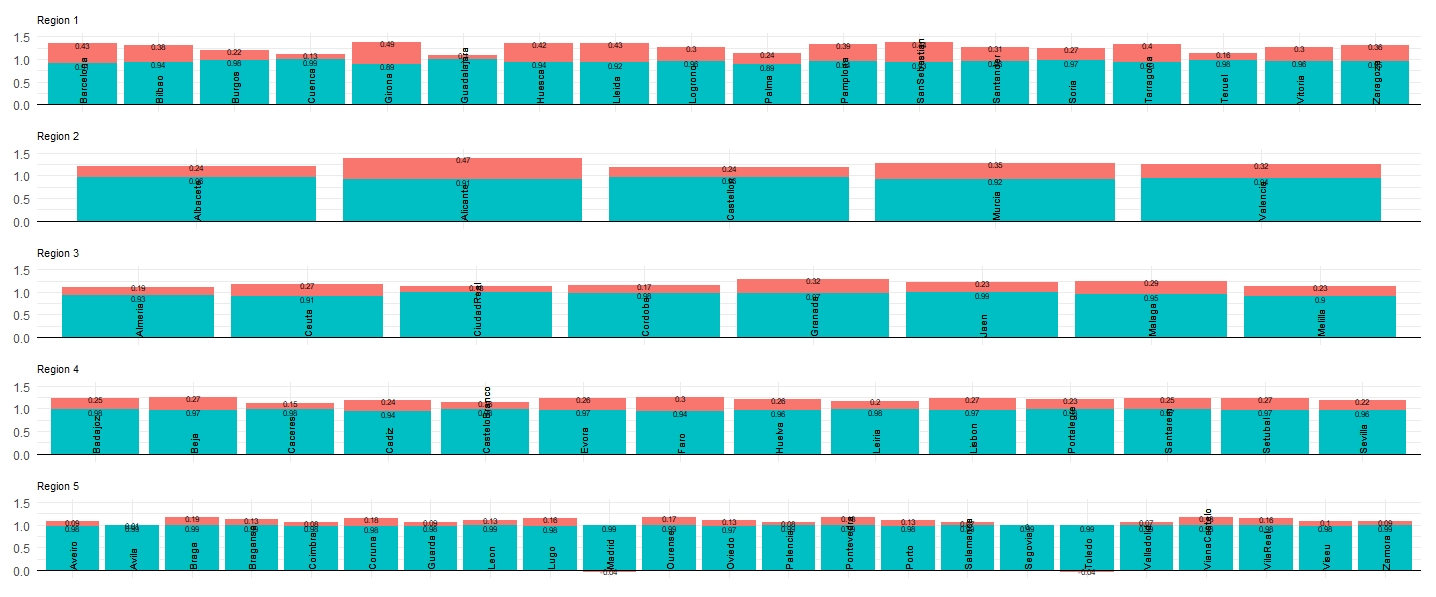}
\end{center}
\end{figure}

\begin{figure}[]
\caption{Map of the Iberian Peninsula with the global (left) and regional (right) loadings of the centre temperature in each location. The darker colors represent larger values.}
\label{fig:factors_centre_map}
\begin{center}
\includegraphics[width=0.45\textwidth]{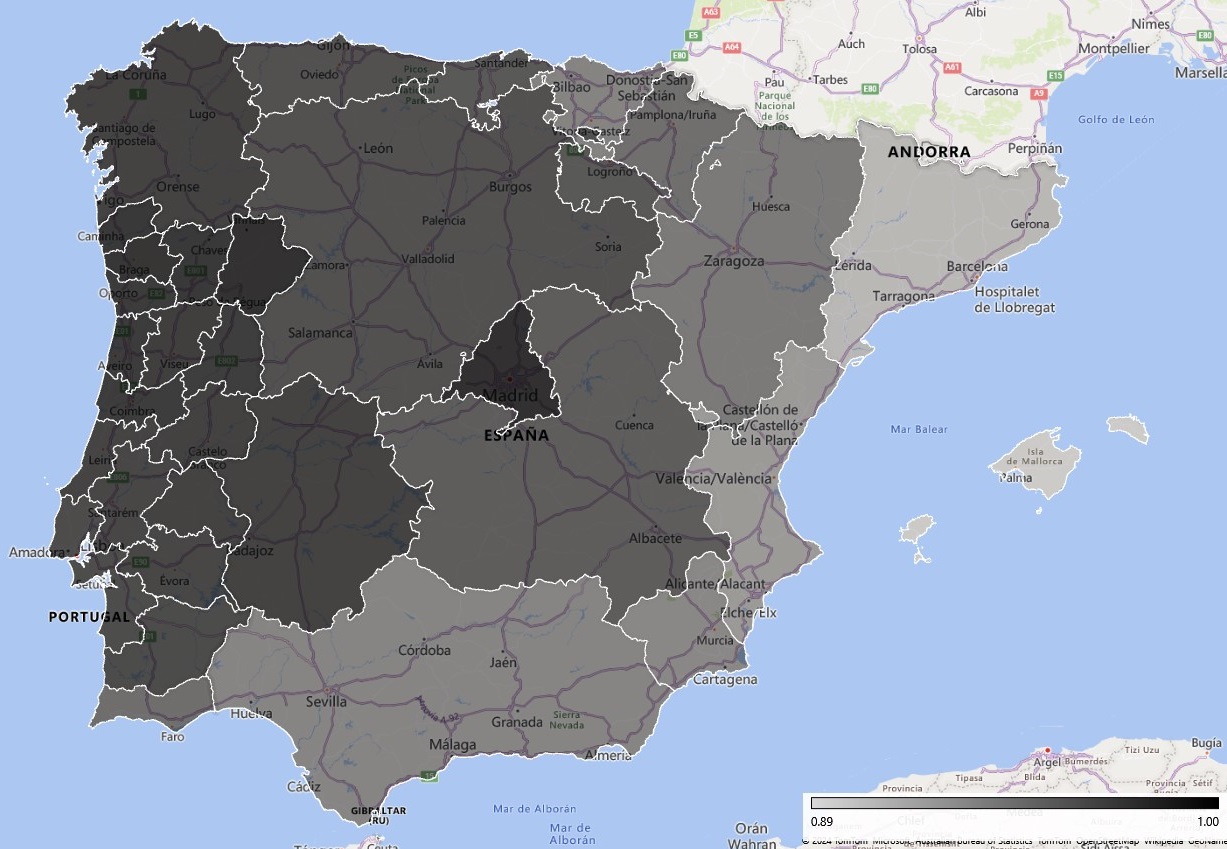}
\includegraphics[width=0.45\textwidth]{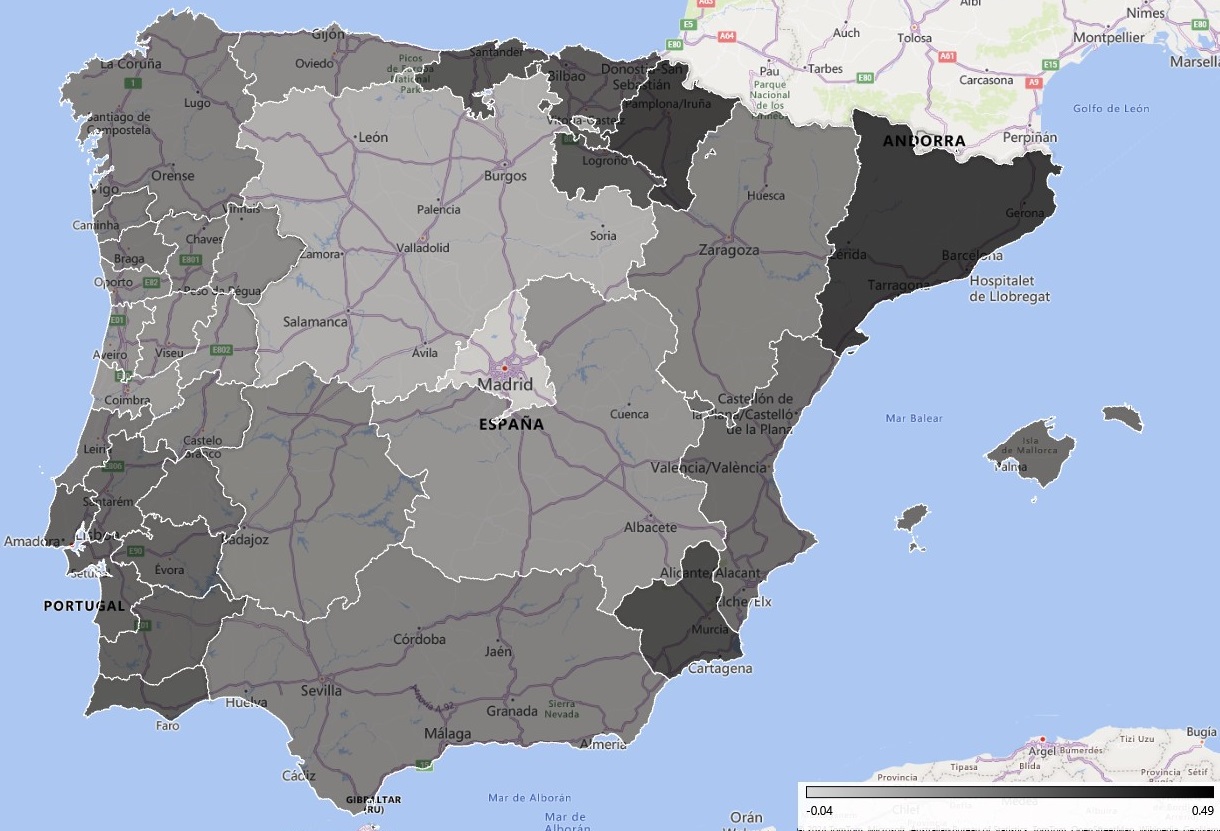}
\end{center}
\end{figure}

Consider now the estimated common factors of the centre temperature corresponding to each of the five regions described above. Figure \ref{fig:regional_factors} plots such common regional factors together with their corresponding loadings. The common regional factors do not show any further trend being stationary. Furthermore, the loadings of the regional factors are relatively small when compared with the loadings of the common global trend. This is especially the case in Region 5, while in Region 1, the weight of the regional factor is relatively larger when compared with that of the global common trend; see also Figure \ref{fig:factors_centre_map} that represents the map of the Iberian Peninsula with the global and regional factors at each location.      

\subsection{Multi-level DFM for log-range temperature}

Finally, we consider the joint modelling of the log-range temperature at the 68 locations in the Iberian Peninsula. The ML-DFM is fitted by assuming a global factor common to all regions, modelled as a random walk, and one stationary regional factor common to the locations in each of the three regions determined in Figure \ref{fig:correlation_LR}.

\begin{figure}[]
\caption{First row: Global common factor of log-range temperature during the entire sample period in black, and (standardised) deseasonalised monthly log-range temperature at 68 locations in the Iberian Peninsula in lightblue (top row). Second row: Estimated loading in each location.
}
\label{fig:global_factor_range}
\begin{center}
\includegraphics[width=0.95\textwidth]{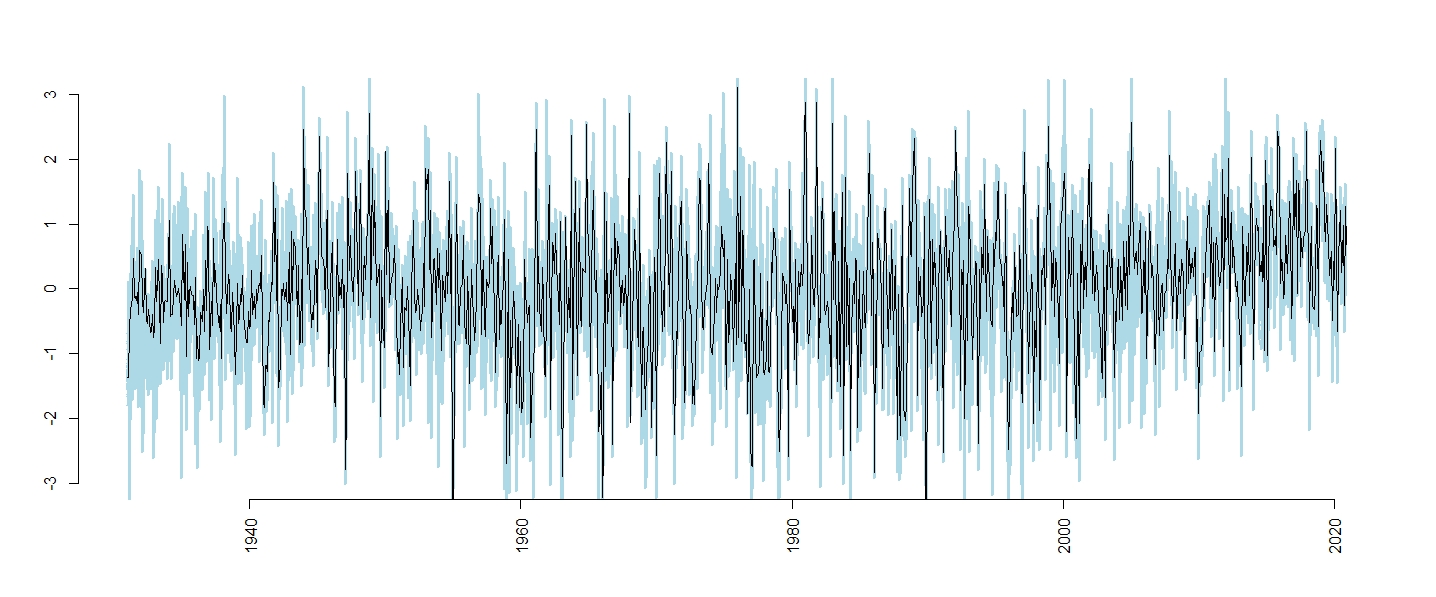}
\includegraphics[width=0.65\textwidth]{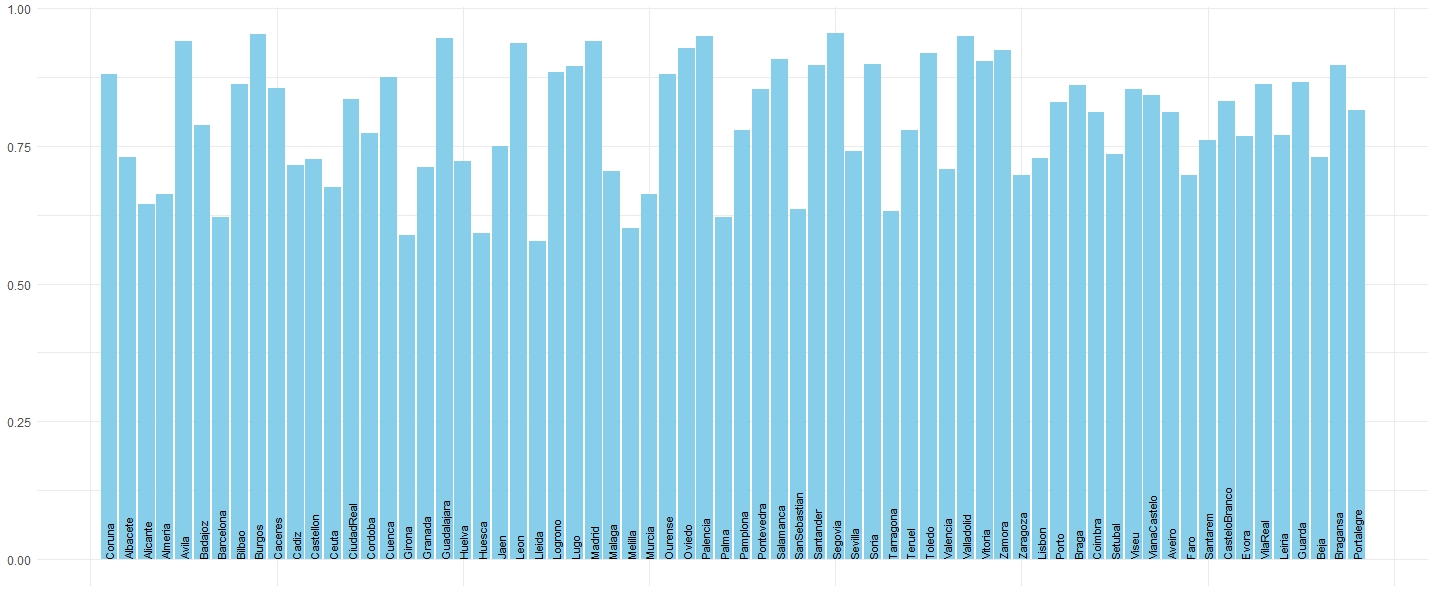}
\end{center}
\end{figure}

Figure \ref{fig:global_factor_range} plots the common global factor of the log-range temperature in the Iberian Peninsula. It shows a pronounced increase in the distance between the minimum and maximum temperature over the last twenty years, which might be attributed to the fact that the maximum temperature is increasing faster. Figure \ref{fig:global_factor_range} also show that the loadings of the global factor can vary between 0.6 in Lleida and 0.96 in Burgos, although they are positive and large in all locations.

\begin{figure}[]
\caption{Regional common factors for log-range temperature during the full sample period (top panel) together with the corresponding loadings of the regional factors (red) and the global factor (blue) (bottom panel) .}
\label{fig:regional_factors_range}
\begin{center}
\includegraphics[width=0.80\textwidth]{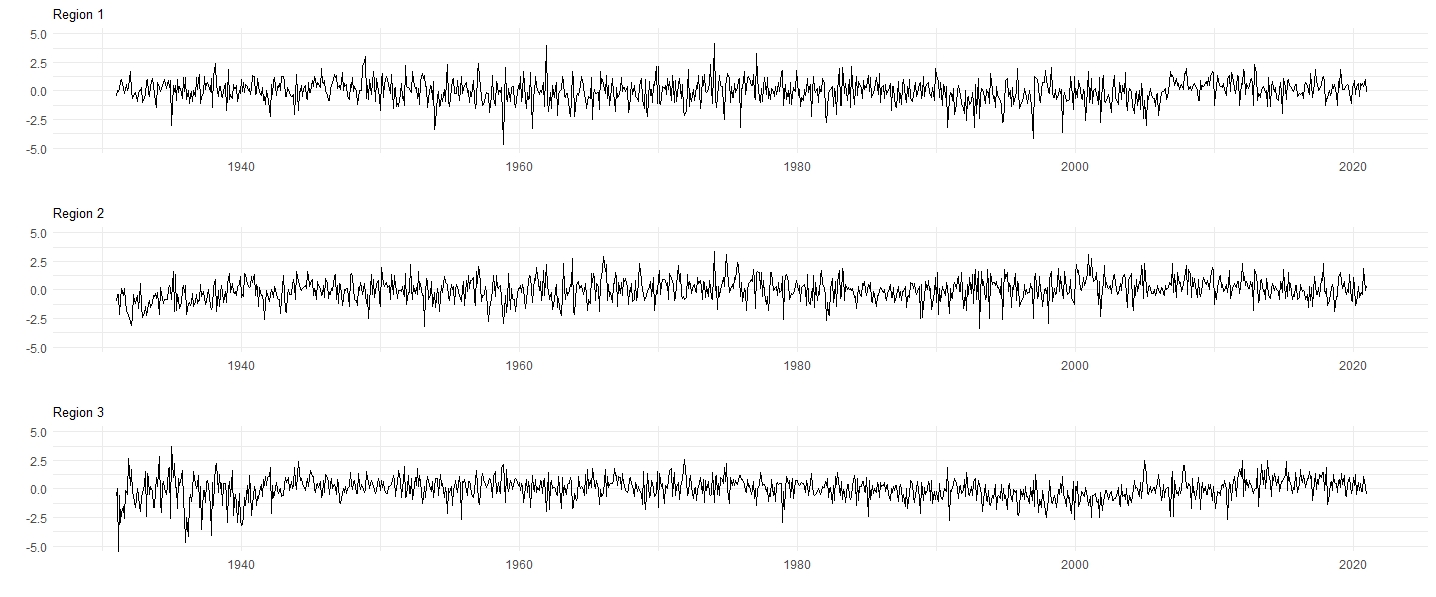}
\includegraphics[width=0.80\textwidth]{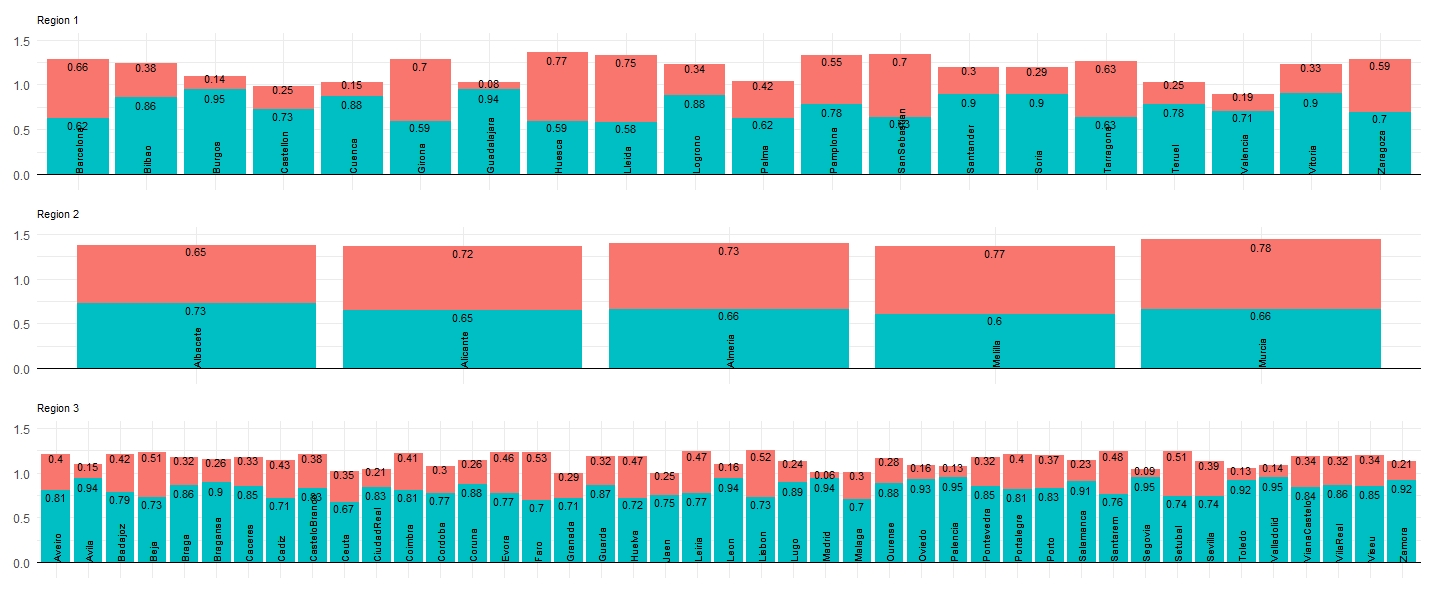}
\end{center}
\end{figure}

\begin{figure}[]
\caption{Map of the Iberian Peninsula with the global (left) and regional (right) loadings of the log-range temperature in each location. The darker colors represent larger values.}
\label{fig:factors_range_map}
\begin{center}
\includegraphics[width=0.45\textwidth]{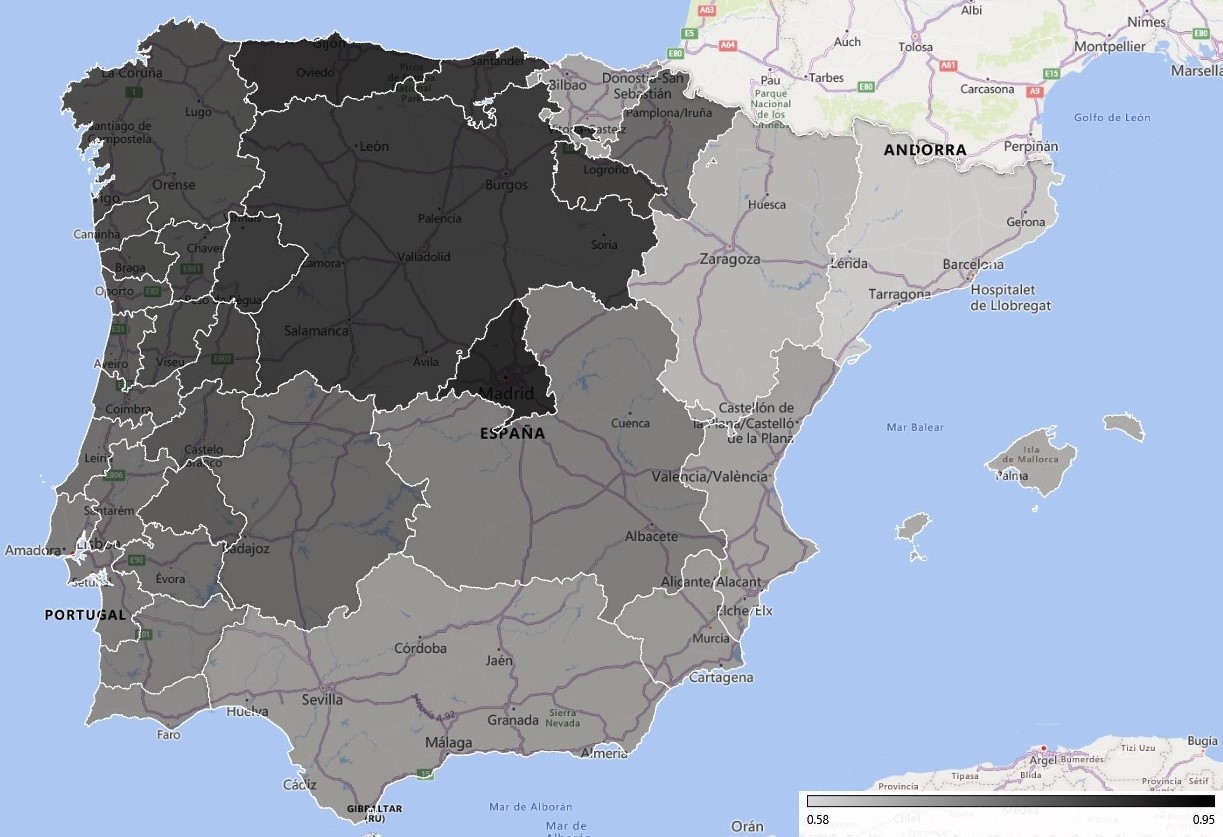}
\includegraphics[width=0.45\textwidth]{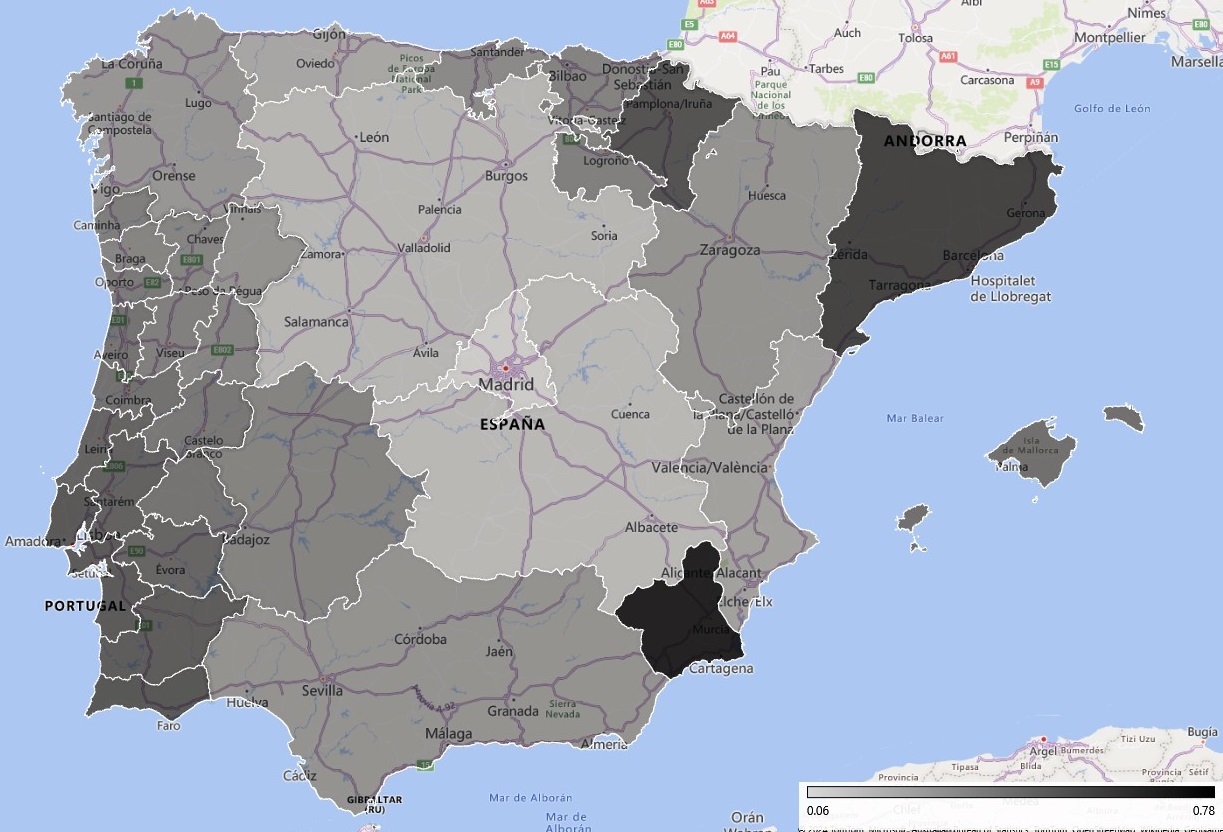}
\end{center}
\end{figure}

To conclude, Figure \ref{fig:regional_factors_range} plots the regional factors. Although these factors are stationary, they seem to have an increase in the level during the last decade, mainly in regions 1 and 3. When looking at the loadings of the regional factors, plotted in the same figure, we can observe that they are relatively large when compared with the loadings of the global factors. Regional factors have an important weight in explaining the evolution of the differences between maximum and minimum temperature in the Iberian Peninsula; see also Figure \ref{fig:factors_range_map}.

\section{Conclusions}
\label{section:conclusions}

In this paper, we propose using unobserved component models to represent the dynamic evolution of centre and log-range temperature observed monthly at different locations in the Iberian Peninsula over the last century. By doing so, we contribute to the literature on climate change in two main ways. First, from a methodological point of view, we use the information not only of the centre (average) temperature but also information associated to extreme (minimum and maximum) temperatures. Also, the methodology implemented in this paper allows the incorporation of seasonal patterns and the distinction between deterministic and stochastic components in temperatures. Finally, the methodology can be implemented to represent centre and log-range temperature not only at a given location but also at a large number of locations. Consequently, it allows the simultaneous analysis of the common components and the heterogeneity of climate change and the analysis of the relative importance of both elements. This methodology can obviously be implemented to analyse temperature in other areas of the world.

Second, in the particular application to model temperatures in the Iberian Peninsula, we conclude that, at each particular location, the centre and log-range temperature can be modelled separately. Furthermore, they are characterised by strong seasonal stochastic components, which are similar but different in magnitude at different locations evolving very smoothly over time. We also conclude that the evolution of the centre temperature can be represented by a smooth integrated random walk, which implies changes in the slope of the trend with a recent increase at some particular locations. However, the evolution of the distance between minimum and maximum temperatures is characterised by an evolving level modelled as a random walk.

Finally, when jointly modelling centre and log-range deseasonalised temperature at several locations, we show that the centre temperature are characterised by a strong global component with relatively weaker regional components. However, the distance between the minimum and maximum temperature is characterised by global and regional components with similar weights.       





\section*{\large{References}}

Bai, J. and S. Ng (2005), Tests for skewness, kurtosis, and Normality for time series data, \textit{Journal of Business \& Economic Statistics}, 23(1), 49-60. 

Baillie, R.T. and S.K. Chung (2002), Modeling and forecasting from trend-stationary long memory models with applications to climatology, \textit{Journal of Forecasting}, 18(2), 215-226.

Banbura, M., D. Giannone and L. Reichlin (2011), Nowcasting, in Clements, M.P. and D.F. Hendry (eds.), \textit{Oxford Handbook of Economic Forecasting}, Oxford University Press.

Barbosa, S.M., M.G. Scotto and A. Alonso (2011), Summarizing changes in air temperature over Central Europe by quantile regression and clustering, \textit{Natural Hazards and Earth System Sciences}, 11, 3227-3233. 




Binelli, C., M. Loveless and B.F. Schaffner (2023), Explaining perceptions of climate change in the US, \textit{Political Research Quarterly}, 76(1), 365-380.

Bloomfield, P. (1992), Trends in global temperature, \textit{Climatic Change}, 21, 1-16.

Bogalo, J., P. Poncela and E. Senra (2024), Understanding fluctuations through multivariate circulant singular spectrum analysis, \textit{Expert Systems with Applications}, 123827.





Busetti, F. and A.C. Harvey (2003), Seasonality tests, \textit{Journal of Business \& Economic Statistics}, 21(3), 420-436.


Campbell, S.D. and F.X. Diebold (2005), Weather forecasting for weather derivatives, \textit{Journal of the American Statistical Association}, 100(469), 6-16.


 Chang, Y., R.K. Kaufmann, C.S. Kim, J.I. Millers, J.Y. Park and S. Park (2020), Evaluating trends in time series of distributions: A spatial fingerprint of human effects on climate, \textit{Journal of Econometrics}, 214(1), 274-294.

Chen, L. J. Gao and F. Vahid (2022), Global temperatures and greenhouse gases: A common features approach, \textit{Journal of Econometrics}, 230(2), 240-254.






Coggin, T.D. (2012), Using econometric methods to test for trends in the HadCRUT3 global and hemispheric data, \textit{International Journal of Climatology}, 32(2), 315-320.

Cohen, J.L., J.C. Furtado, M. Barlow, V.A. Alexeev and J.E. Cherry (2012), Asymmetric seasonal temperature trends, \textit{Geophysics Research Letters}, 39, L04705.



Delle Chiaie, S., L. Ferrara and D. Giannone (2022), Common factors of commodity prices, \textit{Journal of Applied Econometrics}, 37(3), 461--476.

Deng, Q. and Z. Fu (2019), Comparison of methods for extracting annual cycle with changing amplitude in climate series, \textit{Climate Dynamics}, 52, 5059-5070.



 Diebold, F.X. and G.D. Rudebusch (2022a), On the evolution of U.S. temperature dynamics, in Chudik, A., C. Hsiao and A. Timmermann (eds.), \textit{Essays in Honor of Hashem Pesaran: Prediction and Macro Modeling}, Advances in Econometrics, 43A, Emerald Publishing Limited.

Diebold, F.X. and G.D. Rudebusch (2022b), Probability assessments of an ice-free Arctic: Comparing statistical and climate model projections, \textit{Journal of Econometrics}, 231(2), 520-534.

Diebold, F.X. and A.S. Senhadji (1996), The uncertain unit root in real GNP: Comment, \textit{American Economic Review}, 86(5), 1291-1298.

Doz, C., D. Giannone and L. Reichlin (2011), A two-step estimator of large approximate dynamic factor models based on Kalman filtering, \textit{Journal of Econometrics}, 164(1), 188-205. 

Doz, C., D. Giannone and L. Reichlin (2012), A quasi-maximum likelihood approach for large, approximate dynamic factor models, \textit{Review of Economics and Statistics}, 94(4), 1014-1024.

Dupuis, D.J. (2012), Modelling waves of extreme temperature: the changing tails of four cities, \textit{Journal of the American Statistical Association}, 107(497), 24-39.

Dupuis, D.J. (2014), A model for nighttime minimum temperatures, \textit{Journal of Climate}, 27(19), 7207-7229.



Estrada, F. and P. Perron (2017), Extracting and analyzing the warming trend in global and hemispheric temperatures, \textit{Journal of Time Series Analysis}, 38, 711-732. 

Estrada, F. and P. Perron (2021), Disentangling the trend in the warming of urban areas into global and local factors, \textit{Annals of the New York Academy of Sciences}, 1504, 230-246.

Estrada, F., C. Gay and A. S\'anchez (2010), A reply to ``Does temperature contain a stochastic trend? Evaluating conflicting statistical results" by Kaufmann \textit{et al}., \textit{Climatic Change}, 101(3), 407-414.

Estrada, F., D. Kim and P. Perron (2021), Spatial variations in the warming trend and the transition to more severe weather in midlatitudes, \textit{Scientific Reports}, 11(145).

Estrada, F. P. Perron and B. Mart\'inez-L\'opez (2013), Statistically derived contributions of diverse human influences to twentieth-century temperature changes, \textit{Nature Geoscience}, 6(12), 1050-1055.

Fatichi, S., S.M. Barbosa, E. Caporali and M.E. Silva (2009), Deterministic versus stochastic trends: Detection and challenges, \textit{Journal of Geophysical Research}, 114, D18121.


Friedrich, M., S. Smeekes and J.-P. Urbain (2020), Autoregressive wild bootstrap inference for nonparametric trends, \textit{Journal of Econometrics}, 214(1), 81-109.

Friedrich, M., E. Beutner, H. Reuvers, S. Smeekes, J.-P. Urbain, W. Bader, B. Franco, B. Lejeune and E. Mahieu (2020), A statistical analysis of time trends in atmospheric ethane, \textit{Climatic Change}, 162, 105-125.


Gadea Rivas, M.D. and J. Gonzalo (2020), Trends in distributional characteristics: Existence of global warming, \textit{Journal of Econometrics}, 214, 153-174.

Gadea Rivas, M.D. and J. Gonzalo (2022), A tale of three cities: climate heterogeneity, \textit{SERIEs}, 13, 475-511.

Gadea-Rivas, M.D., J. Gonzalo and A. Ramos (2023), Trends in temperature data: Micro-foundations of their nature, arXiv:2312.06379.

Gao, J. and K. Hawthorne (2006), Semiparametric estimation and testing of the trend of temperature series, \textit{Econometrics Journal}, 9(2), 332-355.

Gao, J., O. Linton and B. Peng (2024), A non-parametric panel model for climate data with seasnal and spatial variation, \textit{Journal of the Royal Statistical Society. Series A-Statistics in Society}, 187(1), 158-177.

Gay, G.C., F. Estrada and A. S\'anchez (2009), Global and hemispheric temperatures revisited, \textit{Climatic Change}, 94(3-4), 333-349.



Gil-Alana, L.A. (2005), Statistical modelling of the temperatures in the Northern hemisphere using fractional integration techniques, \textit{Journal of Climate}, 18(24), 5357-5369.

Gil-Alana, L.A. (2008), Time-trend estimation with breaks in temperature time series, \textit{Climatic Change}, 89, 325-337.
 
 Gonz\'alez-Rivera, G., Y. Luo and E. Ruiz (2020), Prediction regions for interval-valued time series, \textit{Journal of Applied Econometrics}, 35(4), 373-390.

Good, S., G. Corlett, J. Remedios, E. Noyes and D. Llewellyn-Jones (2007), The global trend in sea surface temperature from 20 years of advanced very high resolution radiometer data, \textit{Journal of Climate}, 20, 1255-1264.


 Harris, I., T.J. Osborn, P. Jones and D. Lister (2020), Version 4 of the CRUTS monthly high-resolution gridded multivariate climate dataset, \textit{Scientific Data}, 7(109).

Harvey, A.C. (1989), \textit{Forecasting, Structural Time Series Models and the Kalman Filter}, Cambridge University Press, Cambridge.

Harvey, A.C. (1997), Trends, cycles and autoregression, \textit{Economic Journal}, 107, 192-201.

Harvey, A.C. (2001), Testing in unobserved components models, \textit{Journal of Forecasting}, 20(1), 1-19.


Harvey, A.C. and A. Jaeger (1993), Detrending, stylized facts and the business cycle, \textit{Journal of Applied Econometrics}, 8, 231-247.

Harvey, A.C. and T. Mills (2003), Modelling trends in central England temperatures, \textit{Journal of Forecasting}, 22, 35-47.

Hausfather, Z., M.J. Menne, C.N. Williams, T. Masters, R. Broberg and D. Jones (2013), Quantifying the effect of urbanization in U.S. historical climatology network temperature records, \textit{Journal of Geophysical Research: Atmospheres}, 118, 481-494.




Helske, J. (2017), KFAS: Exponential family state space models in R, \textit{Journal of Statistical Software}, 78(10).

 Hillebrand, E. and T. Proietti (2017), Phase changes and seasonal warming in early instrumental temperature records, \textit{Journal of Climate}, 30(17), 6795-6821.

Hindrayanto, I., J.A.D. Aston, S.J. Koopman and M. Ooms (2013), Modeling trigonometric seasonal components for monthly economic time series, \textit{Applied Economics}, 45(21), 3024-3034.

Hodrick R.J. and E.C. Prescott (1997), Postwar US business cycles: an empirical investigation, \textit{Journal of Money, Credit and Banking}, 24,1-16.

Holt, M.T. and T. Ter\"asvirta (2020), Global hemispheric temperatures and co-shifting: A vector shifting-mean autoregressive analysis, \textit{Journal of Econometrics}, 214(1), 198-215.

 Hsiang, S. and R.E. Koop (2018), An economist's guide to climate change science, \textit{Journal of Economic Perspectives}, 32, 3-32.


 IPCC (2014), AR5 Synthesis report: Climate Change 2014. Contribution of working groups I, II and III to the Fifth Assessment Report of the Intergovernmental Panel on Climate Change. [Core Writing Team, R.K. Pachauri and L.A. Meyer (eds.)]. IPCC, Geneva, Switzerland.

 IPCC (2023), AR6 Synthesis report: Climate Change 2022. Contribution of working groups I, II and III to the Sixth Assessment Report of the Intergovernmental Panel on Climate Change. [Core Writing Team, Aldunce, P. \textit{et al}.]. IPCC, Geneva, Switzerland.

Katz, R.W. and B.G. Brown (1992), Extreme events in a changing climate: Variability is more important than averages, \textit{Climatic Change}, 21, 289-302.

Kaufmann, R.K. and D.I. Stern (1997), Evidence from human influence on climate from hemispheric temperature relations, \textit{Nature}, 388(6637), 39-44.

 Kaufmann, R.K., H. Kauppi and J.H. Stock (2010), Does temperature contain a stochastic trend? Evaluating conflicting statistical results, \textit{Climatic Change}, 101, 395-405.

Kaufmann, R.K., H. Kauppi, M.L. Mann and J.H. Stock (2013), Does temperature contain a stochastic trend: linking statistical results to physical mechanisms, \textit{Climatic Change}, 118, 729-743.

Kaufmann, R.K., M.L. Mann, S. Gopal, J.A. Liederman, P.D. Howe, F. Pretis, X. Tang and M. Gilmore (2017), Spatial heterogeneity of climate change as an experimental basis for skepticism, \textit{Proceedings of the National Academy of Sciences, PNAS}, 114(1), 67-71.

Kew, S.F., S.Y. Philip, G. Jan van Oldenborgh, G. van der Schrier, F.E. Otto and R. Vautard (2019), The exceptional summer heat wave in Southern Europe 2017, \textit{Bulletin of the American Meteorological Society}, 100(1), S49-S53.

Kim, D., T. Oka, F. Estrada and P. Perron (2020), Inference related to common breaks in a multivariate system with joined segmented trends with applications to global and hemispheric temperatures, \textit{Journal of Econometrics}, 214, 130-152.

 Kruse-Becher, R. (2023), Adaptive now- and forecasting of global temperatures under smooth structural changes, mimeo.




Maia, A.L.S. and F.A.T. de Carvalho (2011), Holt's exponential smoothing and neural network models for forecasting interval-valued time series, \textit{International Journal of Forecasting}, 27(3), 740-759.

Mangat, M.K. and E. Reschenhofer (2020), Frequency-domain evidence for climate change, \textit{Econometrics}, 8(3), 28.

 Marotta, F. and H. Mumtaz (2023), Vulnerability to climate change: Evidence from a dynamic factor model, Oxford Smith School of Enterprise and the Environment Working Paper no. 23-06. 


Megdhaug, L., M.B. Stolpe, E.M. Fisher and R. Knutti (2017), Reconciling controversies about the ``global warming hiatus'', \textit{Nature}, 545(7652), 41-47.

Meng, X. and J.W. Taylor (2022), Comparing probabilistic forecasts of the daily minimum and maximum temperature, \textit{International Journal of Forecasting}, 38, 267-281.

 Miller, J.I. and K. Nam (2020), Dating Hiatuses: A statistical model of the recent slowdown in global warming and the next one, \textit{Earth System Dynamics}, 11, 1123-1132.


 Mitchell, T.D. and P.D. Jones (2005), An improved method of constructing a database of monthly climate observations and associated high-resolution grids, \textit{International Journal of Climatology}, 25(6), 693-712.


Nyblom, J. and A.C. Harvey (2000), Testing against smooth stochastic trends, \textit{Journal of Applied Econometrics}, 16(3), 415-429.


Palmer, T. and R. Stevens (2019), The scientific challenge of understanding and estimating climate change, \textit{Proceedings of the National Academy of Sciences}, 116(49), 24390-24395.

Pezzulli, S., D. Stephenson and A. Hannachi (2005), The variability of seasonality, \textit{Journal of Climate}, 18, 71-88.

Phella, A., V.J. Gabriel and L.F. Martins (2024), Predicting tail risks and the evolution of temperatures, \textit{Energy Economics}, 131, 107286.

Phillips, P.C.B. (2005), Challenges of trending time series econometrics, \textit{Mathematics and Computers in Simulation}, 68(5), 401-416.

Phillips, P.C.B. (2010), The mysteries of trend, \textit{Macroeconomic Reviews}, 9(2), 82-89.


Pretis, F. and M. Allen (2013), Breaks in trends, \textit{Nature Geoscience}, 6(12), 992-993.

Pretis, F. and D.F. Hendry (2013), Comment on ``Polynomial cointegration tests of anthropogenic impact on global warming'' by Beenstock \textit{et al}. (2012)-some hazards in econometric modelling of climate change, \textit{Earth System Dynamics}, 4, 375-384.

Pretis, F., M.L. Mann and R.K. Kaufmann (2015), Testing competing models of the temperature hiatus: assessing the effects of conditioning variables and temproal uncertainties through sample-wide break detection, \textit{Climatic Change}, 131, 705-718.



Proietti, T. and E. Hillebrand (2015), Seasonal changes in central England temperatures, \textit{Journal of the Royal Statistical Society, Series A (Statistics and Society)}, 180(3), 769-791.

Qu, M., J. Wan and X. Hao (2014), Analysis of diurnal air temperature range in the continental United States, \textit{Weather and Climate Extremes}, 4, 86-95.

Rao, B.B. (2010), Deterministic and stochastic trends in time series models: a guide for the applied economist, \textit{Applied Economics}, 42(17), 2193-2202.





Rodr\'iguez-Caballero, C. and M. Caporin (2019), A multilevel factor approach for the analysis of CDS commonality and risk contribution, \textit{Journal of International Financial Markets, Institutions \& Money}, 63, 101--144.

Schmidt, G.A., D.T. Shindell and K. Tsigaridis (2014), Reconciling warming trends, \textit{Nature Geoscience}, 7(3), 158-160.


Scotto, M.G., S.M. Barbosa and A.M. Alonso (2011), Extreme value and cluster analysis of European daily temperature series, \textit{Journal of Applied Statistics}, 38(12), 2793-2804.

Seidel, D.J. and J.R. Lanzante (2004), An assessment of three alternatives to linear trends for characterizing global atmospheric temperature changes, \textit{Journal of Geophysical Research: Atmospheres}, 109(D14).


Stern, D.I. and R.K. Kaufmann (2000), Detecting a global warming signal in hemispheric temperature series: A structural time series analysis, \textit{Climate Change}, 47, 411-438.


Ventosa-Santaulalia, D., D.R. Heres and L.C. Mart\'inez-Hern\'andez (2014), Long-memory and the sea-level temperature relationship: A fractional cointegration approach, \textit{PloS One}, 9(11), e113439.

Vera-Vald\'es, J.E. (2021), Temperature anomalies. Long-memory and aggregation, \textit{Econometrics}, 9(1), 9.

Visser, H. and J. Molenaar (1995), Trend estimation and regression analysis in climatological time series: An application of structural time series models and the Kalman filter, \textit{Journal of Climate}, 8(5), 969-979.

Vogelsang, T.J. and P.H. Franses (2005), Are winters getting warmer?, \textit{Environmental Modelling Software}, 20, 1449-1455.

Vose, R.S., D.R. Easterling and B. Gleason (2005), Maximum and minimum temperature trends for the globe: An update through 2004, \textit{Geophysical Research Letters}, 32, L23822.


 Wang, Z., Y. Jiang, H. Wan, J. Yan and X. Zhang (2021), Toward optimal fingerprint in detection and attribution of changes in climate extremes, \textit{Journal of the American Statistical Association}, 116(533), 1-13.

Wijngaard, J.B., A.M.G. Klein Tauk and G.P. K\"onnen (2003), Homogeneity of 20th century European daily temperature and precipitations series, \textit{International Journal of Climatology}, 23, 679-692.

Woodward, W.H. and H.L. Gray (1993), Global warming and the problem of testing for trend in time series data, \textit{Journal of Climate}, 6, 953-962. 




Xu, W., Q. Li, X.L. Wang, S. Yao, L. Cao and Y. Feng (2013), Homogenization of Chinese daily surface air temperatures and analysis of trends in the extreme temperature indices, \textit{Journal of Geophysical Research: Atmospheres}, 118(17), 9708-9720.

Zaval, L. E.A. Keenan, E.J. Johnson and E.U. Weber (2014), How warm days increase belief in global warming, \textit{Nature Climate Change}, 4(2), 143-147.


Zheng, X. and R.E. Basher (1999), Structural time series models and trend detection in global and regional temperature series, \textit{Journal of Climate}, 12, 2347-2358.

\appendix
\section*{\large{Supplementary Material}}
\addcontentsline{toc}{section}{Supplementary material}
\renewcommand{\thesubsection}{\Alph{subsection}}


\subsection{STATE SPACE MODEL}
\label{appendix:SSM}

\renewcommand{\theequation}{A.\arabic{equation}}
\renewcommand{\thetable}{A.\arabic{table}}
\renewcommand{\thefigure}{A.\arabic{figure}}

\setcounter{equation}{0}
\setcounter{table}{0}
\setcounter{figure}{0}

Consider model \eqref{SSM} for the centre and log-range temperature, $X_t=\left[C_t, R_t \right] $. This model can be written in state space form as follows
\begin{subequations}
	\label{Model}
	\begin{align}
		X_t & =Z\alpha_t+\varepsilon_t, \label{Model1}\\
		\alpha_t & =T\alpha_{t-1}+\eta_t,\label{Model2}
	\end{align}
\end{subequations}
where $\alpha_t=\left( \mu_{1t}, \mu_{2t}, \beta_{1t}, \beta_{2t}, \gamma^{(1)}_{1t}, \gamma^{(1)}_{2t}, \gamma^{*(1)}_{1t}, \gamma^{*(1)}_{2t}, \gamma^{(2)}_{1t}, \gamma^{(2)}_{2t}, \gamma^{*(2)}_{1t}, \gamma^{*(2)}_{2t}..., \gamma^{(6)}_{1t}, \gamma^{(6)}_{2t} \right)^{\prime}$ is the $26 \times 1$ vector of unobserved states. The $2 \times 26$ observation matrix $Z$ is given by
\begin{equation}
	\label{Z}
	Z=\left[ \begin{smallmatrix*}[r]
		1&0&0&0&-1&0&0&0&-1&0&0&0&-1&0&0&0&-1&0&0&0&-1&0&0&0&-1&0\\
		0&1&0&0&0&-1&0&0&0&-1&0&0&0&-1&0&0&0&-1&0&0&0&-1&0&0&0&-1
	\end{smallmatrix*}\right], 
\end{equation}
while the $26 \times 26$ transition matrix $T$ is given by
\begin{equation}
	\label{T}
	T=\left[ \begin{smallmatrix}
		1&0&1&0&0&0&0&0&0&0&0&0&\cdots&0&0\\
		0&1&0&1&0&0&0&0&0&0&0&0&\cdots&0&0\\
		0&0&1&0&0&0&0&0&0&0&0&0&\cdots&0&0\\
		0&0&0&1&0&0&0&0&0&0&0&0&\cdots&0&0\\
		0&0&0&0&cos\lambda_1&0&sin\lambda_1&0&0&0&0&0&\cdots&0&0\\
		0&0&0&0&0&cos\lambda_1&0&sin \lambda_1&0&0&0&0&\cdots&0&0\\
		0&0&0&0&-sin\lambda_1&0&cos\lambda_1&0&0&0&0&0&\cdots&0&0\\
		0&0&0&0&0&-sin\lambda_1&0&cos\lambda_1&0&0&0&0&\cdots&0&0\\
		0&0&0&0&0&0&0&0&cos\lambda_2&0&sin\lambda_2&0&\cdots&0&0\\
		0&0&0&0&0&0&0&0&0&cos\lambda_2&0&sin \lambda_2&\cdots&0&0\\
		0&0&0&0&0&0&0&0&-sin\lambda_2&0&cos\lambda_2&0&\cdots&0&0\\
		0&0&0&0&0&0&0&0&0&-sin\lambda_2&0&cos\lambda_2&\cdots&0&0\\
		\vdots & \vdots &\vdots &\vdots &\vdots &\vdots &\vdots &\vdots &\vdots &\vdots &\vdots &\vdots &\ddots & \vdots & \vdots \\
		0&0&0&0&0&0&0&0&0&0&0&0&\cdots&cos\lambda_6&0\\
		0&0&0&0&0&0&0&0&0&0&0&0&\cdots&0&cos\lambda_6
	\end{smallmatrix} \right].
\end{equation}
Finally, the covariance matrix of $\varepsilon_t$ is assumed to be constant over time and given by
\begin{equation}
	\label{H}
	H=\begin{bmatrix}
		\sigma^2_{1\varepsilon} & \sigma_{12\varepsilon}\\
		\sigma_{12\varepsilon} & \sigma^2_{2\varepsilon},
	\end{bmatrix}
\end{equation}
while the covariance matrix of $\eta_t$ is given by
\begin{equation}
	\label{Q}
	Q=\left[ \begin{smallmatrix}
		\sigma^2_{\nu1}&\sigma_{\nu12}&0&0&0&0&0&0&0&0&0&0&\cdots&0&0\\
		\sigma_{\nu12}&\sigma^2_{\nu2}&0&0&0&0&0&0&0&0&0&0&\cdots&0&0\\
		0&0&\sigma^2_{\zeta1}&\sigma_{\zeta12}&0&0&0&0&0&0&0&0&\cdots&0&0\\
		0&0&\sigma_{\zeta12}&\sigma^2_{\zeta1}&0&0&0&0&0&0&0&0&\cdots&0&0\\
		0&0&0&0&\sigma^2_{\omega1}&\sigma_{\omega12}&0&0&0&0&0&0&\cdots&0&0\\
		0&0&0&0&\sigma_{\omega12}&\sigma^2_{\omega2}&0&0&0&0&0&0&\cdots&0&0\\
		0&0&0&0&0&0&\sigma^2_{\omega1}&\sigma_{\omega12}&0&0&0&0&\cdots&0&0\\
		0&0&0&0&0&0&\sigma_{\omega12}&\sigma^2_{\omega2}&0&0&0&0&\cdots&0&0\\
		0&0&0&0&0&0&0&0&\sigma^2_{\omega1}&\sigma_{\omega12}&0&0&\cdots&0&0\\
		0&0&0&0&0&0&0&0&\sigma_{\omega12}&\sigma^2_{\omega2}&0&0&\cdots&0&0\\
		0&0&0&0&0&0&0&0&0&0&\sigma^2_{\omega1}&\sigma_{\omega12}&\cdots&0&0\\
		0&0&0&0&0&0&0&0&0&0&\sigma_{\omega12}&\sigma^2_{\omega2}&\cdots&0&0\\
		\vdots & \vdots &\vdots &\vdots &\vdots &\vdots &\vdots &\vdots &\vdots &\vdots &\vdots &\vdots &\ddots & \vdots & \vdots \\
		0&0&0&0&0&0&0&0&0&0&0&0&\cdots&\sigma^2_{\omega1}&\sigma_{\omega12}\\
		0&0&0&0&0&0&0&0&0&0&0&0&\cdots&\sigma_{\omega12}&\sigma^2_{\omega2}
	\end{smallmatrix}\right].
\end{equation}

Consider now the DFM for the deseasonalised centre temperature observed at $N$ locations, denoted as $C_t=\left(c_{1t},...,c_{Nt} \right) ^{\prime}$. Assume that the centres are grouped per region with $c_{1t},...,c_{N_1t}$ being the centres at time $t$ in the first region, $c_{N_1+1t},...,c_{N_1+N_2t}$ being those corresponding to the second region, $c_{N_1+N_2+1t},...,c_{N_1+N_2+N_3t}$ being the centres corresponding to the third region, and $c_{N_1+N_2+N_3+1t},...,c_{N_1+N_2+N_3+N_4t}$ and\newline $c_{N_1+N_2+N_3+N_4+1t},...,c_{N_1+N_2+N_3+N_4+1,...,c_{Nt}}$ being the centres in the fourth and fifth regions, respectively. Assuming also that global common factor, $F_{gt}$, is an integrated random walk, the regional factors, $F_{s1t},...,F_{s5t}$, are stationary AR(1) processes, and the idiosyncratic noises, $\varepsilon_t=\left(\varepsilon_{1t},...,\varepsilon_{Nt} \right)^{\prime}$, are cross-sectionally uncorrelated white noises, the state space form of the multi-level DFM is given by
\begin{align}
	\label{eq:centermeasurement}
	\left[ \begin{smallmatrix}
		c_{1t}\\
		c_{2t}\\
		\vdots \\
		c_{N_1t} \\
		c_{N_1+1t}\\
		c_{N_1+2t}\\
		\vdots\\
		c_{N_1+N_2t}\\
		c_{N_1+N_2+1t}\\
		c_{N_1+N_2+2t}\\
		\vdots\\
		c_{N_1+N_2+N_3t}\\
		c_{N_1+N_2+N_3+1t}\\
		c_{N_1+N_2+N_3+2t}\\
		\vdots\\
		c_{N_1+N_2+N_3+N_4t}\\
		c_{N_1+N_2+N_3+N_4+1t}\\
		c_{N_1+N_2+N_3+N_4+2t}\\
		\vdots\\
		c_{Nt}\\
	\end{smallmatrix} \right] =\left[ \begin{smallmatrix}
		p_{11}& 0 & p_{21}&0&0&0&0\\
		p_{12}&0&p_{22}&0&0&0&0\\
		\vdots & \vdots & \vdots & \vdots & \vdots & \vdots & \vdots\\
		p_{1N_1}& 0& p_{2N_1}& 0 & 0 & 0 & 0\\
		p_{1N_1+1} &0& 0 & p_{31}&0&0&0\\
		p_{1N_1+2}&0& 0 & p_{32}&0&0&0\\
		\vdots & \vdots & \vdots & \vdots & \vdots & \vdots & \vdots\\
		p_{1N_1+N_2}& 0 &0 & p_{3N_2}&0&0&0\\
		p_{1N_1+N_2+1} & 0&0 & 0 & p_{41} & 0 & 0\\
		p_{1N_1+N_2+2} & 0 &0& 0 & p_{42} & 0 & 0\\
		\vdots & \vdots & \vdots & \vdots & \vdots & \vdots & \vdots\\
		p_{1N_1+N_2+N_3} & 0 & 0& 0& p_{4N_3} & 0 & 0\\
		p_{1N_1+N_2+N_3+1} & 0 & 0& 0 & 0 & p_{51} & 0\\
		p_{1N_1+N_2+N_3+2} & 0 & 0& 0 & 0 & p_{52} & 0\\
		\vdots & \vdots & \vdots & \vdots & \vdots & \vdots & \vdots\\
		p_{1N_1+N_2+N_3+N_4} & 0 &0 &  0 & 0 & p_{5N_4} & 0\\
		p_{1N_1+N_2+N_3+N_4+1} & 0&  0 &  0 & 0 & 0 & p_{61} \\
		p_{1N_1+N_2+N_3+N_4+2} & 0 & 0&  0 & 0 & 0 & p_{62} \\
		\vdots & \vdots & \vdots & \vdots & \vdots & \vdots & \vdots\\
		p_{1N} & 0 & 0& 0 & 0 & 0 & p_{6N_5} \\
	\end{smallmatrix}\right] \left[\begin{smallmatrix}
		F_{gt} \\
		\beta_t\\
		F_{s1t} \\
		F_{s2t} \\
		F_{s3t} \\
		F_{s4t} \\
		F_{s5t}
	\end{smallmatrix} \right] + 
	\varepsilon_{t},
\end{align}
where $\varepsilon_{t}$ has diagonal covariance matrix $\Sigma_{\varepsilon}= dig \left( \sigma^2_{\varepsilon1},...,\sigma^2_{\varepsilon N} \right)$. Finally, the transition equation is given by
\begin{equation}
	\label{eq:centertransition}
	\left[\begin{smallmatrix}
		F_{gt} \\
		\beta_{t}\\
		F_{s1t} \\
		F_{s2t} \\
		F_{s3t} \\
		F_{s4t} \\
		F_{s5t}
	\end{smallmatrix} \right]=  \left[ \begin{smallmatrix}
		1 & 1 & 0 & 0 & 0 & 0 & 0 \\
		0 & 1 & 0 & 0 & 0 & 0 & 0 \\
		0 & 0 & \phi_1 & 0 & 0 & 0 & 0 \\
		0 & 0 & 0 & \phi_2 & 0 & 0 & 0 \\
		0 & 0 & 0 & 0 & \phi_3 & 0 & 0 \\
		0 & 0 & 0 & 0 & 0 & \phi_4 & 0 \\
		0 & 0 & 0 & 0 & 0 & 0 & \phi_5 \\
	\end{smallmatrix} \right] \left[\begin{smallmatrix}
		F_{gt-1} \\
		\beta_{t-1}\\
		F_{s1t-1} \\
		F_{s2t-1} \\
		F_{s3t-1} \\
		F_{s4t-1} \\
		F_{s5t-1}
	\end{smallmatrix} \right] +  \eta_t,
\end{equation} 
where $\eta_t=\left( 0, \xi_t, \eta_{1t},...,\eta_{5t} \right)^{\prime}$, with covariance matrix $ \Sigma_{\eta} = diag \left[ 0, \sigma^2_{\xi}, \sigma^2_{\eta 1},..., \sigma^2_{\eta 5}\right]$.

If the parameters in the state space model above were known, one can run the Kalman filter to estimate the factors. However, these parameters are unknown and should be estimated. Inspired by the two-step procedure of Doz, Giannone and Reichlin (2011), we propose to estimate them using PC to extract the factors from the multilevel DFM. Note that, given that the regional factors and the idiosyncratic components are stationary, it does not need to differentiate the original observations to estimate using PC. The procedure is as follows. First, the first PC is extracted from the original system of centres, obtaining, $\hat{F}_{gt}^{(PC)}$ and its corresponding loadings, $\hat{p}_{1i}^{(PC)}, i=1,...,N$. Then, using the residuals obtained by subtracting the first common component, $\hat{u}_{it}^{(PC)}=c_{it}-\hat{p}_{1i}^{(PC)} \hat{F}_{gt}^{(PC)}$, the regional factors and their loadings are estimated, $\hat{F}^{(PC)}_{sjt}$ and $\hat{p}_{j+1i}^{(PC)}$, for $j=1,...5$ and $i=1,...,N_j$. Finally, one can also obtain estimates of the idiosyncratic components by subtracting the common component, which for the centres in region $j$ are given by $\hat{\varepsilon}_{it}^{(PC)}=c_{it}-\hat{p}_{1i}^{(PC)} \hat{F}_{gt}^{(PC)}-\hat{p}^{(PC)}_{j+1i}\hat{F}_{sjt}^{(PC)}$. Then, the loadings in the measurement equation in (\ref{eq:centermeasurement}) are substituted by the corresponding PC estimates $\hat{p}_{ji}^{(PC)}$, while the idiosyncratic variances in the main diagonal of $\Sigma_{\varepsilon}$ are estimated by $\hat{\sigma}^2_{\varepsilon i}=\frac{1}{T} \sum_{t=1}^T \left( \hat{\varepsilon}_{it} - \bar{\hat{\varepsilon}}_{i} \right)^2$, where $\bar{\hat{\varepsilon}}_{i}=\frac{1}{T} \sum_{t=1}^T\hat{\varepsilon}_{it}$. Finally, the autoregressive parameters and variances of the transition equation in (\ref{eq:centertransition}) corresponding to the regional factors are estimated by regressing $\hat{F}^{(PC)}_{sjt}$ on a constant and $\hat{F}^{(PC)}_{sjt-1}$, obtaining, $\hat{\phi}_j$ and $\hat{\sigma}^2_{\eta j}$, while $\hat{\sigma}^2_{\xi}$ is the sample variance of $\triangle ^2 \hat{F}_{gt}^{(PC)}$.  
\newpage

\subsection{Joint modelling of centre and log-range temperature at selected locations of the Iberian Peninsula}

\renewcommand{\theequation}{B.\arabic{equation}}
\renewcommand{\thetable}{B.\arabic{table}}
\renewcommand{\thefigure}{B.\arabic{figure}}

\setcounter{equation}{0}
\setcounter{table}{0}
\setcounter{figure}{0}

Table \ref{tab:estimation_1} reports the estimated parameter when the state space model (\ref{Model1}) is estimated to the system of centre/log-range temperature at each location.

\begin{table}[ht!]
	\caption{Estimation results of the joint state space model fitted to maximum and minimum temperatures in four locations in the Iberian Peninsula.}
	\label{tab:estimation_1}
	\begin{center}
		\resizebox{460pt}{!}{
			\begin{tabular}{l c c c c c c c c c c }
				\hline
				& Center & Log-range & Cov & Corr &&& Center & Log-range & Cov & Corr\\
				\hline
				&\multicolumn{4}{c}{Barcelona} &&& \multicolumn{4}{c}{Coru\~{n}a}\\
				\hline
				Measurement & 1.302 & 0.005 & 0.015 & 0.186 &&& 1.096 & 0.017 & 0.021 & 0.154 \\
				Level & $9.29\times 10^{-23}$ & $2.48 \times 10^{-5}$ & $1.21 \times 10^{-16}$ & 0.003 &&& $7.12 \times 10^{-14}$ & $5.67 \times 10^{-5}$ & $-5.65 \times 10^{-13}$ & -0.000\\
				Slope & $1.92 \times 10^{-36}$ & $1.60 \times 10^{-10}$ & $1.73 \times 10^{-23}$ & 0.987 &&& $8.33 \times 10^{-8}$ & $1.40 \times 10^{-11}$ & $-6.11 \times 10^{-10}$ & -0.566\\
				Seasonal 1 & 1.67 $\times 10^{-4}$ & $1.16 \times 10^{-6}$ & 1.62 $\times 10^{-7}$ & 0.012 &&& $3.99 \times 10^{-7}$ & $3.75 \times 10^{-6}$ & $-1.34 \times 10^{-9}$ & -0.001\\
				Seasonal 2 & 1.47 $\times 10^{-40}$ & $1.57 \times 10^{-10}$ & $1.52 \times 10^{-25}$ & 1.000 &&& $1.50 \times 10^{-44}$ & $4.48 \times 10^{-12}$ & $-2.59 \times 10^{-28}$ & -0.999\\
				\hline
				& \multicolumn{4}{c}{Madrid} &&& \multicolumn{4}{c}{Seville}\\
				\hline
				Measurement & 1.062 & 0.006 & 0.011 & 0.138 &&& 0.979 & 0.006 & 0.004 & 0.052\\
				Level & $1.47\times 10^{-19}$ & $1.60\times 10^{-5}$ & $2.81 \times 10^{-16}$ & 0.004 &&& $1.47 \times 10^{-19}$ & $3.39 \times 10^{-5}$ & $1.07 \times 10^{-14}$ & 0.005 \\
				Slope & $9.99\times 10^{-8}$ & $2.78 \times 10^{-10}$ & $5.19 \times 10^{-9}$ & 0.98 &&& $1.51 \times 10^{-7}$ & $7.89 \times 10^{-10}$ & $1.09 \times 10^{-8}$ & 0.999\\
				Seasonal 1 & $4.25\times 10^{-4}$ & $2.07 \times 10^{-6}$ & $3.39 \times 10^{-7}$ & 0.011 &&& $3.10 \times 10^{-4}$ & $1.45 \times 10^{-6}$ & $4.91 \times 10^{-7}$ & 0.023\\
				Seasonal 2 & $3.29\times 10^{-33}$ & $2.70 \times 10^{-10}$ & $9.42 \times 10^{-22}$ & 0.999 &&& $4.45 \times 10^{-36}$ & $7.78 \times 10^{-10}$ & $5.89 \times 10^{-23}$ & 1.002
			\end{tabular}
		}
	\end{center}
\end{table}

\begin{figure}[]
	\begin{center}
		\begin{subfigure}[h]{0.45\textwidth}
			\includegraphics[width=0.85\textwidth]{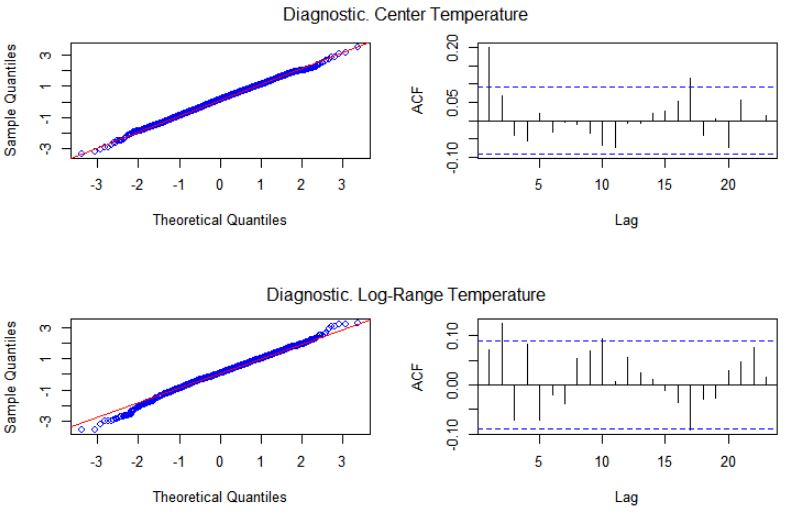}
			\caption{Barcelona}
		\end{subfigure}
		\begin{subfigure}[h]{0.45\textwidth}
			\includegraphics[width=0.85\textwidth]{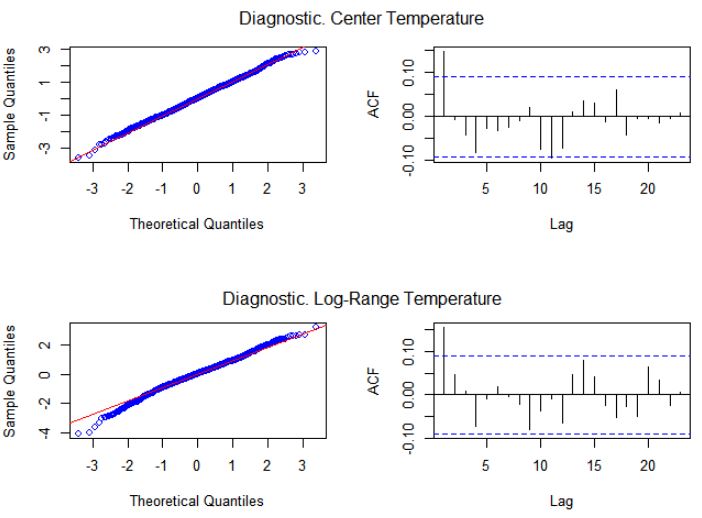}
			\caption{Coru\~{n}a}
		\end{subfigure}
		\begin{subfigure}[h]{0.45\textwidth}
			\includegraphics[width=0.85\textwidth]{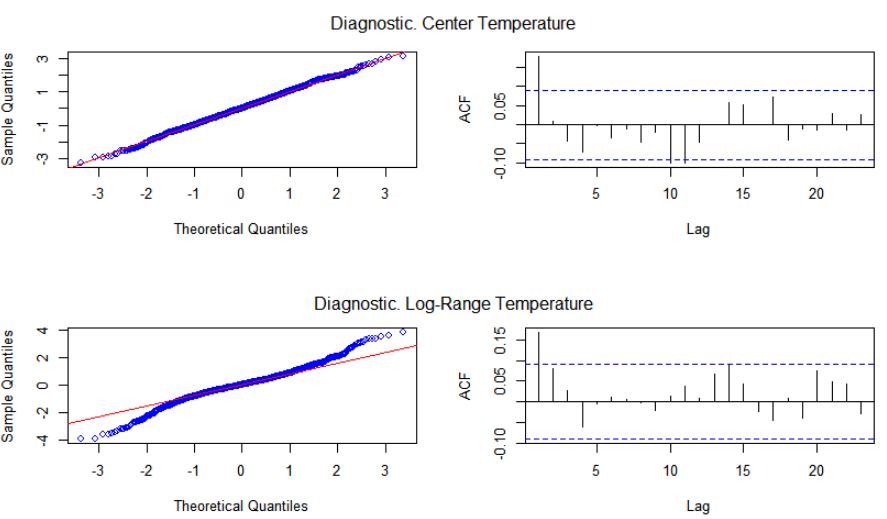}
			\caption{Madrid}
		\end{subfigure}
		\begin{subfigure}[h]{0.45\textwidth}
			\includegraphics[width=0.85\textwidth]{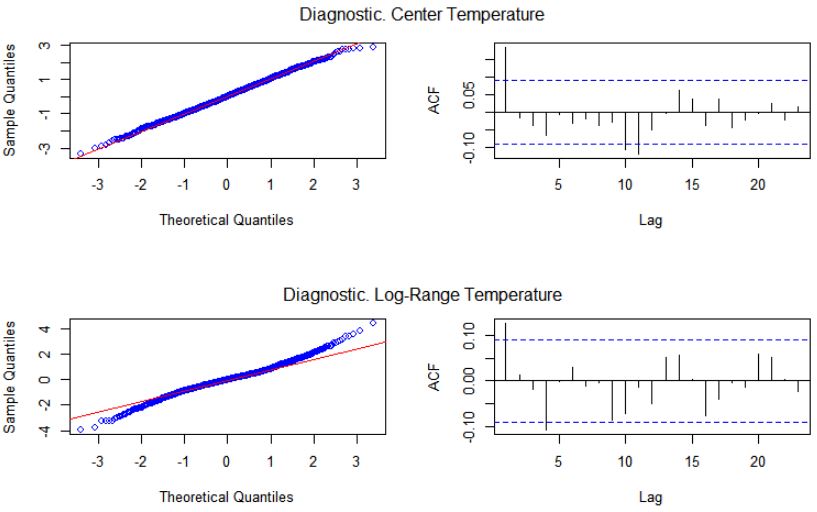}
			\caption{Seville}
		\end{subfigure}
		\caption{Diagnostics of standardized residuals of the SSM model for center (top row) and log-range (bottom row) temperature: Q-Q plot (left column) and autocorrelations (right column).}
		\label{fig:Diag}
	\end{center}
\end{figure}

The estimates reported in Table \ref{tab:estimation_1} show that, in Coru\~{n}a, Madrid and Seville, $\hat{\sigma}^2_{\eta_1}$ is approximately zero while $\hat{\sigma}^2_{\xi_1}$ is small. Consequently, the center temperature in these three locations can be well represented by a Integrated Random Walk according to which the trend is smoothly evolving over time. The behaviour of the trend of the centre temperature is slightly different in Barcelona with the estimates of both variances being close to zero implying a deterministic trend. Furthermore, Table \ref{tab:estimation_1} shows that $\hat{\sigma}^2_{\omega^{(1)}_1}$ is rather small while $\hat{\sigma}^2_{\omega^{(2)}_1}$ is approximately zero, implying a stochastic evolution of the seasonal pattern of the centre temperature. Figure \ref{fig:Diag}, which plots the QQ plots and estimated autocorrelations of the corresponding standardised residuals, does not show any sign of misspecification.

When looking at the results corresponding to the log-range temperature, Table \ref{tab:estimation_1} shows that, regardless of the location, all the estimated variances, although small, imply non-negligible signal-to-noise ratios. Therefore, both the trend and seasonal components of log-range temperature are stochastic. The diagnosis of the standardised residuals show that, although mostly uncorrelated, they are characterised by a distribution with heavy tails. 

Finally, a final remarkable result observed in Table \ref{tab:estimation_1} is that the estimated covariances between the shocks of the centre and log-range temperature imply correlations close to zero, when the variances are different from zero. An important consequence of this result is that the centre and log-range temperature can be modelled separately.

\clearpage
\end{document}